\documentclass[aps,prx,superscriptaddress,showpacs,reprint]{revtex4-1}
\usepackage{amsmath}
\usepackage{amssymb}
\usepackage{setspace}
\usepackage{graphicx}
\usepackage{subfigure}
\usepackage{braket}
\usepackage{mathrsfs}
\usepackage[colorlinks = true,linkcolor = red,urlcolor  = blue,citecolor = blue,anchorcolor = blue]{hyperref}

\newcommand{\nn}{\nonumber \\}
\newcommand{\angstrom}{\text{\normalfont\AA}}
\begin{document}

%%%%%%%%%%%%%%%%%%%%%%%%%%%%
\title{Andreev bound states versus Majorana bound states in quantum
dot-nanowire-superconductor hybrid structures: Trivial versus topological zero-bias conductance peaks}
\author{Chun-Xiao Liu}
\affiliation{Condensed Matter Theory Center and Joint Quantum Institute and Station Q Maryland, 
Department of Physics, University of Maryland, College Park, Maryland 20742-4111, USA}

\author{Jay D. Sau}
\affiliation{Condensed Matter Theory Center and Joint Quantum Institute and Station Q Maryland,
Department of Physics, University of Maryland, College Park, Maryland 20742-4111, USA}

\author{Tudor D. Stanescu}
\affiliation{Condensed Matter Theory Center and Joint Quantum Institute and Station Q Maryland,
Department of Physics, University of Maryland, College Park, Maryland 20742-4111, USA}
\affiliation{Department of Physics and Astronomy, West Virginia University, Morgantown, WV 26506}

\author{S. Das Sarma}
\affiliation{Condensed Matter Theory Center and Joint Quantum Institute and Station Q Maryland,
Department of Physics, University of Maryland, College Park, Maryland 20742-4111, USA}

\date{\today}

\begin{abstract}
Motivated by an important recent experiment [Deng \textit{et al}., \href{http://science.sciencemag.org/content/354/6319/1557}{Science $\bold{354}$, 1557} (2016)], we theoretically consider the interplay between Andreev and Majorana bound states in disorder-free quantum dot-nanowire semiconductor systems with proximity-induced superconductivity in the presence of spin-orbit coupling and Zeeman spin splitting (induced by an external magnetic field).  The quantum dot induces Andreev bound states in the superconducting nanowire which show complex behavior as a function of magnetic field and chemical potential, and the specific question is whether two such Andreev bound states can come together forming a robust zero-energy topological Majorana bound state. We find generically that the Andreev bound states indeed have a high probability of coalescing together producing near-zero-energy midgap states as Zeeman splitting and/or chemical potential are increased, but this mostly happens in the nontopological regime below the topological quantum phase transition although there are situations where the Andreev bound states could indeed come together to form a zero-energy topological Majorana bound state. The two scenarios (two Andreev bound states coming together to form a nontopological almost-zero-energy Andreev bound state or to form a topological zero-energy Majorana bound state) are difficult to distinguish just by tunneling conductance spectroscopy since they produce essentially the same tunneling transport signatures. We find that the ``sticking together" propensity of Andreev bound states  to produce an apparent stable zero-energy midgap state is generic in class D systems in the presence of superconductivity, spin-orbit coupling, and magnetic field, even in the absence of any disorder. We also find that the conductance associated with the coalesced zero-energy nontopological Andreev bound state is non-universal and could easily be $2e^2/h$ mimicking the quantized topological Majorana zero-bias conductance value. We suggest experimental techniques for distinguishing between trivial and topological zero-bias conductance peaks arising from the coalescence of Andreev bound states.
\end{abstract}

\maketitle

%%%%%%%%%%%%%%%%%%%%%%%%%%%%

\section{introduction}\label{sec:introduction}
The great deal of current interest~\cite{Nayak2008Non-Abelian, DasSarma2015Majorana, Alicea2012New, Elliott2015Colloquium, Stanescu2013Majorana, Leijnse2012Introduction, Beenakker2013Search} in Majorana zero modes (MZMs) or Majorana fermions focusing on semiconductor-superconductor hybrid structures~\cite{Sau2010Generic, Lutchyn2010Majorana, Oreg2010Helical, Sau2010Non} arises mainly from the significant experimental progress~\cite{Mourik2012Signatures, Das2012Zero, Deng2012Anomalous, Churchill2013Superconductor, Finck2013Anomalous, Albrecht2016Exponential, Chen2016Experimental, Zhang2016Ballistic, Deng2016Majorana} made in the subject during the last five years.  In particular, proximity-induced superconductivity in spin-orbit-coupled semiconductor nanowires can become topological with localized MZMs in the wire if the system has a sufficiently large Zeeman spin splitting overcoming the induced superconducting gap.  Such MZMs, being zero-energy midgap states, should produce quantized zero-bias conductance peaks (ZBCPs) associated with perfect Andreev reflection in tunneling measurements~\cite{Sengupta2001Midgap, Akhmerov2009Electrically, Law2009Majorana, Flensberg2010Tunneling}. Indeed, experimentally many groups have observed such zero-bias conductance peaks in tunneling measurements on nanowire-superconductor hybrid structures although the predicted precise and robust quantization (with a conductance value $2e^2/h$) has been elusive. Many reasons have been provided to explain the lack of precise ZBCP quantization~\cite{Lin2012Zero, DasSarma2016How, Liu2017Role}, but alternative scenarios, not connected with MZMs, for the emergence of the ZBCP have also been discussed in the literature~\cite{Lee2012Zero, Liu2012Zero, Bagrets2012Class, Pikulin2012Zero, Lee2014Spin}. Whether the experimentally observed ZBCPs in semiconductor-superconductor hybrid structures arise from MZMs or not remains a central question in spite of numerous publications and great experimental progress in the subject during the 2012-2017 five-year period.

\begin{figure}[h]
\includegraphics[width=0.47\textwidth]{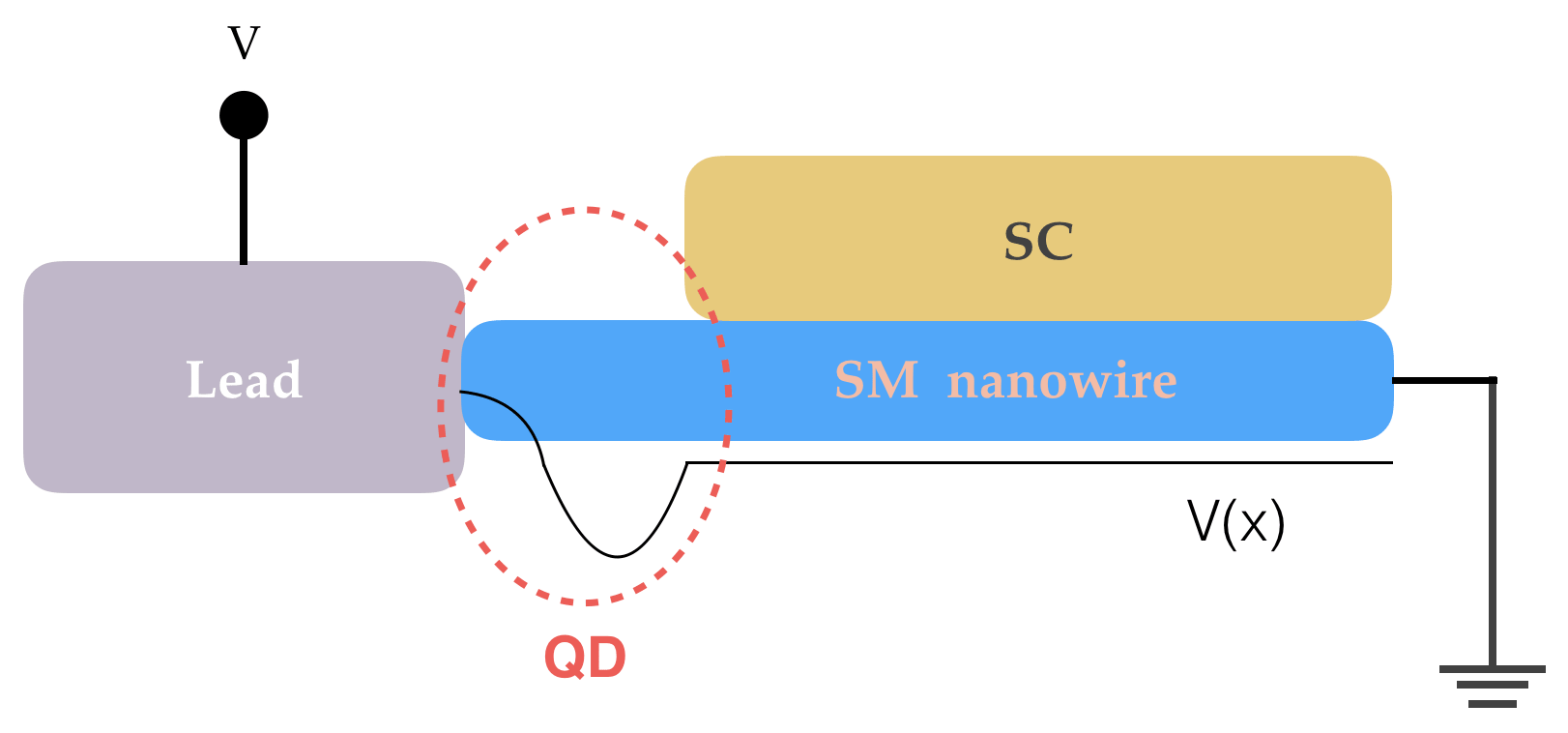}
\caption{(color online). A schematic plot of the junction composed of lead and quantum dot-nanowire-superconductor hybrid structure, which represents the actual system setup in Deng \textit{et al}. experiment~\cite{Deng2016Majorana}. A semiconductor (SM) nanowire is mostly covered by a parent $s$-wave superconductor (SC). One fraction of the nanowire is not covered by the superconductor and is subject to a confinement potential. This part (encircled by the red dash line) between the lead and the superconducting nanowire is called quantum dot (QD) in this paper. Figs.~\ref{fig:summary}-\ref{fig:proximitized} are results based on this configuration. Later we also consider situations where a part or the whole of the dot is covered by the superconductor making the whole hybrid structure superconducting. Note that the quantum dot here is strongly coupled to the nanowire and may not exhibit any Coulomb blockade behavior.}
\label{fig:schematic}
\end{figure}

A key experimental paper by Deng \textit{et al}. has recently appeared in the context of ZBCPs in semiconductor-superconductor hybrid systems~\cite{Deng2016Majorana}, which forms the entire motivation for the current theoretical work. In their work, Deng \textit{et al}. studied tunneling transport through a hybrid system composed of a quantum dot-nanowire-superconductor, where no superconductivity (SC) is induced in the quantum dot (i.e., the superconductivity is induced only in the nanowire). In Fig.~\ref{fig:schematic}, we provide a schematic of the experimental system, where the dot simply introduces a confining potential at one end of the nanowire which is covered by the superconductor to induce the proximity effect. Such a quantum dot may naturally be expected to arise because of the Fermi energy mismatch of the lead and the semiconductor much in the way a Schottky barrier arises in semiconductors. Reducing the potential barrier at the lead-semiconductor interface to produce a strong conductance signature likely requires the creation of a quantum dot as shown in Fig.~\ref{fig:schematic}. Thus a quantum dot might be rather generic in conductance measurements, i.e., one may not have to introduce a real quantum dot in the system although such a dot did exist in the set-up of Ref.~\cite{Deng2016Majorana}. The quantum dot may introduce Andreev bound states (ABSs) in the nanowire, and the specific issue studied in depth by Deng \textit{et al}. is to investigate how these Andreev bound states behave as one tunes the Zeeman spin splitting and the chemical potential in the nanowire by applying a magnetic field and a gate potential respectively. It is also possible that the ABSs in the Deng \textit{et al}. experiment arise from some other potential fluctuations in the nanowire itself which is akin to having quantum dots inside the nanowire arising from uncontrolled potential fluctuations associated with impurities or inhomogeneities.  (We consider both cases, the dot being outside or inside the nanowire, in this work.) The particular experimental discovery made by Deng \textit{et al}., which we theoretically examine in depth, is that Andreev bound states may sometimes come together with increasing Zeeman splitting (i.e., with increasing magnetic field) to coalesce and form zero-energy states which then remain zero-energy states over a large range of the applied magnetic field, producing impressive ZBCPs with relatively large conductance values $\sim0.5e^2/h$. Deng \textit{et al}. speculate that the resulting ZBCP formed by the coalescing ABSs is a direct signature of MZMs, or in other words, the ABSs are transmuting into MZMs as they coalesce and stick together at zero energy. It is interesting and important to note that the sticking together property of the ABSs at zero energy depends crucially on the gate voltage in Deng \textit{et al}. experiment, and for some gate voltage, the ABSs repel away from each other without coalescing at zero energy and at still other gate voltages, the ABSs may come together at some specific magnetic field, but then they separate out again with increasing magnetic field producing a beating pattern in the conductance around zero bias. Our goal in the current work is to provide a detailed description of what may be transpiring in the Deng \textit{et al}. experiment within a minimal model of the dot-nanowire-superconductor structure elucidating the underlying physics of ABS versus MZM in this system. In addition, we consider situations where the quantum dot is, in fact, partially (or completely) inside the nanowire (i.e., the dot itself is totally or partially superconducting due to proximity effect), which may be distinct from the situation in Deng \textit{et al}. experiment~\cite{Deng2016Majorana} where the quantum dot is not likely to be proximitized by the superconductor although any potential inhomogeneity inside the wire would act like a quantum dot in general for our purpose. Specific details of how the ABSs arise in the nanowire are not important for our theory as most of the important new qualitative features we find are generic as long as ABSs are present in the nanowire.

It may be important here to precisely state what we mean by a ``quantum dot'' in the context of our theory and calculations. The ``quantum dot'' for us is simply a potential fluctuation somewhere in or near the wire which produces Andreev bound states in the system.  This ``quantum dot'', being strongly coupled to the nanowire (perhaps even being completely inside the nanowire or arising from the Schottky barrier at the tunnel junction), does not have to manifest any Coulomb blockade as ordinary isolated quantum dots do. In fact, our theory does not include any Coulomb blockade effects because the physics of ABS transmuting into MZM or not is independent of Coulomb blockade physics (although the actual conductance values may very well depend on the Coulomb energy of the dot). The situation of interest to us is when the confined states in the dot extend into the nanowire (or are entirely inside the nanowire) so that they become Andreev bound states. In situations like this, perhaps the expression ``quantum dot'' is slightly misleading (since there may or may not be any Coulomb blockade here), but we use this expression anyway since it is convenient to describe the physics of Andreev bound states being discussed in our work.

\begin{figure*}[!htb]
\includegraphics[width=0.99\textwidth]{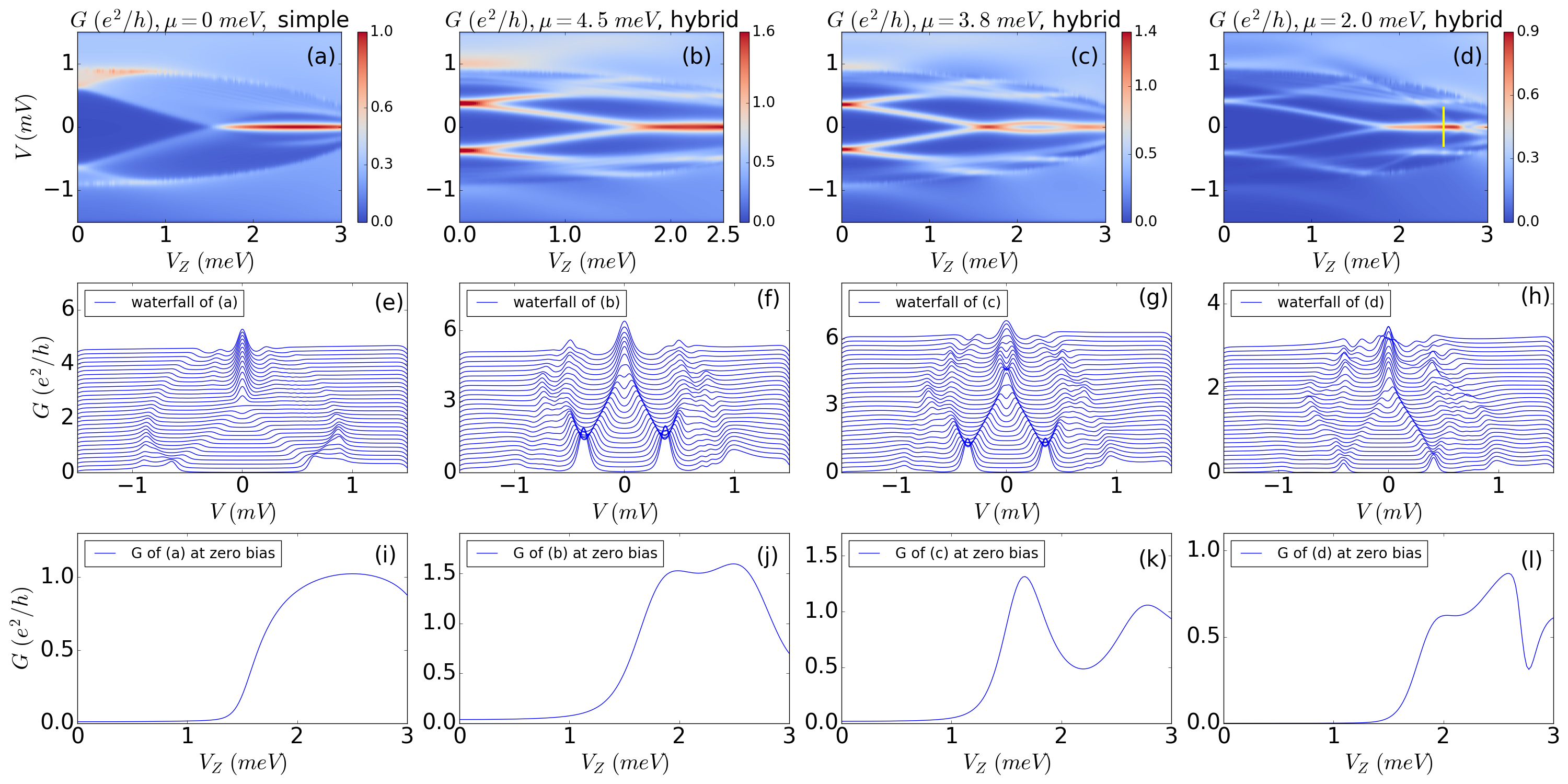}
\caption{(color online). Differential conductance through four nanowire systems with the same dissipation $\Gamma=0.01~$meV, temperature $T=0.02~$meV and tunnel barrier of height $10~$meV and width $20~$nm. (a): a simple nanowire without quantum dot at chemical potential $\mu=0~$meV. A ZBCP from MZM forms after TQPT at $V_Z =1.5~$meV, since the SC pairing at low bias is renormalized to $1.5~$meV due to the self-energy term. (b): a hybrid structure with $\mu=4.5~$meV. Two ABSs come together at $V_Z\sim1.5~$meV and remain stuck at zero energy up to $V_Z = 2.5~$meV (and beyond) although the system is non-topological($V_Z < \mu$). (c): a hybrid structure with $\mu=3.8~$meV. Two ABSs come together at $V_Z\sim1.5~$meV then split at a somewhat higher Zeeman field, but coming together again at $V_Z\sim3~$meV. Again, this is all in the non-topological regime since $V_Z < \mu$. (d): a hybrid structure with $\mu=2.0~$meV. Two ABSs first stick together at $V_Z\sim1.8~$meV which is in the trivial regime (i.e. $V_Z < \mu$), but then the ZBCP continues all the way to the topological regime ($V_Z > 2.5~$meV, marked by the yellow vertical line), with the ZBCP value remaining $\sim e^2/h$ throughout. Note that nothing special happens to the ZBCP feature across the yellow line indicating TQPT. Calculations here include self-energy renormalization by the parent superconductor thus renormalizing the bare induced gap so that $V_Z=1.5~$meV is the TQPT point rather than $0.9~$meV as it would be without any renormalization. Panels (e)-(h) correspond respectively to panels (a)-(d) showing ``waterfall'' diagrams of the conductance against bias voltage for various Zeeman splitting--each line corresponds to a $0.1~$meV shift in $V_Z$ increasing vertically upward. Similarly, panels (i)-(l) correspond respectively to panels (a)-(d) showing the calculated zero-bias conductance in each case as a function of Zeeman splitting.}
\label{fig:summary}
\end{figure*}

It may be useful to provide a succinct summary of our main findings already in this introduction before providing the details of our theory and numerics. We show our most important findings in Fig.~\ref{fig:summary} (all obtained by assuming the dot leading to ABSs to be entirely outside the nanowire), where we show our calculated differential conductance in the dot-nanowire-superconductor system as a function of Zeeman splitting energy ($V_Z$) and the source-drain voltage ($V$) in the nanowire for a fixed chemical potential in each panel (which, however, varies from one panel to the next). The four panels indicate the four distinct generic results which may arise depending on the values of chemical potential and Zeeman splitting (with all other parameters, e.g., bulk superconducting gap, spin-orbit coupling, tunnel barrier, temperature, dissipative broadening, etc. being fixed throughout the four panels). We start by reminding that the topological quantum critical point separating trivial and topological phases in this system is given by the critical Zeeman splitting $V_{Zc}=\sqrt{\mu^2 + \Delta^2}$, where $\mu$ and $\Delta$ are respectively the chemical potential and the proximity-induced superconducting gap in the nanowire~\cite{Sau2010Generic, Lutchyn2010Majorana, Oreg2010Helical}--thus $V_Z<\mu$ automatically implies a trivial phase where no MZM can exist. Fig.~\ref{fig:summary}(a) shows the well-studied result of the ZBCP arising from the MZM as the system enters the topological superconducting phase with the topological quantum phase transition point being at $V_{Zc}=1.5~$meV with the chemical potential being zero, $\mu=0$. (We note that our calculations include the self-energy renormalization effect by the parent superconductor which renormalizes the superconducting gap, as discussed in Sec.~\ref{sec:numerical} of the manuscript.) This result is obtained without any quantum dot (or ABS) being present, and is the generic well-known theoretical result for the simple nanowire in the presence of induced superconductivity, Zeeman splitting, and spin-orbit coupling as predicted in Refs.~\cite{Sau2010Generic, Lutchyn2010Majorana, Oreg2010Helical, Sau2010Non}. We provide this well-known purely nanowire (with no dot, and consequently, no ABS) result only for the sake of comparison with the other three panels of Fig.~\ref{fig:summary} where ABS physics is present because of the presence of the quantum dot. In Fig.~\ref{fig:summary}(b), the chemical potential is increased to $\mu=4.5~$meV with the nonsuperconducting quantum dot being present at the end of the nanowire. Here, the two ABSs come together around $V_Z=1.5~$meV and remain stuck to zero energy up to $V_Z=2.5~$meV (and beyond) although the system is nontopological throughout the figure (as should be obvious from the fact that $V_Z<\mu$ throughout). Thus, ABSs coalescing and sticking at zero energy for a finite range of magnetic field is not necessarily connected with MZMs or topological superconductivity. It should be noted that the ZBCP value in~\ref{fig:summary}(b) is close to $2e^2/h$, but this has nothing to do with the MZM quantization, and we find that the ZBCP arising from coalescing ABSs could have any non-universal value. In Fig.~\ref{fig:summary}(c), we change the chemical potential to $\mu=3.8~$meV, resulting in the two ABSs coming together at $V_Z \sim 1.5~$meV, and then splitting at a somewhat higher magnetic field, but coming together again at $V_Z\sim 3.0~$meV with the ZBCP value varying from $e^2/h$ to $1.5 e^2/h$. Again, this is all in the nontopological regime since $V_Z < \mu$ throughout the figure. Finally, in Fig.~\ref{fig:summary}(d) we show the result for $\mu=2~$meV, where the two ABSs first stick together at $V_Z\sim1.8~$meV which is in the trivial regime (i.e. $V_Z<\mu$), but then the ZBCP continues all the way to the topological regime ($V_Z>2.5~$meV, marked by the yellow vertical line), with the ZBCP value remaining $> e^2/h$ throughout. Interestingly, although there is a topological quantum phase transition (TQPT) in Fig.~\ref{fig:summary}(d) at the yellow line, nothing remarkable happens in the ZBCP--it behaves essentially the same in the trivial and the topological regime! We note that the specific value of the ZBCP in each panel depends on parameters such as temperature and tunnel barrier, and can be varied quite a bit, but their relative values are meaningful and show that the ZBCP in the trivial and the topological regime may have comparable strength, and no significance can be attached (with respect to the existence or not of MZMs in the system) based just on the existence of zero-bias peaks and their conductance values. Thus, stable zero-bias conductance peak is necessary for MZMs, but the reverse is untrue--the existence of stable ZBCP does not by itself imply the existence of MZMs. Note that we are employing the simplest possible model with no disorder at all, and as such our findings are completely different from the disorder-induced class D peak discussed in Refs.~\cite{Liu2012Zero, Bagrets2012Class, Pikulin2012Zero}. This is consistent with the semiconductor nanowire in Ref.~\cite{Deng2016Majorana} being ballistic or disorder-free, and hence the ABS-MZM physics being discussed in our work has nothing whatsoever to do with the physics of `class D peaks' discussed in Refs.~\cite{Liu2012Zero, Bagrets2012Class, Pikulin2012Zero} where disorder plays the key role in producing effectively an antilocalization zero bias peak.

For the sake of completeness, we also show in Fig.~\ref{fig:summary} panels (e)-(l) as the details of the calculated results with Figs.~\ref{fig:summary}(e)-(h) and Figs.~\ref{fig:summary}(i)-(l) corresponding respectively to those in Figs.~\ref{fig:summary}(a)-(d). Figs.~\ref{fig:summary}(e)-(h) show the ``waterfalls'' cuts of the actual conductance in Figs.~\ref{fig:summary}(a)-(d) with each line corresponding to a different magnetic field (increasing vertically). Figs.~\ref{fig:summary}(i)-(l) show the calculated zero-bias conductance, corresponding to panels~\ref{fig:summary} (a)-(d) respectively, as a function of Zeeman splitting. The main message of Figs.~\ref{fig:summary} (e)-(l) is the same as in Figs.~\ref{fig:summary} (a)-(d), i.e., ZBCPs arising from the zero-sticking of trivial ABSs look very similar to those arising from topological MZMs. In particular, Figs.~\ref{fig:summary} (a)/(b) as well as Figs.~\ref{fig:summary} (e)/(f) and Figs.~\ref{fig:summary}(i)/(j) look qualitatively identical although one set of these results belongs to topological MZM (Figs.~\ref{fig:summary}(a), (e) and (i)) and the other set (Figs.~\ref{fig:summary} (b), (f), and (j)) to trivial ABS.  Similarly, the TQPT in Figs.~\ref{fig:summary}(d), (h), and (l) does not manifest itself in any striking way for it to be discerned without already knowing its existence a priori. We emphasize that Figs.~\ref{fig:summary}(e) and (f) look essentially identical qualitatively although the ZBCP in Fig.~\ref{fig:summary}(e) arises from the MZM and in Fig.~\ref{fig:summary}(f) from coalesced ABSs. Similarly, the dependence of the zero-bias conductance on $V_Z$ could be quite similar in these two cases too (Figs.~\ref{fig:summary}(i) and (j)).

We note that the results of Fig.~\ref{fig:summary} are produced for a nominal temperature $\sim 200~$mK which is higher than the fridge temperature ($\sim 40~$mK) where typical experiments are done. The main reason for this is that finite temperature smoothens fine structures in the calculated conductance spectra arising from energy levels in the nanowire which are typically not seen experimentally. Having a finite temperature does not in any way affect the existence or not of the zero mode or any of our conclusions. We add that the electron temperature in semiconductor nanowires is typically much larger ($>100~$mK) than the fridge temperature, and $T=0.02~$meV may not be completely inappropriate even for the realistic system although our reason for including this finite $T$ is purely theoretical.

The importance of our results as summarized in Fig.~\ref{fig:summary} is obvious. In particular, the coalescing of ABSs and their sticking together near zero energy with a fairly strong ZBCP is generic (as we will explain in the Sec.~\ref{sec:numerical}) in the trivial regime of the magnetic field and chemical potential, and equally importantly, there is no special feature in the ZBCP itself for one to discern whether such a coalesced ZBCP is in the topological or trivial regime just based on tunneling conductance measurements. The generic occurrence of almost-zero-energy modes has previously been attributed as a property of quantum dots in symmetry class D~\cite{Mi2014X} in the presence of random disorder whereas our theory by contrast is manifestly in the clean disorder-free limit. In fact, as our Fig.~\ref{fig:summary}(d) indicates, the ZBCP may very well form in the trivial regime and continue unchanged into the topological regime with nothing remarkable happening to it as the magnetic field sweeps through the topological quantum phase transition! Experimental tunneling spectroscopy, by itself, might find it difficult to distinguish MZMs from accidental zero-energy ABSs just based on the observation of the ZBCP (even when the ZBCP conductance $\sim2e^2/h$) since experimentally one simply does not know where the topological quantum phase transition point is in the realistic nanowires. The good thing is that our results indicate that it is possible that some of the Deng \textit{et al}. ZBCPs~\cite{Deng2016Majorana} may be topological, but it is also possible that all of them are trivial ZBCPs. We simply do not know based just on tunneling conductance measurements that have been performed so far.

We mention that there have been earlier indications that ABSs (or in general, low energy fermionic subgap states) may manifest ZBCP features indistinguishable from MZM-induced zero-bias peak behavior~\cite{Kells2012Near, Mi2014X, Lee2014Spin, Stanescu2014Nonlocality, Moore2016Majorana, Chiu2017Conductance}. In particular, it was shown by a number of authors that the presence of a smooth varying potential background in the nanowire could produce multiple MZMs along the wire (and not just the two pristine MZMs localized at the wire ends), which could lead in some situations to trivial ZBCPs in tunneling measurements mimicking MZM-induced ZBCPs~\cite{Kells2012Near, Stanescu2014Nonlocality, Stanescu2014Nonlocality, Moore2016Majorana}. The fact that small quantum dot systems could have ABS-induced ZBCPs was experimentally established by Lee \textit{et al}.~\cite{Lee2014Spin}. Our work, however, specifically addresses the quantum dot-nanowire-superconductor system, showing that the recent observation by Deng \textit{et al}. of ABSs coalescing together near zero energy and then remaining stuck at zero energy for a finite range of magnetic field by itself cannot be construed as evidence for ABSs combining to form MZMs--the ZBCP in such situations may very well arise from accidental coalesced ABSs which happen to beat or stay near zero energy. Clearly, more work is necessary in distinguishing ABS-induced trivial zero modes from MZMs in nanowire-superconductor hybrid structures. There has been other recent theoretical work~\cite{Flensberg2017Note, Schuray2017Fano, Clarke2017Experimentally, Prada2017Measuring} on trying to understand the Deng \textit{et al}. experimental work of Ref.~\cite{Deng2016Majorana} using alternative approaches assuming that the experimental ZBCPs form in the topological regime (i.e., lying above the TQPT point in the magnetic field).

We emphasize that although our initial goal motivating this work was to understand the experiment of Ref.~\cite{Deng2016Majorana} where transmutation of ABS into MZM is claimed in disorder-free ballistic nanowires, we have stumbled upon a generic result of substantial importance in the current search for topological Majorana modes in nanowires (and perhaps in other solid state systems too, where ABSs may arise). This generic result is that the combined effect of spin-orbit coupling and spin splitting could lead to subgap Andreev bound states generically sticking around midgap in a superconductor, and these nontopological `alomost-zero' energy modes are virtually indistinguishable from topological Majorana zero modes using  tunneling spectroscopy. Our result implies that considerable caution is now necessary in searching for MZMs in nanowires since the mere observation of ZBCPs even in clean systems is insufficient evidence for the existence of MZMs.

The paper is organized as follows: In Sec.~\ref{sec:minimal} we give the minimal theory describing the quantum dot-nanowire-superconductor hybrid structures. In Sec.~\ref{sec:numerical}, we introduce the numerical method and calculate the tunneling differential conductance in simple and hybrid structure systems. In Sec.~\ref{sec:analytic}, analytical low-energy spectra of hybrid structures are calculated to provide insightful information about the corresponding zero-bias conductance behavior. In Sec.~\ref{sec:distinguish}, we consider the effect of strongly changing dot confinement on the zero-bias behavior of the ABS, contrasting it with that of MZM, providing one possible experimental avenue for distinguishing between trivial and topological ZBCPs. In Sec.~\ref{sec:dot}, we calculate the differential conductance for hybrid structures where the quantum dot has partial or complete induced superconductivity(i.e., it is a strongly coupled part of the nanowire itself). In Sec.~\ref{sec:understanding} we discuss how our quantum dot-induced ABS results connect with the corresponding results in the case of smooth confinement at the wire ends and can be understood using the reflection matrix theory. Sec.~\ref{sec:conclusion} concludes our work with a summary and open questions. A number of appendices provide complementary detailed technical results not covered in the main text of the paper.

%%%%%%%%%%%%%%%%%%%%%%%%%%%%

\section{minimal theory}\label{sec:minimal}
We calculate the differential tunnel conductance $G=dI/dV$ through a junction of a normal lead and the quantum dot-nanowire-superconductor hybrid structure, as shown in the schematic Fig.~\ref{fig:schematic}. We use the following Bogoliubov-de Gennes (BdG) Hamiltonian as the non-interacting low-energy effective theory for the nanowire~\cite{Sau2010Generic, Lutchyn2010Majorana, Oreg2010Helical}
\begin{align}
&\hat{H} = \frac{1}{2} \int dx \hat{\Psi}^{\dagger}(x) H_{NW} \hat{\Psi}(x), \nn
&H_{NW} = \left( -\frac{\hbar^2}{2m^*} \partial^2_x -i \alpha_R \partial_x \sigma_y - \mu \right)\tau_z + V_Z \sigma_x + \Delta_0 \tau_x,
\label{eq:H_NW}
\end{align}
where $\hat{\Psi} = \left( \hat{\psi}_{\uparrow}, \hat{\psi}_{\downarrow},  \hat{\psi}^{\dagger}_{\downarrow}, -\hat{\psi}^{\dagger}_{\uparrow}  \right)^T $, and $\sigma_{\mu}(\tau_{\mu})$ are Pauli matrices in spin (particle-hole) space, $m^*$ is the effective mass, $\alpha_R$ spin-orbit coupling, $V_Z$ the Zeeman spin splitting energy, $\Delta_0$ the induced superconducting gap. In some discussions and calculated results we also replace the superconducting pairing term by a more complex self-energy term to mimic renormalization effects by the parent superconductor~\cite{Stanescu2010Proximity}, which will be elaborated later. The normal lead by definition does not have induced SC, thus the lead Hamiltonian is 
\begin{align}
H_{lead} = \left( -\frac{\hbar^2}{2m^*} \partial^2_x -i \alpha_R \partial_x \sigma_y - \mu + E_{lead} \right)\tau_z + V_Z \sigma_x,
\label{eq:H_lead}
\end{align}
where an additional on-site energy $E_{lead}$ is added representing a gate voltage. The quantum dot Hamiltonian is
\begin{align}
H_{QD} = \left( -\frac{\hbar^2}{2m^*} \partial^2_x -i \alpha_R \partial_x \sigma_y + V(x) - \mu \right)\tau_z + V_Z \sigma_x ,
\label{eq:H_QD}
\end{align}
where $V(x)=V_D \cos(\frac{3\pi x}{2l})$ is the confinement potential. (We have ensured that other models for confinement potential defining the dot do not modify our results qualitatively.) The quantum dot size $l$ is only a fraction of the total nanowire length $L$. The quantum dot is non-SC at this stage although later (in Sec.~\ref{ssec:proximitized} and Sec.~\ref{sec:dot}) we consider situations where the dot could have partial or complete induced superconductivity similar to the nanowire. Whether the quantum dot exists or not, there is always a barrier potential between the lead and the hybrid nanowire system. Multi-sub-band effects are introduced by constructing a second nanowire with different chemical potential. An infinitesimal amount of dissipation $i\Gamma$ is also added into the nanowire Hamiltonian Eq.~\eqref{eq:H_NW} for the sake of smoothening the conductance profile without affecting any other aspects of the results~\cite{DasSarma2016How, Liu2017Role}. We emphasize that there is no disorder in our model distinguishing it qualitatively from earlier work~\cite{Liu2012Zero, Bagrets2012Class, Pikulin2012Zero} where class D zero bias peaks in this context arise from disorder effects. Given this quantum dot-nanowire model, our goal is to calculate the low lying energy spectrum and the differential conductance of the system varying the chemical potential and the Zeeman splitting in order to see how any dot-induced ABSs behave. The specific goal is to see if we can qualitatively reproduce the key features of the Deng \textit{et al.} experiment in a generic manner without fine-tuning parameters.  Our goal is not to demand a quantitative agreement with the experimental data since too many experimental parameters are unknown(confinement potential, chemical potential, tunnel barrier, superconductor-semiconductor coupling, spin-orbit coupling, effective mass, Lande $g$-factor, etc.), but we do want to see whether ABSs coalesce generically and whether such coalescence around zero energy automatically implies a transmutation of ABSs into MZMs.

%%%%%%%%%%%%%%%%%%%%%%%%%%%%

\section{numerical results for tunnel conductance}\label{sec:numerical}

The goal of our current work is to understand the interplay between Andreev and Majorana bound states in quantum dot-nanowire-superconductor hybrid structures, and to answer the specific question whether two Andreev bound states can coalesce forming a zero-energy bound state leading to a stable ZBCP in the tunnel conductance (as observed in Ref.~\cite{Deng2016Majorana}). This motivates all the calculations in this section. In Sec.~\ref{ssec:simple}, we calculate the differential conductance of a set of nanowires without any quantum dot for the sake of making comparison with the situation where ABS physics is dominant due to the presence of the quantum dot.(We emphasize, as mentioned already in Sec.~\ref{sec:introduction}, that our ``quantum dot'' is simply a prescription for introducing ABS into the physics of the hybrid structure, and is not connected with Coulomb blockade or any other physics one associates with isolated quantum dots.) In Sec.~\ref{ssec:hybrid}, the differential conductance of quantum dot-nanowire-superconductor hybrid structures is calculated as a function of Zeeman field or chemical potential for various parameter regimes. Near-zero-bias peaks similar to the Deng \textit{et al}. experimental data are obtained, and the topology and quantization properties of these peaks are carefully studied. In Sec.~\ref{ssec:TV}, topological visibility~\cite{DasSarma2016How, Liu2017Role} is calculated for both Andreev and Majorana-induced ZBCPs as a theoretical tool discerning the two cases, i.e., to explicitly check whether a zero-energy state is trivial or topological. Of course, in our simulations, we explicitly know the location of the TQPT and can read off the topological or trivial nature of a particular ZBCP simply by knowing the Zeeman field, the chemical potential, and the induced gap. The topological visibility calculation provides an additional check, which simply verifies that a ZBCP arising below (above) the TQPT is trivial ABS (topological MZM), as expected.

For clarification we first provide definitions of two frequently used terms in the rest of this paper: simple nanowire and hybrid structure. A simple nanowire, which by definition does not have any ABS, is defined as a semiconductor nanowire without quantum dot ( i.e., the usual system already extensively studied in the literature~\cite{Lin2012Zero, DasSarma2016How, Liu2017Role}). A hybrid structure, the opposite of a simple nanowire, may have ABS in it, and is defined as a quantum dot-nanowire-superconductor system. The hybrid structure qualitatively mimics the Deng \textit{et al}. system of Ref.~\cite{Deng2016Majorana} (see Fig.~\ref{fig:schematic}). For results presented in this section the chemical potential and the on-site energy are uniform throughout the nanowire since the quantum dot is explicitly outside the nanowire with the dot being non-SC whereas the nanowire being SC (due to proximity effect). Note that although the dot is considered outside the nanowire, any bound state wavefunction in the dot may extend well inside the nanowire (thus making it an ABS) depending on system parameters.

Differential conductance is calculated using the S matrix method, which is a universal method in mesoscopic physics. Numerical implementation of the S matrix method is carried out in this section through KWANT~\cite{Kwant}, which is a Python package for calculating the S matrix of scattering regions in tight-binding models. The model defined in Sec.~\ref{sec:minimal} is particularly well-suited to the KWANT methodology of calculating the S matrix. We discretize Eqs.~\eqref{eq:H_NW}-\eqref{eq:H_QD} into a one-dimensional lattice chain and extract the differential conductance from the corresponding S matrix~\cite{Blonder1982Transition, Setiawan2015Conductance}. Since the calculational technique is well-established, here we focus on presenting and discussing our results, referring the reader to the literature for the details~\cite{DasSarma2016How, Liu2017Role, Kwant, Blonder1982Transition, Setiawan2015Conductance}. The new aspect of our work is to introduce the quantum dot (see Fig.~\ref{fig:schematic}) in the problem and calculate the S matrix exactly for the combined dot-nanowire system.

For the results presented in this section we choose the following representative parameter values for the quantum dot-nanowire system. Effective mass is chosen to be $m^* = 0.015m_e$, along with induced superconducting gap $\Delta_0=0.9~$meV (we present some results for a smaller SC gap later), nanowire length $L \simeq1.3~\mu$m, Zeeman energy $V_Z$[meV] $= 1.2B$[T] where $B$ in Tesla is the applied magnetic field and spin-orbit coupling $\alpha_R=0.5~$eV$\angstrom$~\cite{Liu2017Role}. (Note that this induced bare gap will be renormalized by self-energy corrections.) The gate voltage in the lead is $E_{lead}=-25~$meV. The confinement potential in the quantum dot has a strength $V_D=4~$meV and length $l = 0.3~\mu$m. (We have varied the dot parameters to ensure that our qualitative results are generic, i.e., the qualitative physics discussed in our work does not arise from some special choice of the dot confinement details.) The default value of the barrier between lead and nanowire has height $E_{barrier}=10~$meV and width $l_{barrier}=20~$nm. Note that there is nothing special about these numbers and no attempt is made to get any quantitative agreement with any experimental data since the applicable parameters (even quantities as basic as the effective mass and the $g$-factor) for the realistic experimental systems are unknown. Our goal here is a thorough qualitative understanding and not quantitative numerical agreement with experimental data. We also leave out disorder and/or soft gap effects since these are not central to our study of ABS versus MZM physics in hybrid systems. Introducing these effects is straightforward, but the results become much less transparent.

\subsection{Simple nanowire}\label{ssec:simple}

\begin{figure*}[!t]
\includegraphics[width=0.99\textwidth]{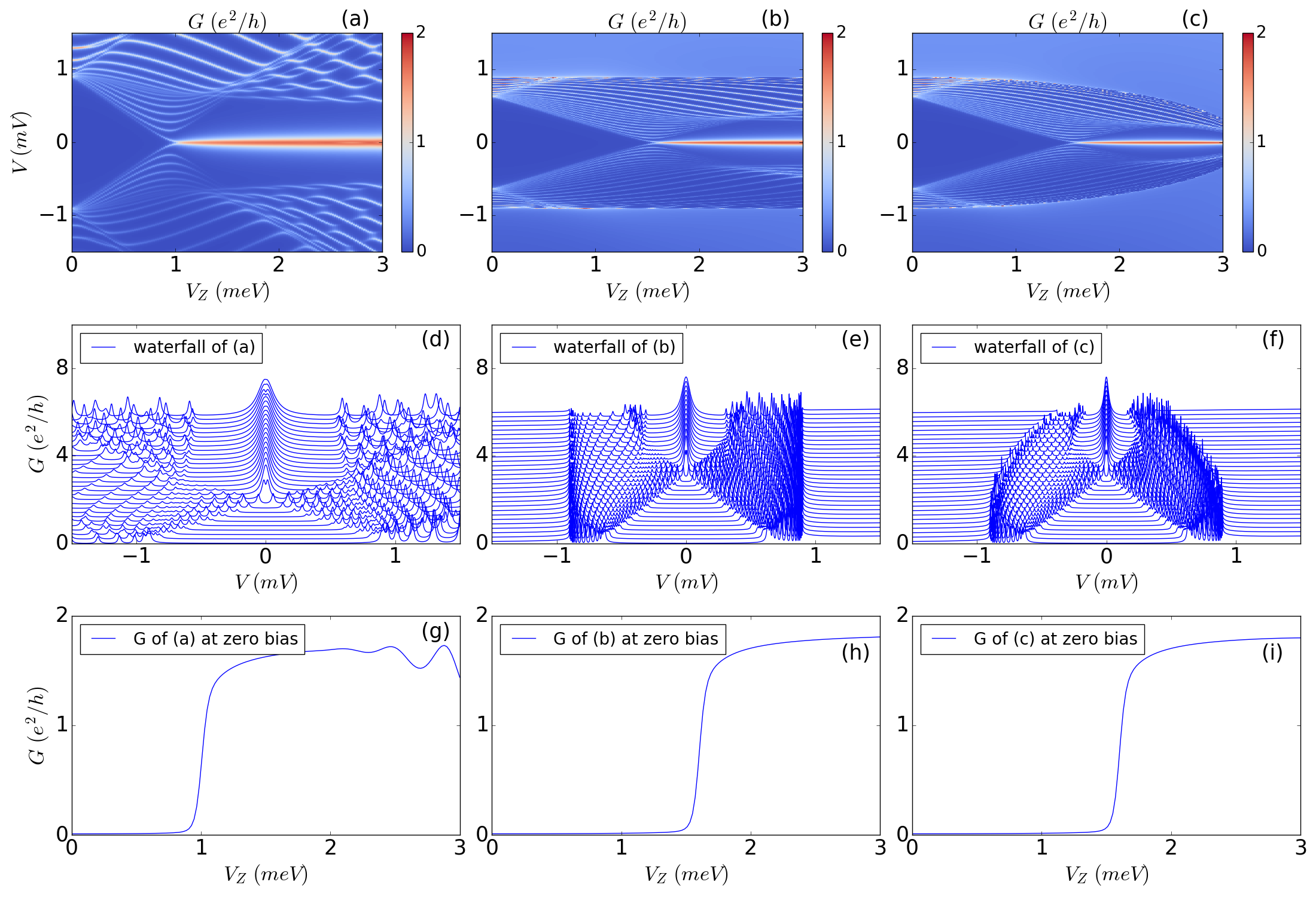}
\caption{(color online). Differential conductance through three simple nanowires (no quantum dot) with chemical potential $ \mu=0~$meV and length $L \simeq 1.3~\mu$m at zero temperature. The three nanowires are different in the way of introducing proximity superconducting effect. (a): constant $s$-wave SC pairing; (b): self-energy with constant parent SC pairing potential; (c): self-energy with the parent SC pairing decreasing with Zeeman field. Note that in panels (b) and (c) the self-energy effect renormalizes the induced gap to the tunnel coupling value $\lambda=1.5~$meV so that TQPT is at $V_Z=1.5~$meV (and not the bare gap value $0.9~$meV). In panel (a), by contrast, there is no self-energy correction, and hence the TQPT is at $V_Z=0.9~$meV. In panels (d)-(f) we show the ``waterfalls'' corresponding to panels (a)-(c), respectively. In panels (g)-(i) we plot the zero-bias conductance as a function of $V_Z$ corresponding to panels (a)-(c) respectively. We note that although we only show $\mu=0$ results here for the simple nanowire, the corresponding results for all finite $\mu$ look identical to the results shown here except for the TQPT point shifting to larger values of $V_Z$ consistent with the well-known theory (i.e., TQPT being given by $\sqrt{\Delta^2 + \mu^2}$).}
\label{fig:simple}
\end{figure*}

We first focus on simple nanowires without any quantum dots. There are no ABSs in this case by construction, and any ZBCP can only arise from MZMs in our model. The corresponding conductance has been well studied~\cite{Lin2012Zero, DasSarma2016How, Liu2017Role}. However, we still present our numerical simulations for such simple nanowire systems for two reasons. First, we will compare Andreev and Majorana-induced conductances later in this paper, and therefore it is important to have the pure MZM results in simple nanowires (i.e., without any ABS) for our specific parameter values. Second, the proximity effect (with or without self-energy effects) discussed in the simple model is generic and is applicable to the situation with quantum dot. The conductance of three simple nanowire systems is shown in Fig.~\ref{fig:simple}. All of them use a one-band model with chemical potential $\mu=0~$meV. The difference lies in the way of introducing the proximity SC effect. In the first case (Fig.~\ref{fig:simple}(a)), a phenomenological constant $s$-wave SC pairing is introduced and thus its Hamiltonian is exactly the minimal model defined by Eq.~(\ref{eq:H_NW}). In the other two cases (Figs.~\ref{fig:simple}(b) and (c)), degrees of freedom in the SC are microscopically integrated out, giving rise to a self-energy term in the semiconductor nanowire~\cite{Stanescu2010Proximity, Sau2010Robustness, Reeg2017Transport}
\begin{align}
\Sigma(\omega) = - \lambda \frac{\omega \tau_0 + \Delta_0\tau_x}{\sqrt{\Delta^2_0 - \omega^2}},
\label{eq:SE}
\end{align}
where $\lambda$ has the dimension of energy and is proportional to the tunnel coupling between the parent superconductor and the semiconductor nanowire. We choose $\lambda=1.5~$meV throughout this work. In the low energy limit $\omega \to 0$, the self-energy term goes to the simple form of $s$-wave SC pairing but with a renormalized SC pairing amplitude $-\lambda \tau_x$. Therefore in both cases, the Hamiltonian becomes energy-dependent including the substrate-induced self-energy term
\begin{align}
H(\omega) = \left( -\frac{\hbar^2}{2m^*} \partial^2_x -i \alpha_R \partial_x \sigma_y - \mu \right)\tau_z + V_Z \sigma_x + \Sigma(\omega).
\label{eq:H_SE}
\end{align}
In the third case (Fig.~\ref{fig:simple}(c)), not only is a self-energy term introduced, the bulk SC gap also has $V_Z$-dependence, i.e., $\Delta_0$ in Eq.~(\ref{eq:SE}) is replaced by
\begin{align}
\Delta(V_Z) = \Delta_0 \sqrt{1-(V_Z/V_{Zc})^2},
\label{eq:DeltaVZ}
\end{align}
where $V_{Zc}$ represents the critical magnetic field beyond which the bulk superconductivity is destroyed. (We introduce such a field-dependent SC gap since this appears to be case often in the nanowire experiments.) Then the Hamiltonian becomes
\begin{align}
H(\omega) = \left( -\frac{\hbar^2}{2m^*} \partial^2_x -i \alpha_R \partial_x \sigma_y - \mu \right)\tau_z + V_Z \sigma_x + \Sigma(\omega, V_Z).
\label{eq:H_SEVZ}
\end{align}

Our reason for introducing a self-energy in the problem is to include the renormalization effects by the parent superconductor to some degree~\cite{Stanescu2017Proximity}. This is not essential for studying the ABS-MZM story in itself, but the calculated transport properties agree better with experiment in the presence of the self-energy corrections. In spite of the three different ways of introducing proximity SC effect, the calculated differential conductance at low energies (small bias voltage) shows universal behavior for the simple nanowire--a ZBCP forms right after gap closing, indicating the TQPT. This ZBCP is obviously associated with the MZMs at the ends of the nanowire. The ZBCP is quantized at $2e^2/h$ because of the nanowire being in a topological superconducting phase and is robust against variations in the tunnel barrier, chemical potential and other parameters at zero temperature. Here in our simulation, however, the peak value is slightly below the quantized value $2e^2/h$ because a small amount of dissipation ($\Gamma=0.01~$meV) has been added for data smoothening. Although the three results are universal and identical in Fig.~\ref{fig:simple} for  the low energy regime near the ZBCP, in the high energy regime (large bias voltage) conductance shows qualitative differences with or without self-energy. In addition, the TQPT point may shift due to self-energy corrections as the induced SC gap is renormalized by the tunnel coupling $\lambda$ in Eq.~(\ref{eq:SE}). The calculated conductance in Fig.~\ref{fig:simple}(a) has clear patterns at large bias voltage, while in Fig.~\ref{fig:simple}(b) and (c), the calculated conductance at $eV > \Delta$ is smooth and featureless. This featureless conductance can be understood by the smearing of the spectral function due to nanowire electrons tunneling into the quasiparticle continuum in the parent superconductor. Thus, the continuum (i.e., electron-hole) behavior above the SC gap is different in Fig.~\ref{fig:simple} with and without self-energy although the below-gap behavior near zero energy is essentially the same in all three approximations (except for a shift of TQPT to a higher critical $V_Z$ due to the self-energy renormalization). We note, however, that in Fig.~\ref{fig:simple}(a) there is some evidence for the MZM-overlap induced ZBCP oscillations~\cite{Cheng2009Splitting, DasSarma2012Splitting} at the highest magnetic field values ($V_Z > 2.5~$meV) which is more obvious in Fig.~\ref{fig:simple}(g) at the highest $V_Z$ values. The edge of the quasiparticle continuum in Fig.~\ref{fig:simple}(b) stays at a fixed bias voltage due to constant $\Delta_0$, while the edge shrinks in Fig.~\ref{fig:simple}(c) due to a decrease of the field-dependent SC gap $\Delta(V_Z)$ as a function of Zeeman field. In the Deng \textit{et al}. experimental data~\cite{Deng2016Majorana}, we clearly see the quasiparticle continuum edge shrinking with Zeeman field and the conductance is featureless outside the SC gap, which leads us to believe that a self-energy term for describing proximity superconducting effect and a $V_Z$-dependent bulk SC gap $\Delta(V_Z)$ are necessary physical ingredients for correctly describing the higher energy features. Thus in all the calculations in the rest of the main paper, the proximity effect will be introduced by a self-energy term and the SC bulk gap will be $\Delta(V_Z)$, unless explicitly stated otherwise. Here for the simple nanowire case, we only show the conductance of one-band models, while relegating the corresponding conductance of two-sub-band models in the Appendix~\ref{app:2b} for completeness. We note that both the self-energy effect and the two-sub-band effect are necessary only for the qualitative agreement between our conductance calculations and the experimental data away from the midgap zero-energy regime. If we are only interested in the zero-energy behavior of ABS and MZM, the minimal model of Eq.~(\ref{eq:H_NW}) without any self energy or two-sub-band effect is perfectly adequate.

\subsection{Quantum dot-nanowire-superconductor hybrid structures}\label{ssec:hybrid}
In nanowire tunneling experiments quantum dot physics is quite generic, and it may appear at the interface between the nanowire and the lead due to Schottky barrier effects as mentioned in Sec.~\ref{sec:introduction}, since all that is needed is a small potential confinement region in between the lead and the wire which is non-SC. In our model, the only role played by the quantum dot potential is to introduce ABSs in the nanowire, and hence, if an experiment observes in-gap ABS in the superconducting nanowire, we model that by a ``quantum dot'' strongly coupled to the nanowire. In this subsection, we calculate the differential conductance of generic hybrid structures, for which the Hamiltonian is a combination of quantum dot Eq.~\eqref{eq:H_QD} and nanowire Eq.~\eqref{eq:H_SEVZ}. Only one-band model with the self-energy is presented in the main text, while two-sub-band models and constant $s$-wave proximity pairing cases are discussed in Appendix~\ref{app:hybrid_const}. We also present the energy spectra of hybrid structures with or without Zeeman spin splitting and spin-orbit coupling in Appendix~\ref{app:with_without}. In the main text of this section, we mainly show our calculated tunneling conductance results.

\subsubsection{Scan of Zeeman field}\label{sssec:scanVZ}

\begin{figure*}[!htb]
\includegraphics[width=0.99\textwidth]{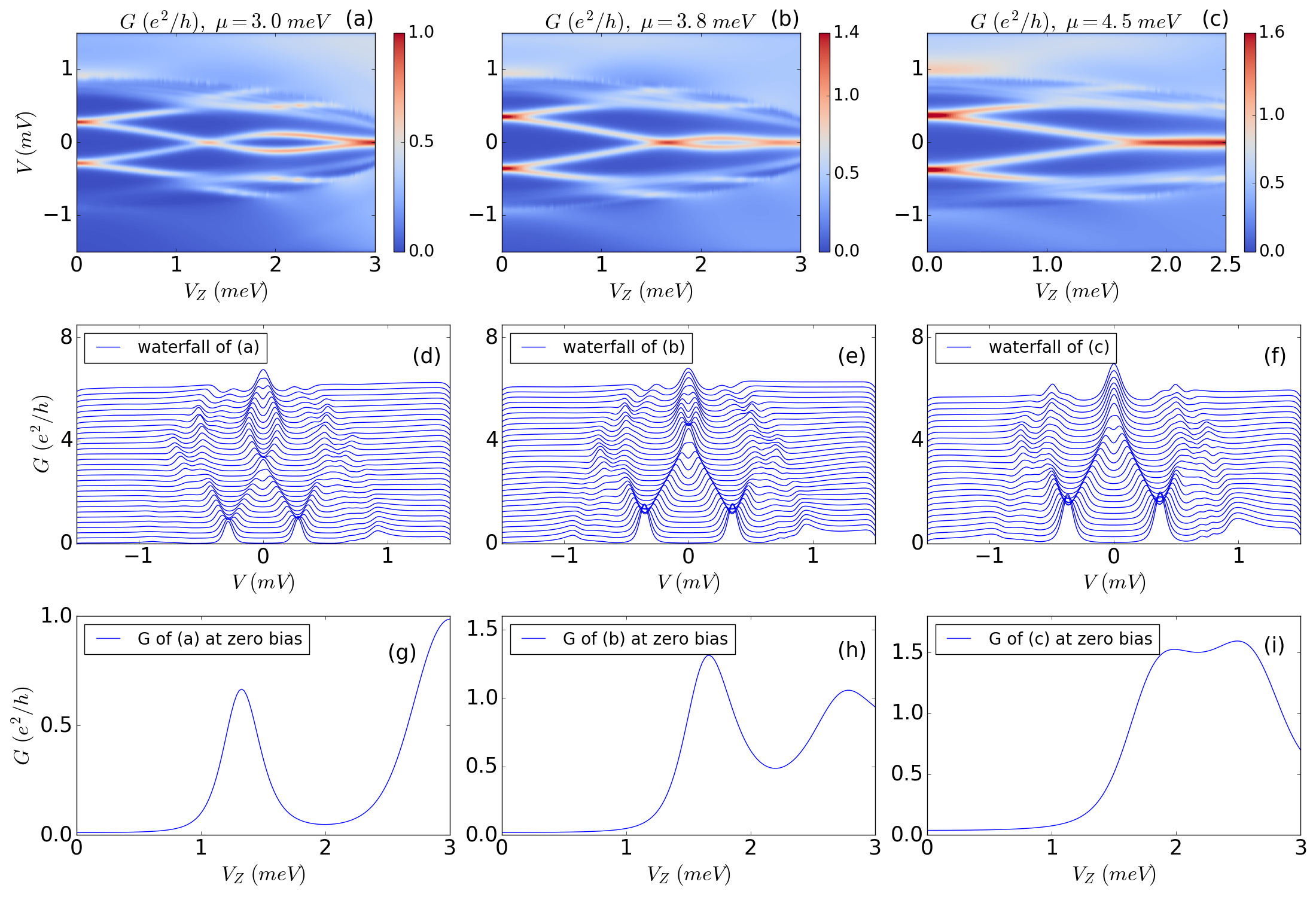}
\caption{(color online). The calculated differential conductance through the dot-nanowire hybrid structure as a function of Zeeman field at various fixed chemical potentials ($\mu= 3.0,~3.8,~4.5~$meV) at $T=0.02~$meV. In all three panels (a)-(c), a pair of ABS conductance peaks at positive and negative bias voltage tend to come close to each other when the Zeeman field is turned on. At finite Zeeman field ($\sim1.5~$meV), in (a) and (b), these two ABS peaks cross zero bias and beat, while in (c) they stick with each other. However these near-zero-energy peaks, especially the ZBCP formed by sticking of two ABSs in (c), are all topologically trivial ABS peaks because $V_Z < \sqrt{\mu^2 + \Delta^2}$ with the Zeeman splitting explicitly being less than the critical value necessary for the TQPT. In panels (d)-(f) we show ``waterfall'' plots of conductance line cuts for different $V_Z$ (increasing vertically upward by $0.1~$meV for each line) corresponding to panels (a)-(c) respectively, whereas in panels (g)-(i) we show the calculated zero-bias conductance in each case corresponding to panels (a)-(c) respectively. Note that these results include self-energy renormalization correction for the proximity effect.}
\label{fig:scanVZ}
\end{figure*}

The calculated differential conductance through the dot-nanowire hybrid structure as a function of Zeeman field at various fixed chemical potentials($\mu= 3.0,~3.8,~4.5~$meV) is shown in Fig.~\ref{fig:scanVZ}. Finite temperature $T=0.02~$meV is introduced by a convolution between zero-temperature conductance and derivative of Fermi-Dirac distribution: $G_T(V) = - \int dE G_0(E) f^{'}_T(E-V)$. In each panel of Fig.~\ref{fig:scanVZ}, a pair of ABS-induced conductance peaks at positive and negative bias voltage tend to come close to each other when the Zeeman field is turned on. At finite Zeeman field ($\sim1.5~$meV), these two ABS peaks either cross zero bias and beat (Figs.~\ref{fig:scanVZ}(a) and (b)) or stick with each other near zero energy(Fig.~\ref{fig:scanVZ}(c)), all of which are similar to the observations in the Deng \textit{et al}. experiment~\cite{Deng2016Majorana}. However these near-zero-energy peaks, especially the ZBCP formed by sticking of two ABSs, are all topologically trivial ABS peaks in Fig.~\ref{fig:scanVZ} because $V_Z < \sqrt{\mu^2 + \Delta^2}$ with the Zeeman splitting explicitly being less than the critical value necessary for the TQPT. We emphasize that experimentally the TQPT critical field is unknown whereas in our theory we know it by definition. If we did not know the TQPT point, there was no way to discern (just by looking at these conductance plots) whether the ZBCP in Fig.~\ref{fig:scanVZ} arises from trivial or topological physics! The generic beating or accidental sticking behavior from the coalesced ABS pair is the consequence of the renormalization of the bound states in the quantum dot in proximity with nanowire in the presence of Zeeman splitting and spin-orbit coupling, which has little to do with topology and Majorana. More detailed discussion of this point will be presented in Sec.~\ref{sec:analytic}. All we emphasize here is that coalescence of ABS pairs into a ZBCP (as in Fig.~\ref{fig:scanVZ}(c)) cannot be construed as ABSs merging into MZMs without additional supporting evidence. In Figs.~\ref{fig:scanVZ}(d)-(f) we provide further details by showing ``waterfalls'' patterns of conductance for increasing $V_Z$ corresponding to the results in Figs.~\ref{fig:scanVZ}(a)-(c), respectively, whereas in Figs.~\ref{fig:scanVZ}(g)-(i) we show the calculated zero-bias conductance as a function of $V_Z$ for results in Figs.~\ref{fig:scanVZ}(a)-(c), respectively.

\subsubsection{Scan of chemical potential}\label{sssec:scanmu}

\begin{figure*}[!htb]
\includegraphics[width=0.99\textwidth]{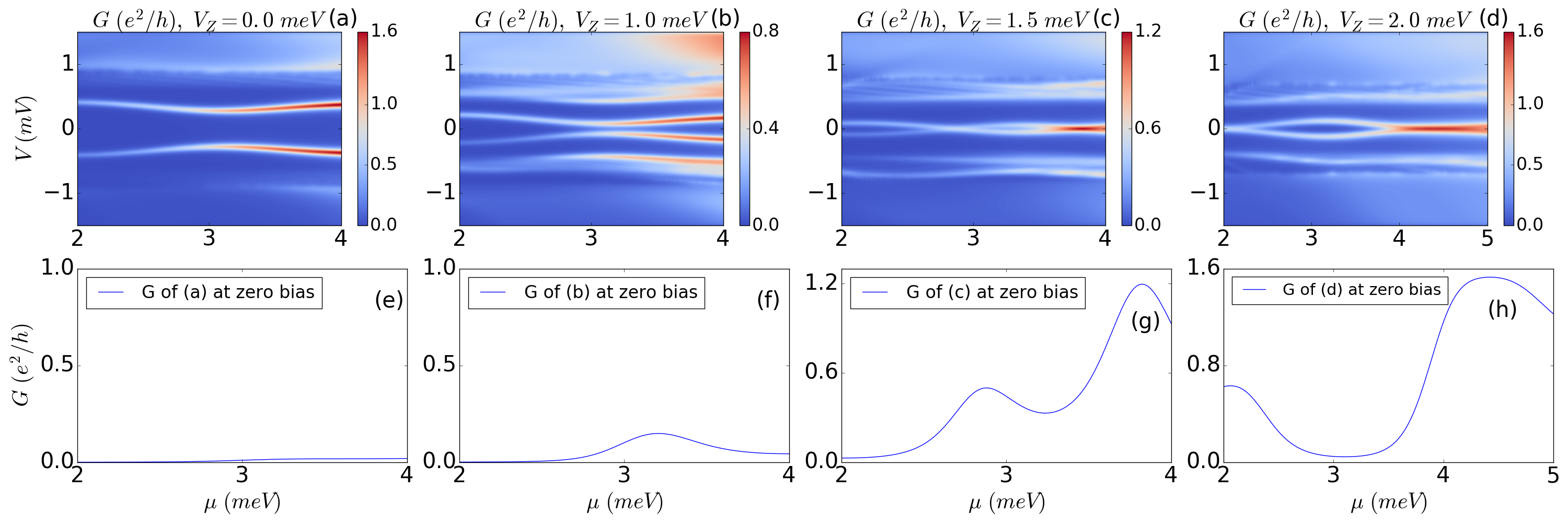}
\caption{(color online). Calculated differential conductance through the hybrid structure as a function of chemical potential at various Zeeman fields at $T=0.02~$meV. In (a) and (b), the ABS conductance peaks repel away from each other without coalescing at zero energy. In (c) the ABS peaks come together at some specific magnetic field, and beat with increasing chemical potential. In (d) ABS peaks beat and stick with each other. However, all of these near-zero-energy peaks are topologically trivial because $V_Z < \sqrt{\mu^2 + \Delta^2}$. In panels (e)-(h) we show the calculated zero-bias conductance corresponding respectively to panels (a)-(d) as a function of chemical potential at fixed $V_Z$.  Note that the TQPT happens here at low $V_Z < 2.0~$meV (not shown).}
\label{fig:scanmu}
\end{figure*}

Calculated differential tunnel conductance through the dot-nanowire hybrid structure as a function of chemical potential at various Zeeman fields at $T=0.02~$meV is shown in Fig.~\ref{fig:scanmu}. In Fig.~\ref{fig:scanmu}(a) and (b), the ABS-induced conductance peaks repel away from each other without coalescing at zero energy. In Fig.~\ref{fig:scanmu}(c) the ABS peaks come together at some specific magnetic field, and beat with increasing chemical potential. In Fig.~\ref{fig:scanmu}(d) ABS peaks beat and stick with each other. All these features are similar to observations in the Deng \textit{et al}. although the relevant variable in the experiment is a gate voltage whose direct relationship to the chemical potential in the wire (our variable in Fig.~\ref{fig:scanmu}) is unknown, precluding any kind of direct comparison with experiment~\cite{Deng2016Majorana}. But all of these near-zero-energy peaks are topologically trivial in our results of Fig.~\ref{fig:scanmu} because $V_Z < \sqrt{\mu^2 + \Delta^2}$ everywhere. We show in Figs.~\ref{fig:scanmu}(d)-(f) the calculated zero-bias conductance corresponding to Figs.~\ref{fig:scanmu}(a)-(c) respectively. Again, sticking together of ABSs at zero energy producing impressive ZBCP peaks are not sufficient to conclude that topological MZMs have formed. In Fig.~\ref{fig:scanmu}, all the results are nontopological!

We note that the ABSs sticking to almost zero energy and producing trivial ZBCPs generically happen only for larger values of chemical potential (as should be obvious from Figs.~\ref{fig:scanVZ} and~\ref{fig:scanmu}) with the ABSs tending to repel away from each other or not quite stick to zero (e.g., Figs.~\ref{fig:scanmu}(a) and (b)) for $\mu<\Delta$.  We find this to be a general trend. Unfortunately, the chemical potential is not known in the experimental samples.

\subsubsection{Generic near-zero-bias conductance features independent of the choice of parameters}\label{sssec:generic}

\begin{figure*}[!thb]
\includegraphics[width=0.99\textwidth]{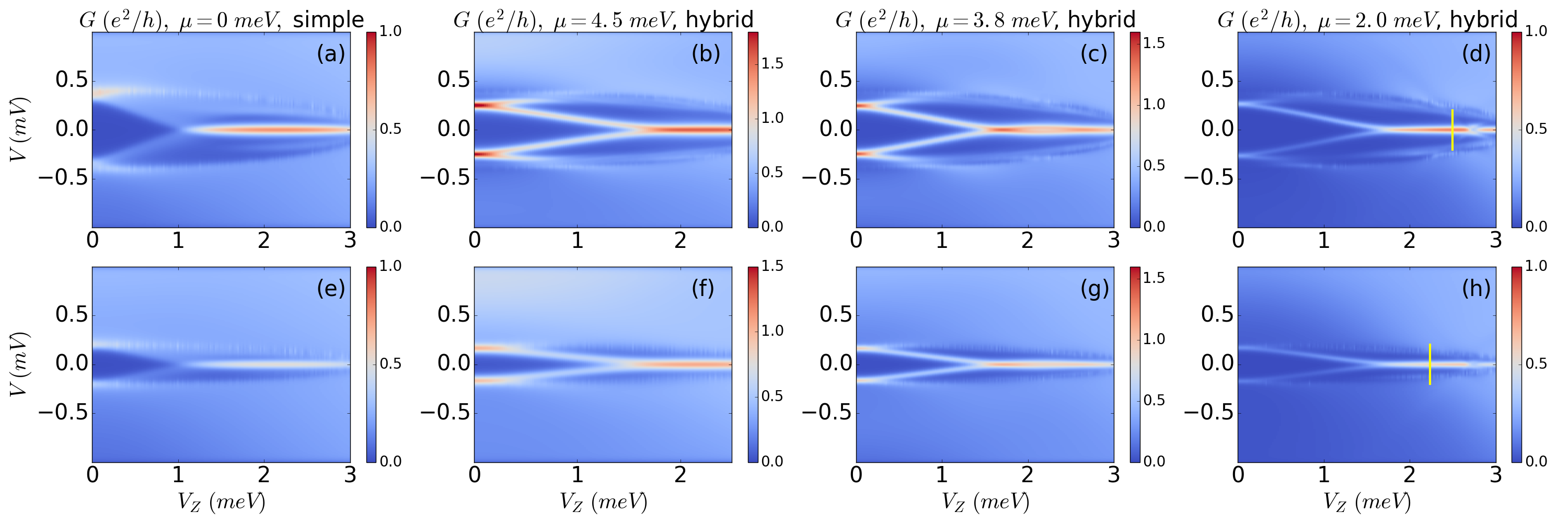}
\caption{(color online). Differential conductance for simple nanowires and hybrid structures with different SC gap parameters. In the upper panels, (a)-(d), the parent superconducting gap is $\Delta_0=0.4~$meV, and the coupling between the nanowire and the parent superconductor is $\lambda=1.5~$meV. In the lower panels, (e)-(h), $\Delta_0=0.2~$meV, and $\lambda=1.0~$meV. These plots should be directly compared with Fig.~\ref{fig:summary}, which shows that all the ABS-induced near-zero-bias conductance features are generic. (In the other figures in this paper, $\Delta_0=0.9~$meV and $\lambda =1.5~$meV.)}
\label{fig:generic}
\end{figure*}

In the previous subsections, we show how topologically trivial ABSs could induce near-zero-bias conductance peaks that are quite similar to MZM-induced ZBCPs. The most important results among them are also summarized in the introduction (Fig.~\ref{fig:summary}). In order to show that all these results are generic, not dependent on the particular choice of parameters, we here present another sets of differential conductance plots (Fig.~\ref{fig:generic}) with different choice of parent superconducting bulk gap $\Delta_0$ and the coupling $\lambda$ between the semiconductor nanowire and the proximitizing superconductor. In the previous discussions, the default values are $\Delta_0 = 0.9~$meV and $\lambda=1.5~$meV. Here in Fig.~\ref{fig:generic}, the upper panels use $\Delta_0 = 0.4~$meV and $\lambda=1.5~$meV, while the lower panels use $\Delta_0 = 0.2~$meV and $\lambda=1.0~$meV. Apart from these different parameters, all other ingredients are kept exactly the same as those in Fig.~\ref{fig:summary} so as to make direct comparison. If we compare Fig.~\ref{fig:summary}(a)-(d) with Fig.~\ref{fig:generic}(a)-(d), we find that the edge of the quasiparticle continuum is determined by the value of $\Delta_0$, while the near-zero-bias conductance behavior looks exactly the same, independent of $\Delta_0$, because the low-energy induced gap is the coupling $\lambda$, not the bare bulk gap $\Delta_0$, as discussed below Eq.~\eqref{eq:SE}. Thus in Fig.~\ref{fig:generic}(e)-(h), the low-energy conductance behavior is changed by a difference choice of $\lambda$ (e.g., the critical Zeeman field for the formation of MZM-induced ZBCP in Fig.~\ref{fig:generic}(e) is smaller than that in Fig.~\ref{fig:generic}(a) ). However, this kind of variation for the near-zero-bias ABS-induced conductance peaks due to the change of $\Delta_0$ and $\lambda$ is perturbative, as shown in Fig.~\ref{fig:generic}(f)-(h) with respect to either Fig.~\ref{fig:generic}(b)-(d) or Fig.~\ref{fig:summary}(b)-(d). The way to understand this observation is that ABSs are bound states localized in the quantum dot, with some wavefunction leakage into the proximitized nanowire, and thus the effect of superconducting gap on the ABSs is only perturbative. Thus, ABS-induced ZBCP physics is independent of the SC gap size as long as the gap is not so small as to be comparable with the energy resolution in the experiment (or numerics).  We expect this physics to arise whenever there are ABSs in the system in the presence of spin-orbit-coupling and Zeeman splitting independent of the SC gap size and other details (except that the chemical potential should not be too small). More detailed discussion on this perturbative effect will be presented in Sec.~\ref{sec:analytic}.

\subsubsection{Continuous crossover from ABS to MZM-induced ZBCP}\label{sssec:continuous}

\begin{figure}[!t]
\includegraphics[width=0.49\textwidth]{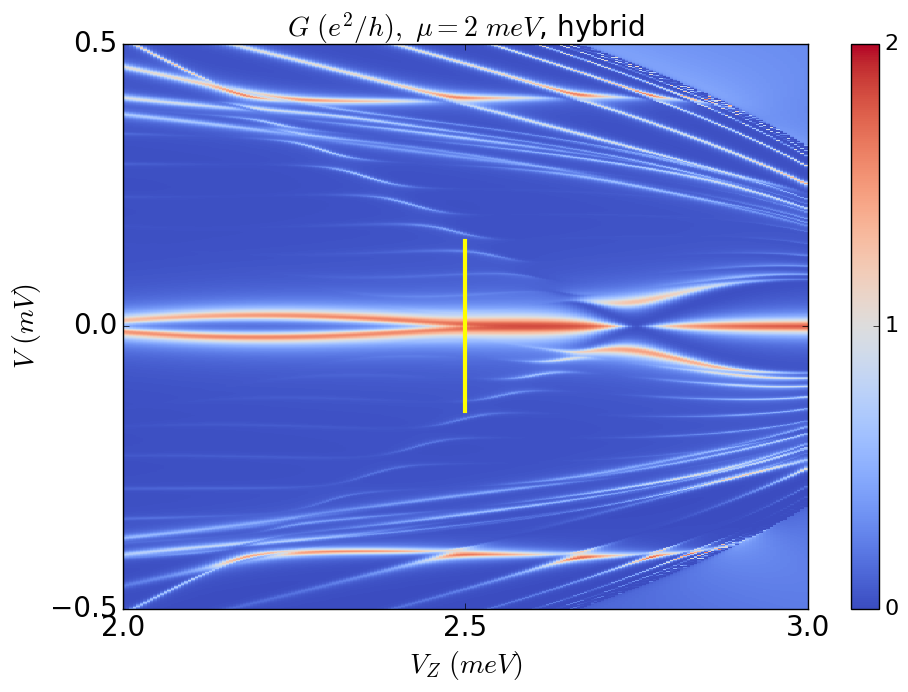}
\caption{(color online). Zoom-in of Fig.~\ref{fig:summary}(d), the conductance for a hybrid structure with $\mu=2~$meV as a function of Zeeman field, where the critical Zeeman field is $V_{Zc}=2.5~$meV, as shown by the vertical yellow line.}
\label{fig:zoomin}
\end{figure}

As already mentioned in the introduction, a topologically trivial ABS-induced near-zero-bias conductance peak can continue all the way to the topologically nontrivial MZM-induced zero-bias conductance peak, with nothing remarkable happening at the TQPT point (Fig.~\ref{fig:summary}(d)). The ABS to MZM transition is in fact a smooth crossover, not that different from what would happen to the MZM itself if one starts from a very short wire with strongly overlapping end-MZMs and then crosses over to exponentially protected well-separated MZMs in the long wire limit simply by increasing the wire length. Here we provide a zoom-in plot of Fig.~\ref{fig:summary}(d) focusing on the vicinity of TQPT, in order to see explicitly how ABSs and MZMs interact with each around around the TQPT. As shown in Fig.~\ref{fig:zoomin}, it is the conductance for a hybrid structure with chemical potential $\mu=2~$meV as a function of Zeeman field and bias voltage. The critical Zeeman field is $V_{Zc}=2.5~$meV, as indicated by the vertical yellow line, to the left (right) of which, the hybrid structure is in topologically trivial (nontrivial) regime. When $V_Z<2.5~$meV, there is ABS near zero-bias, while when $V_Z>2.5~$meV, the MZM-induced ZBCP forms and stays over a large range of Zeeman field. We want to emphasize that the ABS-induced peaks and the MZM-induced peaks are uncorrelated with each other, they do not transmute into each other by any means. This statement is supported by the observation that the near-zero-energy ABS below the formation of the MZM-induced ZBCP in Fig.~\ref{fig:zoomin} still exists at finite energy in the topological regime, and it affects the MZMs by squeezing the width of the ZBCP and lowering its peak value when the their energy separation is small ($\sim 2.7~$meV). Put in another way, those ABSs forming the near-zero-bias conductance peaks never transmute into the MZMs, they exist on their own and may affect the MZMs at some point. All that happens in Fig.~\ref{fig:zoomin} is that the ABS is near zero energy below the TQPT, and once the MZM forms above the TQPT, the ABS moves away from zero energy producing some level repulsion physics with the MZM above the TQPT. We emphasize that there is neither an ABS-MZM transition nor an ABS-MZM transmutation. We note, however, that the level repulsion pushing the ABS away from zero energy in Fig.~\ref{fig:zoomin} actually happens a finite field above the TQPT reflecting crossover nature of the ABS-MZM `transition'.

\subsubsection{Effect of tunnel barrier}\label{sssec:effect}
It has been well established that a zero-temperature ZBCP from MZM has a robust quantized peak value $2e^2/h$ against the variation of tunnel barrier. For peaks from ABSs, however, such robustness is absent, and there is no generic value for the height of ABS peaks - they range from 0 to $4e^2/h$~\cite{Sengupta2001Midgap}. We have checked explicitly that we can get any conductance value associated with the ABS-associated ZBCP by tuning various parameters. In particular, a ZBCP conductance around $2e^2/h$ is quite common from the non-topological ZBCP arising from coalesced ABSs through fine-tuned barrier strength. This dependence of ABS-induced ZBCP on the tunnel barrier strength can be used to check the robustness of any experimentally observed ZBCP. If the ZBCP height is immune to variations in the tunnel barrier, the likelihood is high that the corresponding ZBCPs are induced by topological MZMs.

\subsection{Topological visibility}\label{ssec:TV}

\begin{figure}[!t]
\includegraphics[width=0.49\textwidth]{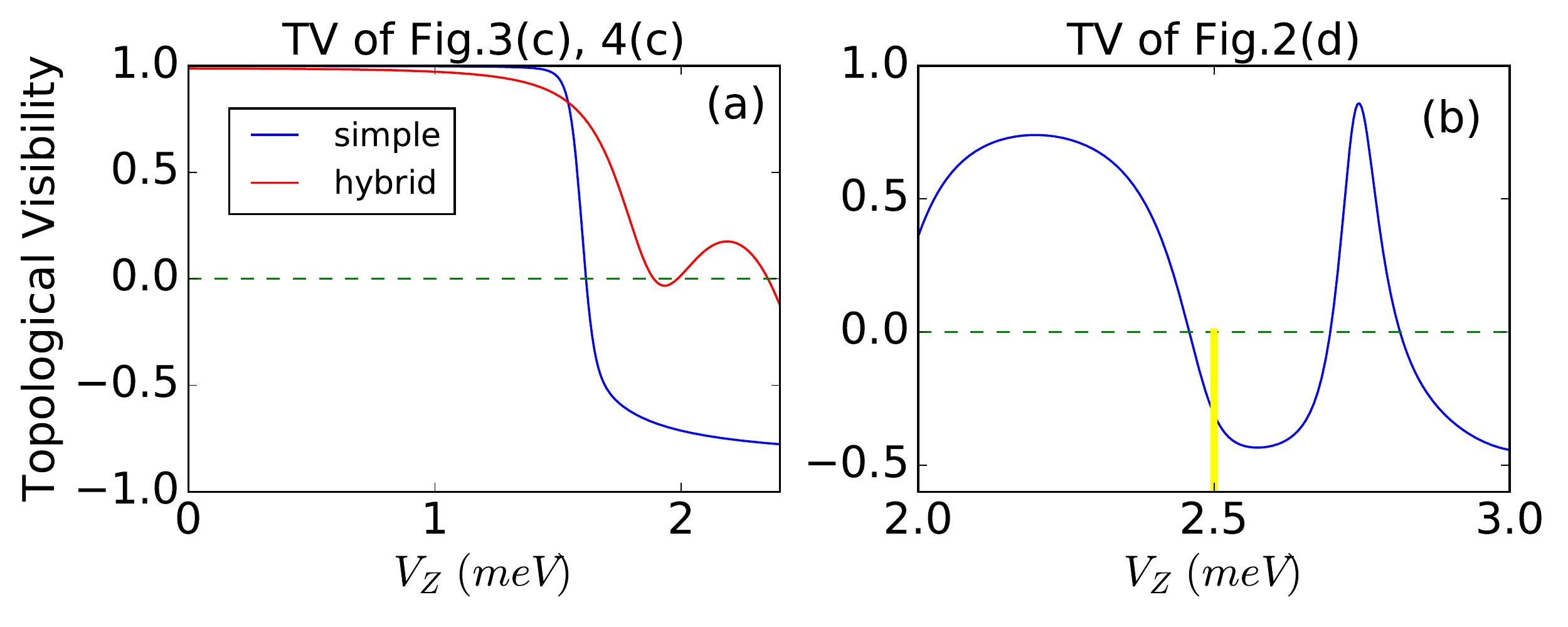}
\caption{(color online). (a) Calculated topological visibility of a simple nanowire and a hybrid structure, whose corresponding conductance is in Fig.~\ref{fig:simple}(c) with $\mu=0~$meV and \ref{fig:scanVZ}(c) with $\mu=4.5~$meV. At small Zeeman field, TV in both cases are close to 1, indicating trivial phases. At large Zeeman field, the TV of the simple nanowire goes down to negative values approaching $-1$ while that of the hybrid structure also goes down but still remains around zero. (b) Calculated TV corresponding to Fig.~\ref{fig:summary}(d) of hybrid structure with $\mu=2.0~$meV sweeping through the TQPT point at $V_{Zc}=2.5~$meV (the vertical yellow line separating the trivial ZBCP from the topological ZBCP). }
\label{fig:TV}
\end{figure}

Based on our numerical simulations, we conclude that it is difficult to differentiate between Majorana and Andreev-induced ZBCPs by merely looking at differential conductance, e.g., Fig.~\ref{fig:simple}(c) and \ref{fig:scanVZ}(c) both show ZBCPs approaching $2e^2/h$ at large Zeeman field. Whether the ZBCPs are topological or not is determined by calculating whether $V_Z$ is larger or smaller than the critical value for the TQPT, i.e., $V_{Zc}=\sqrt{\mu^2 + \Delta^2}$. We can also use another complementary quantity called topological visibility~\cite{DasSarma2016How} to measure the topology of ZBCPs, discerning topological MZM-induced ZBCPs from trivial ABS-induced ZBCPs. Topological visibility (TV) is defined as the determinant of the reflection matrix:
\begin{align}
TV = Det(r),
\end{align}
where the reflection matrix $r$ contains both the normal and the Andreev reflections from the nanowire at zero-bias voltage. Topological visibility is a generalization of topological invariant ($Q$) defined by S matrix at zero-bias voltage, which is $Q=Det(r)=sgn(Det(r))$. The topological invariant takes only binary values as $\pm1$ due to the assumption of particle-hole symmetry and unitarity of the reflection matrix~\cite{DasSarma2016How, Akhmerov2011Quantized}. However, for a finite-length nanowire, the topological invariant always takes the trivial value, i.e., $Q=+1$ because of the Majorana splitting from MZM overlapping making conductance at zero bias always zero even when the topological criteria is satisfied. When an infinitesimal amount of dissipation is added into the nanowire leading to a finite value of zero-bias conductance, $Det(r)$ can take any real value between $-1$ to $+1$ because the unitarity condition is no longer satisfied. Thus TV$=Det(r)$, as a generalization of topological invariant, varies between -1 and $+1$. When the value of TV is close to $-1$, the system is thought of as topologically nontrivial, and otherwise, the system is more topologically trivial. The TV of a simple nanowire and a hybrid structure are shown in Fig.~\ref{fig:TV}(a), whose corresponding conductance is in Fig.~\ref{fig:simple}(c) and \ref{fig:scanVZ}(c) respectively. At small Zeeman field, TVs in both cases are close to 1, indicating trivial phases. At large Zeeman field, the TV of the simple nanowire goes down to negative values approaching $-1$ while that of the hybrid structure also goes down but still remains around zero. This fact indicates that although a pair of ABSs coalesce forming a ZBCP, this peak is topologically trivial, while Majorana-induced ZBCP is topological. Thus merely getting a ZBCP with value close to $2e^2/h$ does not necessarily mean the system enters topological regime. Unfortunately, there is no direct method to measure the topological visibility experimentally. In Fig.~\ref{fig:TV}(b), we show the calculated TV corresponding to Fig.~\ref{fig:summary}(d) where the TQPT is at $V_{Zc}= 2.5~$meV (the vertical yellow line). We notice that TV starts to dive to be more negative for $V_Z > 2.5~$meV consistent with the TQPT separating a trivial ZBCP below and above $V_Z=2.5~$meV, but the result is not absolutely definitive because of the presence of dissipation, finite temperature, gap closing, and Majorana overlap.  These problems may exist in the experimental systems too masking the TQPT and making it difficult to distinguish trivial and topological regimes.

More details on the role of topological visibility in this context can be found in Ref.~\cite{DasSarma2016How}, and we do not show any more TV results in the current paper except to make one remark. The calculated TV is approaching $-1$ (or not) whenever the corresponding Zeeman energy for the zero mode is above (below) the critical TQPT value $V_{Zc}$, thus distinguishing (theoretically) the MZM and ABS zero modes. For our purpose, any apparent zero mode (or almost-zero mode) below (above) the TQPT point (which is exactly known in our theory, but not in the experiment) is considered to be an ABS (MZM) by definition.

We note in this context that the trivial ZBCP in Fig.~\ref{fig:scanVZ}(b) and \ref{fig:scanVZ}(c) may persist to large Zeeman splittings (as in Fig.~\ref{fig:summary}(d)) going beyond the TQPT point ($V_Z>3.8,~4.5~$meV in Fig.~\ref{fig:scanVZ}(b) and \ref{fig:scanVZ}(c) respectively), and then the coalesced ABSs have eventually become MZMs at large enough magnetic field values (see, e.g., Fig.~\ref{fig:summary}(d)). Unfortunately, there is no way to know about such a trivial to topological crossover by looking simply at the ZBCP (without knowing the precise TQPT point), and hence experimentally, one cannot tell whether a coalesced ZBCP is trivial or topological by studying only the ZBCP. One way to distinguish is perhaps careful experimentation varying many experimental parameters (e.g., magnetic field, chemical potential, tunnel barrier, SC gap) to test the stability of the absolute value of the ZBCP against such perturbations. The MZM-induced topological ZBCP should manifest the universal strength of $2e^2/h$ whereas the trivial ABS-induced ZBCP will have non-universal behavior. Another issue which may become important in the experimental context~\cite{Deng2016Majorana} is that the bulk SC gap may collapse in the high magnetic field regime where one expects the MZM to manifest itself.

\subsection{Proximitized quantum dot}\label{ssec:proximitized}

\begin{figure*}[!htb]
\includegraphics[width=0.99\textwidth]{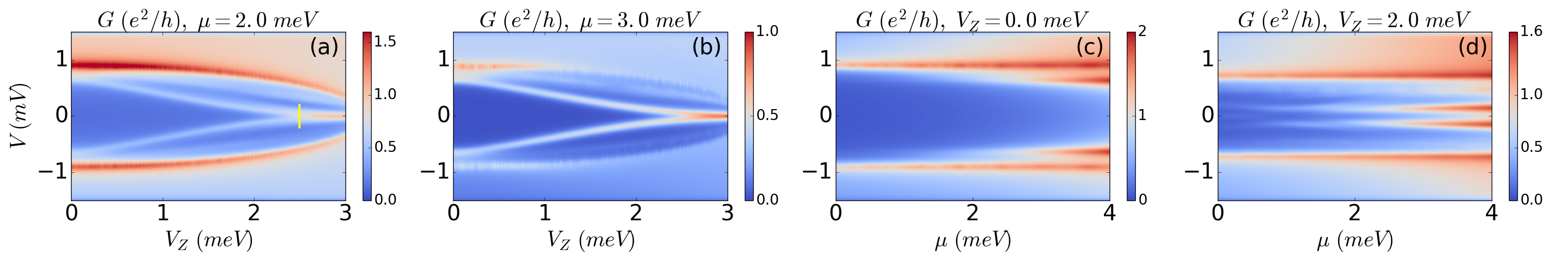}
\caption{(color online). Calculated differential conductance through four hybrid nanowire systems with the quantum dot completely proximitized and with the same amount of dissipation $\Gamma=0.01~$meV and temperature $T=0.02~$meV. (a) and (b): Differential conductance as a function of Zeeman splitting at fixed chemical potential. In (a), the critical Zeeman field is at $V_{Zc} = \sqrt{\Delta^2 + \mu^2} = 2.5~$meV (marked by a yellow vertical line), beyond which the system enters the topological regime. Here $\Delta=\lambda=1.5~$meV due to self-energy renormalizing the SC pairing. In (b), the critical Zeeman field is outside the range of $V_Z$, thus the zero-energy peak here is trivial. (c) and (d): Differential conductance as a function of chemical potential at fixed Zeeman field $V_Z=0$ and $V_Z=2.0~$meV.}
\label{fig:proximitized}
\end{figure*}

All the calculations in the previous subsections are based on hybrid structures with the quantum dot outside the nanowire, i.e., there is no induced superconductivity in the quantum dot at all. In real experimental situations, however, it is possible that unintentional quantum dots may appear inside the SC nanowire, or the parent superconductor may partially or completely proximitize the quantum dot. Another way of saying this is that ABSs may arise in the nanowire from unknown origins where no obvious quantum dots are present. (Such a possibility can never be ruled out although whether it actually happens in a particular experimental system or not would depend on unknown and uncontrolled microscopic details.) We now consider hybrid structures with the quantum dot completely proximitized and calculate the corresponding differential conductance. The calculated differential conductance is shown in Fig.~\ref{fig:proximitized}. Both Fig.~\ref{fig:proximitized}(a) and \ref{fig:proximitized}(b) are differential conductance as a function of Zeeman splitting. In Fig.~\ref{fig:proximitized}(a), the critical Zeeman field is at $V_{Zc} = \sqrt{\Delta^2 + \mu^2} = 2.5~$meV (marked by a yellow vertical line), beyond which the system enters the topological regime. By contrast in Fig.~\ref{fig:proximitized}(b), the critical Zeeman field is outside the range of $V_Z$, thus the zero-bias peak is trivial. But there is no way to differentiate between the two situations by just looking at the ZBCPs. Another intriguing phenomenon in Fig.~\ref{fig:proximitized}(a) and (b) is that the positions of the pair of ABSs at zero Zeeman field are much closer to the induced SC gap than situations where the quantum dot is not proximitized, as shown in Fig.~\ref{fig:scanVZ}. This is because the SC pairing for ABSs in fully proximitized dot is larger than the renormalized SC pairing in unproximitized dot. Thus the gap in the former case is larger and closer to the induced SC gap in the nanowire. Thus the position of ABS peaks at zero Zeeman field can be regarded as a clue to the degree of proximitization of the quantum dot. Such a feature is also manifest in Fig.~\ref{fig:proximitized}(c), where we show the conductance as a function of chemical potential at zero Zeeman field. In contrast with Fig.~\ref{fig:scanmu}(a), now the peaks from ABSs are closer to the induced SC gap. Fig.~\ref{fig:proximitized}(d)shows the differential conductance as a function of chemical potential at a finite Zeeman field $V_Z=2~$meV, where peaks from the two ABSs inside the SC gap are close to zero-energy. We believe that in most experimental situations the ABSs arise from strongly-coupled ``effective'' quantum dots within the nanowire (or from ``dots'' present at the Schottky barrier between the semiconductor nanowire and the normal metallic lead). More detailed results and discussion on hybrid structures with proximitized quantum dot are provided in Sec.~\ref{sec:dot}.

%%%%%%%%%%%%%%%%%%%%%%%%%%%%

\section{Self-energy model of quantum dot}\label{sec:analytic}

So far we have numerically calculated differential conductance through various nanowire systems, showing either ZBCPs or near-zero-bias peaks. Some of these conductance plots, e.g. Figs.~\ref{fig:scanVZ} and \ref{fig:scanmu}, are quite similar (essentially identical qualitatively) to those in the Deng \textit{et al}. experiment~\cite{Deng2016Majorana}, but this is only suggestive as we have no way of quantitatively simulating the experimental devices because of many unknown parameters (not the least of which are the detailed quantum dot characteristics). We have shown explicitly that ABSs could come together and remain stuck at zero energy in the quantum dot-nanowire hybrid system producing trivial ZBCPs which perfectly mimic the topological ZBCPs associated with MZMs in simple nanowires.  This tendency seems to be quite generic at higher chemical potentials whereas at lower chemical potentials the ABSs seem to simply repel each other. This section is devoted to understanding the relevant physics leading to the conductance patterns discussed above. We calculate analytically the energy spectra of hybrid structures, especially focusing on low-energy states, which can provide insightful information about the corresponding zero-bias conductance behavior. We mention that superconductivity, spin-orbit coupling, and Zeeman splitting are all essential ingredients for the ABS physics being discussed here. Thus, the zero-sticking property of trivial ABSs (as a function of magnetic field) is a generic feature of class D superconductors, even without any disorder.

With no loss of generality, we focus on a single hybrid structure with chemical potential $\mu=3.0~$meV using the minimal model of a constant $s$-wave paring potential in this section since the low-energy behavior is not affected by the way proximity SC is introduced. The basic idea here is to see how a self-energy theory of quantum dot bound states, taking explicitly (but perturbatively) into account the SC nanowire as well as Zeeman splitting and spin-orbit coupling, leads naturally to ABS-sticking near zero energy independent of trivial or topological regime one is considering. In other words, the tendency of ABSs coalescing near zero energy is a generic property of class D superconductors and has nothing whatsoever to do with MZMs or TQPT. This is consistent with a previous analysis~\cite{Mi2014X} of so-called Y-shaped resonances that were proposed to occur in generic quantum dots coupled to SCs on the grounds of random matrix theory where the system of interest was random (i.e., had disorder in it in sharp contrast to our disorder-free consideration). The focus of our work here is to expand on the likelihood of this occurrence in a spin-orbit coupled nanowire system in general even without any disorder. The resulting ZBCP may arise from an MZM in the topological regime or an ABS in the trivial regime controlled entirely by the magnetic field where it happens (i.e. whether this field is above or below the critical Zeeman field for the TQPT). What we find is (and show in Sec.~\ref{sec:numerical} in depth) that the trivial ABSs could stick to zero energy for a large range of magnetic field without being repelled away, thus mimicking the expected zero mode behavior of topological MZMs.

\subsection{Exact results from diagonalization}\label{ssec:exact}

\begin{figure*}[!htb]
\includegraphics[width=0.99\textwidth]{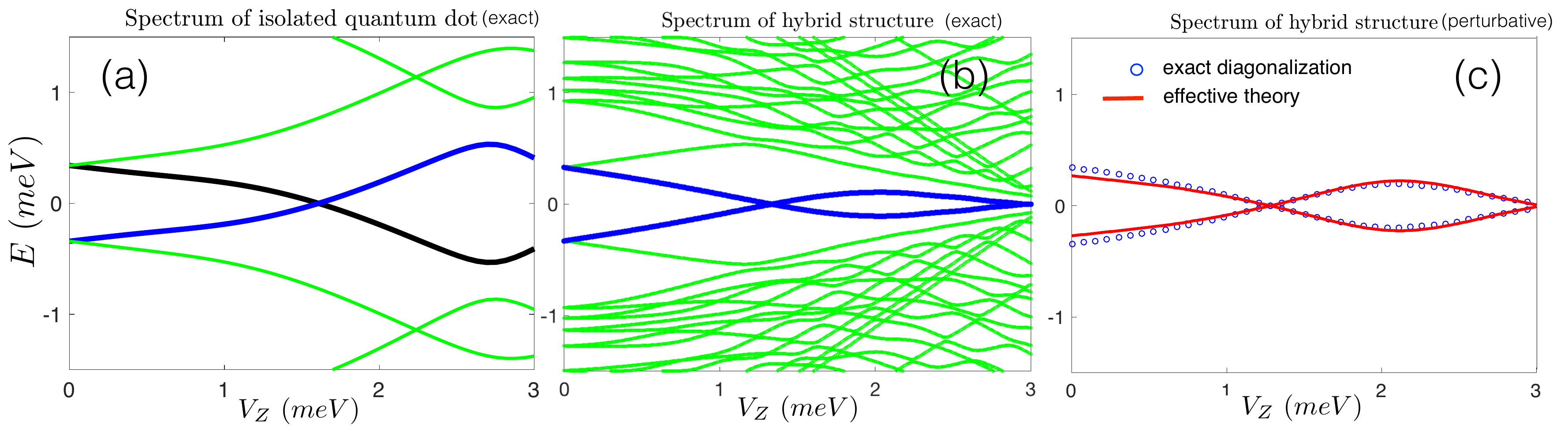}
\caption{(color online). The spectrum of the isolated quantum dot. The blue curve is the spectrum of the bound state whose energy crosses Fermi level as a function of Zeeman field. The black curve is its particle-hole partner, which is redundant in this case since there is no SC pairing in the isolated quantum dot. Green curves are spectra of other bound states that are always well above or below Fermi level.}
\label{fig:analytic}
\end{figure*}

First, we look at the isolated quantum dot system whose Hamiltonian is $H_{QD}$ as shown in Eq.~\eqref{eq:H_QD}. The spectrum is shown in Fig.~\ref{fig:analytic}(a), where the blue curve denotes the bound state whose energy crosses Fermi level as a function of Zeeman field, the black curve denotes its particle-hole partner, which is redundant in this case since there is no SC pairing in the isolated quantum dot, and the green curves are spectra of other bound states that are always well above or below the Fermi level. As Zeeman field increases, the bound state eigen-energy crosses Fermi level, but it is then repelled by neighboring energy states due to spin-orbit coupling, which leads to the beating shape in the spectrum. Note that the theory explicitly must consider both Zeeman and spin-orbit coupling effects. We provide more details on the directly calculated energy spectra of the hybrid system in Appendix~\ref{app:with_without}.

Second, we include the superconducting nanowire and couple it with the quantum dot; they together constitute the hybrid structure. The system is now class D (but with no disorder)--it has superconductivity, Zeeman splitting and spin-orbit coupling. The total Hamiltonian is a combination of Eqs.~\eqref{eq:H_NW} and \eqref{eq:H_QD}
\begin{align}
H_{tot} &= H_{QD} + H_{NW} + H_t, \nn
H_t &= u + u^{\dagger}= \hat{f}^{\dagger}_{\alpha}(-t\delta_{\alpha \beta} + i \alpha_R \sigma^y_{\alpha \beta})\hat{c}_{\beta} + h.c.,
\label{eq:H_tot}
\end{align}
where $H_{QD}$ is the isolated quantum dot, $H_{NW}$ is the superconducting nanowire and $H_t$ is the coupling between them. $\hat{c}$ annihilates an electron at the end of the nanowire adjacent to the dot and $\hat{f}^{\dagger}$ creates an electron at the end of the dot adjacent to the nanowire. By diagonalizing the total Hamiltonian, we obtain the spectrum shown in Fig.~\ref{fig:analytic}(b), where the blue curves are particle-hole pairs that cross Fermi level, while the green curves are states well above or below Fermi level. By focusing on the spectra near Fermi level in Fig.~\ref{fig:analytic}(a) and (b), we see that the effect of the nanowire on the bound states of the quantum dot is that it shifts the spectrum and changes the spectrum curvature. The strong similarity between the spectrum of hybrid structure in Fig.~\ref{fig:analytic}(b) and the differential conductance in Fig.~\ref{fig:scanVZ}(a) indicates that the energy spectrum provides a good perspective on understanding the behavior of differential conductance, which is in general true at low temperature since the low-temperature transport is dominated by contributions from the low-energy states..

\subsection{Approximate results from self-energy theory}\label{ssec:approximate}

The numerical results in the previous subsection show graphically that coupling with the nanowire has a perturbative effect on the energy spectrum of the isolated quantum dot. We now calculate the analytic form of the perturbed spectrum in the quantum dot using an effective theory including perturbative corrections of the quantum dot spectra arising from the superconducting nanowire. The total Hamiltonian is still Eq.~\eqref{eq:H_tot}. We first project $H_{QD}$ onto the subspace spanned by the bound state crossing the Fermi level and its redundant hole partner, thus obtaining
\begin{align}
H^{\text{eff}}_{QD} = E^{\text{eff}}_{QD} \gamma_z,
\label{eq:EeffQD}
\end{align}
where $\gamma$'s are Pauli matrices on the projected two-dimensional subspace. $E^{\text{eff}}_{QD} = E_{QD} - \Delta \mu$, with $E_{QD}$ the bare energy of the bound state in the isolated quantum dot crossing the Fermi level, and $\Delta \mu$ represents the renormalization of the chemical potential due to projecting out all the other states. Then we integrate out the degrees of freedom in the nanowire, leading to an energy-dependent self-energy term in the isolated quantum dot
\begin{align}
\Sigma(\omega) = u( \omega - H_{NW} )^{-1}u^{\dagger},
\end{align}
where $u, u^{\dagger}$ represent the hopping between nanowire and quantum dot. Similarly, we project this self-energy term onto the two-dimensional subspace in quantum dot and get
\begin{align}
F(\omega) = \hat{P} \Sigma(\omega) \hat{P},
\end{align}
where $\hat{P}$ denotes the projection operator. Thus the approximate energy spectrum of the hybrid structure near the Fermi level is given by the roots of
\begin{align}
\text{Det}( \omega - H^{\text{eff}}_{QD} - F(\omega) ) = 0.
\label{eq:det}
\end{align}
The spectrum obtained from this effective theory is shown in Fig.~\ref{fig:analytic}(c), where blue circles represent the exact spectra from diagonalizing the total Hamiltonian in the previous subsection, while the red line is the spectrum obtained from the projected effective theory with the appropriate choice of $\Delta \mu$ in Eq.~\eqref{eq:EeffQD}. The excellent agreement between the exact diagonalization results and the effective theory results demonstrates that the proximity effect from the SC nanowire onto the quantum dot bound states is perturbative renormalization.

We can take one more step to get an analytic expression of the ABS spectra using the low energy assumption $\omega \to 0$. In the nanowire, particle-hole symmetry constrains the form of projected self-energy term $F(\omega)$ to be (Appendix~\ref{app:expansion})
\begin{align}
F(\omega) &= f_0(\omega) \gamma_0 +  f_x(\omega) \gamma_x +  f_z(\omega) \gamma_z \nn
& \simeq \beta_0 \omega \gamma_0 +  \beta_x \omega \gamma_x +  \alpha_z \gamma_z,
\label{eq:F_expand}
\end{align}
where $f_{0,x}$ are odd functions of $\omega$, and $f_{z}$ is an even function of $\omega$. Here, $f$'s are expanded up to their leading order for small $\omega$. Then the leading order solution is given by the approximate root of Eq.~\eqref{eq:det}:
\begin{align}
\omega \simeq \frac{E^{\text{eff}}_{QD} + \alpha_z }{\sqrt{(\beta_0 - 1 )^2 - \beta^2_x}}.
\end{align}
This result indicates that the proximity effect of the nanowire is two-fold: it first shifts the projected spectrum of the isolated quantum dot, and it reduces the curvature (i.e., enhances the effective mass if we focus on the parabolic part) of the spectrum, since numerics show $\beta_0 \gg 1, \beta_x$.

Our finding is that the near-zero conductance peaks in hybrid structures are mainly contributed by the ABSs related to the quantum dot. These ABSs can be regarded as bound states in the quantum dot perturbatively renormalized by the nanowire. ABS spectra show parabolic shapes as a function of Zeeman field with renormalized effective mass and chemical potential. When the parabolic spectrum crosses the Fermi level, the spectrum together with its particle-hole partner manifests a beating pattern around midgap, and if this beating involves small amplitude, the resulting ABS will appear to be stuck at zero energy manifesting a generic ZBCP, which has nothing to do with MZMs. It is simply a low energy fermionic bound state in the SC gap. For the approximately zero-energy ABSs the renormalized effective mass is accidentally huge and the renormalized chemical potential shifts the ABS close enough to zero energy. How close is ``close enough'' depends entirely on the energy resolution of the experiment--all these apparent zero-energy trivial ABS modes are beating around midgap, it is only when this beating happens to be smaller than the resolution, the mode appears stuck at zero energy. Especially when broadening effects from finite temperature and/or intrinsic dissipation are larger than the beating amplitude, near-zero peaks seem to appear stuck at zero energy since the energy resolution is not fine enough to resolve the beating pattern. This makes it essentially impossible to obtain a simple analytic form for the range of magnetic field (i.e., range of $V_Z$) over which the trivial ABSs will remain close to zero--this range is a multidimensional complicated function determined determined by all the parameters of the hybrid system even in our simple perturbative model (chemical potential, magnetic field, induced gap, quantum dot confinement details, experimental resolution around zero bias, temperature, broadening, etc.).

We emphasize that all the four ingredients are essential in the perturbative theory:  quantum dot, superconducting nanowire, spin-orbit coupling, and Zeeman splitting.  What is, however, not necessary is topological $p$-wave superconductivity or Majorana modes. Generically, the ABSs in class D superconducting nanowires may be attracted to the midgap, and once they coalesce there, they will have a tendency to stick to zero energy. The fact that class D superconductors generically allow trivial zero-energy states can also be seen from the known level statistics whose probability distribution has no repulsion from zero energy~\cite{Altland1997Nonstandard} in contrast to the other class superconductors. What we show in our analysis here is that this tendency of D-class peaks to stick to zero energy can happen for simple ABSs arising from single quantum dots, there is no need to invoke disorder as leading to such class D peaks~\cite{Liu2012Zero, Bagrets2012Class, Pikulin2012Zero, Sau2013Density}, and such zero-bias sticking could survive over a large range of magnetic field variation. The disorder-free nature of our theory distinguishes it from earlier work on class D zero-bias peaks which are caused by disorder induced quantum interference~\cite{Liu2012Zero, Bagrets2012Class, Pikulin2012Zero, Sau2013Density}.

Specifically, the ingredients discussed in the previous paragraph produce localized ABSs in the symmetry class D with a large weight at the end. Superconductivity provides particle-hole symmetry and Zeeman splitting breaks time-reversal symmetry in order to place the system in the symmetry class D. Spin-orbit coupling is needed to break spin-conservation without which the system would become two copies of a different symmetry class. Class D is important to induce energy-level repulsion that pushes the lowest pair of ABSs towards zero energy~\cite{Beenakker1997Random}. As seen from Eq.~\eqref{eq:F_expand}, the self-energy from the superconducting nanowire that is in symmetry class D generates the eigenstate repulsion which pushes the ABSs towards zero energy. The tendency of ABSs to stick as the Zeeman field is varied in class D is analogous to the Y-type resonance discussed in the context of superconducting quantum dots~\cite{Mi2014X}.

A definitive prediction of the arguments in the previous paragraph is that the combination of spin-orbit coupling and Zeeman splitting is required to create states that stick to zero energy, which only occurs in symmetry class D. This can be checked explicitly by obtaining the corresponding low-energy spectra in the hybrid quantum dot-nanowire system without Zeeman splitting or without spin-orbit coupling respectively. We carry out these direct numerical simulations and show the corresponding results in Appendix~\ref{app:with_without}, where it can be clearly seen that only the situations with superconductivity, spin-orbit coupling, and Zeeman splitting all being finite allow for the possibility of zero-sticking (and beating) of ABS. Thus, the same ingredients which lead to the existence of MZMs in nanowires (superconductivity, spin-orbit coupling, and Zeeman splitting) also lead to Andreev bound states sometimes producing almost-zero-energy midgap states. This is a most unfortunate situation indeed. This means that confirming the presence of Majoranas through transport measurement might be more complicated than simply observing a robust zero-bias peak. While a ZBCP is indeed necessary, it is by no means sufficient even if the ZBCP value agrees with the expected quantized  conductance of $2e^2/h$. It will also be necessary to vary the tunneling through the quantum dot to reduce it to a quantum point contact which can explicitly be verified to be carrying a single spin-polarized channel in the normal state~\cite{Wimmer2011Quantum}. In addition, it must be ensured that the ZBCP quantization is indeed robust against variations in various system parameters such as tunnel barrier, magnetic field, and chemical potential. In particular, varying the quantum dot confinement through tunable external gate voltage and checking for the stability of the ZBCP may be essential to ensure that the relevant ZBCP indeed arises from MZMs and not ABSs. This is considered in the next section.

%%%%%%%%%%%%%%%%%%%%%%%%%%%%

\section{Distinguishing between trivial and topological zero modes}\label{sec:distinguish}

\begin{figure*}[!htb]
\includegraphics[width=0.99\textwidth]{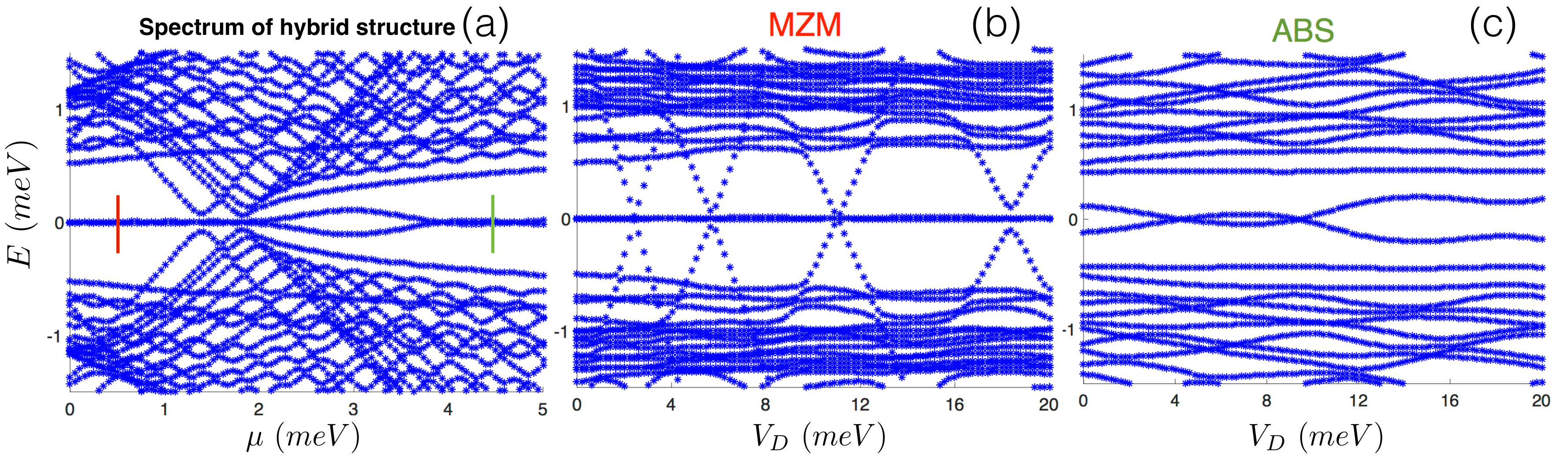}
\caption{(color online).(a) Calculated energy spectrum of a hybrid structure as a function of chemical potential $\mu$ with fixed Zeeman splitting $V_Z=2.0~$meV, dot depth $V_D = 4.0~$meV. Critical chemical potential is at $\mu_c = \sqrt{V^2_Z - \Delta^2} \simeq 1.8~$meV with red (green) lines indicating topological (trivial) zero modes. (b) Fixed chemical potential in the topological regime $\mu=0.5~$meV $<\mu_c$, to see how MZMs vary with the depth of the quantum dot.  (c) Fixed chemical potential in the non-topological regime $\mu=4.5~$meV $>\mu_c$, to see how near-zero-energy ABSs vary with the depth of the quantum dot. We see that MZMs are robust against the change of the dot depth, while ABSs oscillate or split with the change of the dot depth. So varying the dot depth(experimentally changing the gate potential) could be a stability test distinguishing between topological MZMs and non-topological ABSs.}
\label{fig:distinguish}
\end{figure*}

In the previous sections, we numerically show that differential conductance from MZMs and near-zero-energy ABSs may share strong similarities with each other, making them hard to distinguish. Although theoretically one can look at topological criteria or TV to distinguish between the two cases, quantities like chemical potential and TV are hardly known in the real experimental setup. So in order to distinguish ZBCPs arising from topological and nontopological situations, we discuss an alternate experimentally (in principle) accessible method, i.e., to see how the zero modes are affected by the change of the depth of the quantum dot confinement potential. We mentioned before that the phenomenon of the generic existence of trivial ABS-induced zero modes is qualitatively independent of the quantum dot confinement details, but now we are asking a different question. We focus on a fixed hybrid structure with ABS- (or MZM-) induced zero modes, and ask how this specific zero mode and the near-zero-bias differential conductance (comparing the ABS and the MZM cases) react to the change in the depth $V_D$ of the quantum dot confinement potential keeping everything else exactly the same. 

\subsection{Energy spectra for hybrid structures with ABS and MZM-induced zero modes}\label{ssec:E_distinguish}
We show our numerical results in Fig.~\ref{fig:distinguish}. Fig.~\ref{fig:distinguish}(a) is the calculated spectrum as a function of chemical potential at fixed $V_Z=2.0~$meV for $V_D=4~$meV with topological MZM-(or trivial ABS-) induced zero modes at small (large) chemical potential regimes. Now, we ask how this spectrum evolves if we only vary $V_D$ keeping everything else exactly the same. Fig.~\ref{fig:distinguish}(b) presents the MZM spectrum (i.e., at small chemical potential) as a function of dot depth, showing that it is robust against change of dot depth. By contrast, Fig.~\ref{fig:distinguish}(c) shows the ABS spectrum (i.e., large chemical potential) as a function of the dot potential depth, clearly showing that the ABS ``zero mode'' is not stable and oscillates (or splits) as a function of the dot potential. Put in another way, the fact that we see near-zero-energy ABSs is quite accidental for any particular values of Zeeman splitting and chemical potential, which only happens when the dot depth is fine-tuned to be some value, e.g., $V_D=4~$meV so that the energy splitting of the ABS zero mode happens to be smaller than the resolution. So varying the dot depth (e.g., by experimentally changing gate potential) will be a stability test distinguishing topological MZMs and non-topological ABSs. Note that it is possible (even likely) that the original ABS-induced ZBCP will split as the dot potential changes whereas a new trivial zero mode could appear, but the stability (or not) of specific ZBCPs to gate potentials could be a powerful experimental technique for distinguishing trivial and topological ZBCPs. Of course, experimentally tuning the dot potential by an external gate may turn out to be difficult in realistic situations, but modes which are unstable to variations in gate potentials are likely to be trivial ABS-induced ZBCPs.

\subsection{Conductance for hybrid structures with ABS and MZM-induced zero modes}\label{ssec:G_distinguish}

\begin{figure*}[!htb]
\includegraphics[width=0.99\textwidth]{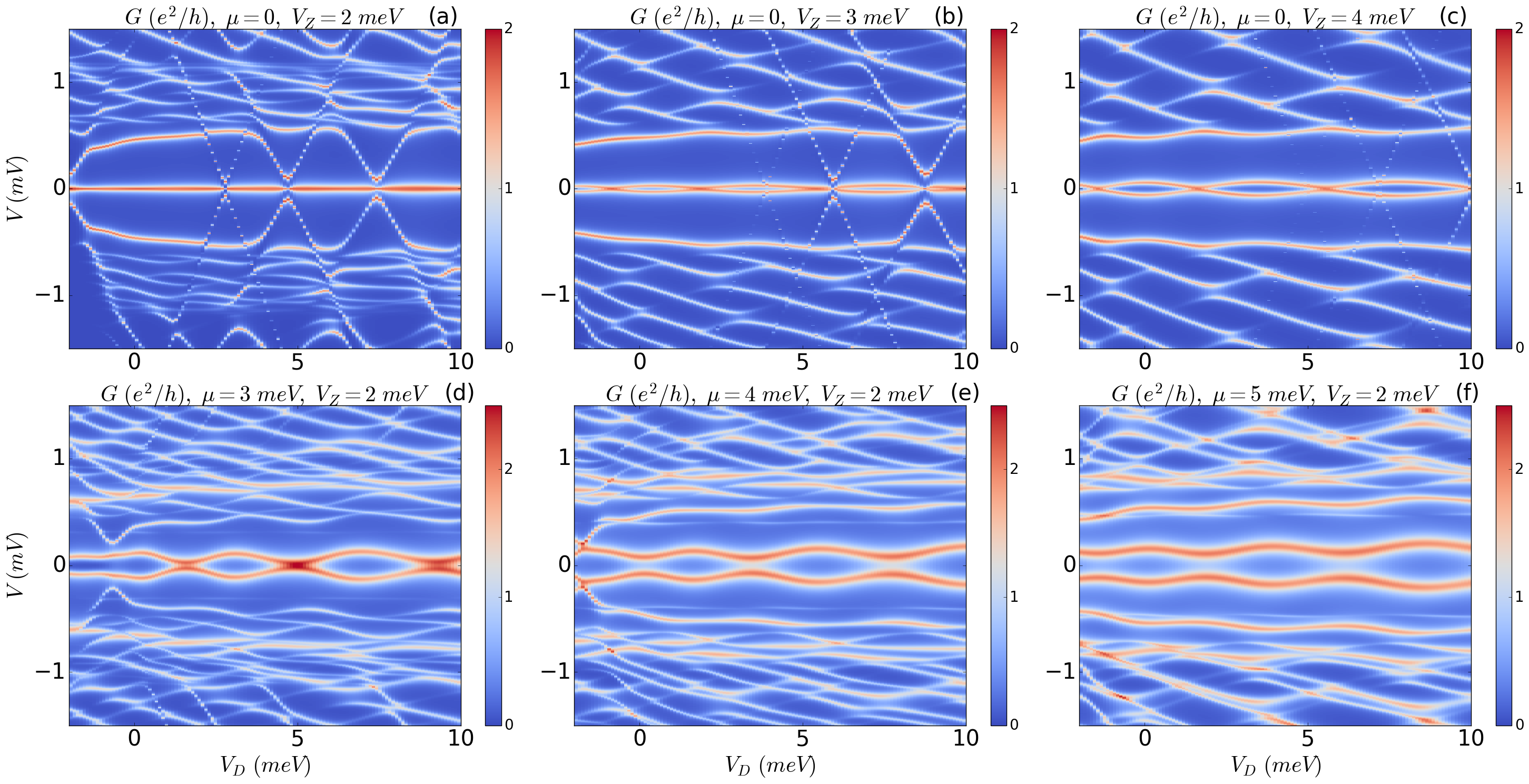}
\caption{(color online). Differential conductance as a function of the dot depth for hybrid structures at various but fixed chemical potential and Zeeman field. In the upper panel (a)-(c), all the hybrid structures are in the topological regime, i.e., all the zero-bias or near-zero-bias conductance peaks are MZM-induced. In the lower panels (d)-(f), all the hybrid structures are topologically trivial, i.e., the zero-bias or near-zero-bias conductance peaks are ABS-induced.}
\label{fig:G_distinguish}
\end{figure*}

We also show the calculated differential conductance through the hybrid structures as a function of the depth of the quantum dot and bias voltage, as shown in Fig.~\ref{fig:G_distinguish}. The conductance color plots in the upper panels (a)-(c) are for topological nanowires, i.e., $V_Z > V_{Zc}=\sqrt{\mu^2+\Delta^2}$, and thus all the zero-bias or near-zero-bias conductance peaks are MZM-induced. Such ZBCPs are stable against the variation of the depth of the quantum dot. With the increase of the Zeeman field, ZBCPs will be split and form Majorana oscillations as a function of the dot depth. By contrast, the conductance color plots in the lower panels (d)-(f) are for topologically trivial nanowires ($V_Z < \mu$), and thus all the near-zero-bias conductance peaks are ABS-induced. These nontopological near-zero-bias peaks also show beating patterns as a function of the dot depth, which is quite similar to the patterns for Majorana oscillations, although the origin is nontopological. But the crucial difference between the two situations is that ABS-induced oscillations are not guaranteed to cross zero bias for a variation of the parameter choice, e.g., increasing chemical potential as shown in (e) and (f), while for MBS-induced oscillations, although the amplitude of oscillation will increase with parameters in the nanowire (e.g., Zeeman field), the oscillation itself is sure to pass through zero-bias voltage. The difference between the two situations rises from the crucial fact that ABS induced ZBCPS are almost zero modes involving (always) some level repulsion whereas the MZM induced ZBCP oscillations arise from the splitting of a true zero mode in the infinite wire limit.

%%%%%%%%%%%%%%%%%%%%%%%%%%%%

\section{Quantum dots as short-range inhomogeneity}\label{sec:dot}

So far, our theoretical analysis (except for Sec.~\ref{ssec:proximitized} and Fig.~\ref{fig:proximitized}) has focused on quantum dots explicitly created at the end of a nanowire (see Fig.~\ref{fig:schematic}). In this case  the quantum dot is normal (i.e. non-superconducting), while the rest of the wire is proximity-coupled to the parent superconductor. However, in general the quantum dot could be unintentional, i.e. the experimentalist may be unaware of its presence near the wire end, and it could be partially or completely covered by the superconductor. For example, such a situation may arise if a potential well with a depth of a few meV forms near the end of the proximitized segment of the wire. Similar phenomenology emerges in the presence of a low (but wide enough) potential barrier.  After all, there is no easy way to rule out shallow potential wells (and low potential barriers) inside the nanowire or near its ends. In this context, we emphasize that a better understanding of the profile of the effective potential along the wire represents a critical outstanding problem. It turns out that all our results obtained so far still apply qualitatively even if the quantum dot is partially or completely inside the nanowire. In these cases we obtain exactly the same type of low-energy ABSs that have a tendency of sticking together near zero energy, thus producing ZBCPs that mimic MZM-induced ZBCPs. We present these results in detail  below. We are providing these results here in order to go all the way from an isolated non-superconducting dot at the wire end (as in the previous sections of this paper) to a situation where the dot is inside the wire and is completely superconducting. We explicitly establish that the main results of the previous sections can be obtained everywhere within this range, i.e. from isolated dots to dots completely inside the nanowire. In fact, this behavior is rather generic in non-homogeneous semiconductor nanowires~\cite{Moore2016Majorana}. Finally,  in this section we pay special attention to the profile of the ZBCPs associated with the almost-zero-energy ABSs. The key question that we want to address is whether or not a quantized ZBCP (i.e., a ZBCP with a peak height of $2e^2/h$) can be used as a hallmark for the Majorana zero modes expected to emerge beyond a certain critical field.

%%%%%%%%%%%%%%%%%%%%%%%%%%%%
\begin{figure}[t]
\begin{center}
\includegraphics[width=0.48\textwidth]{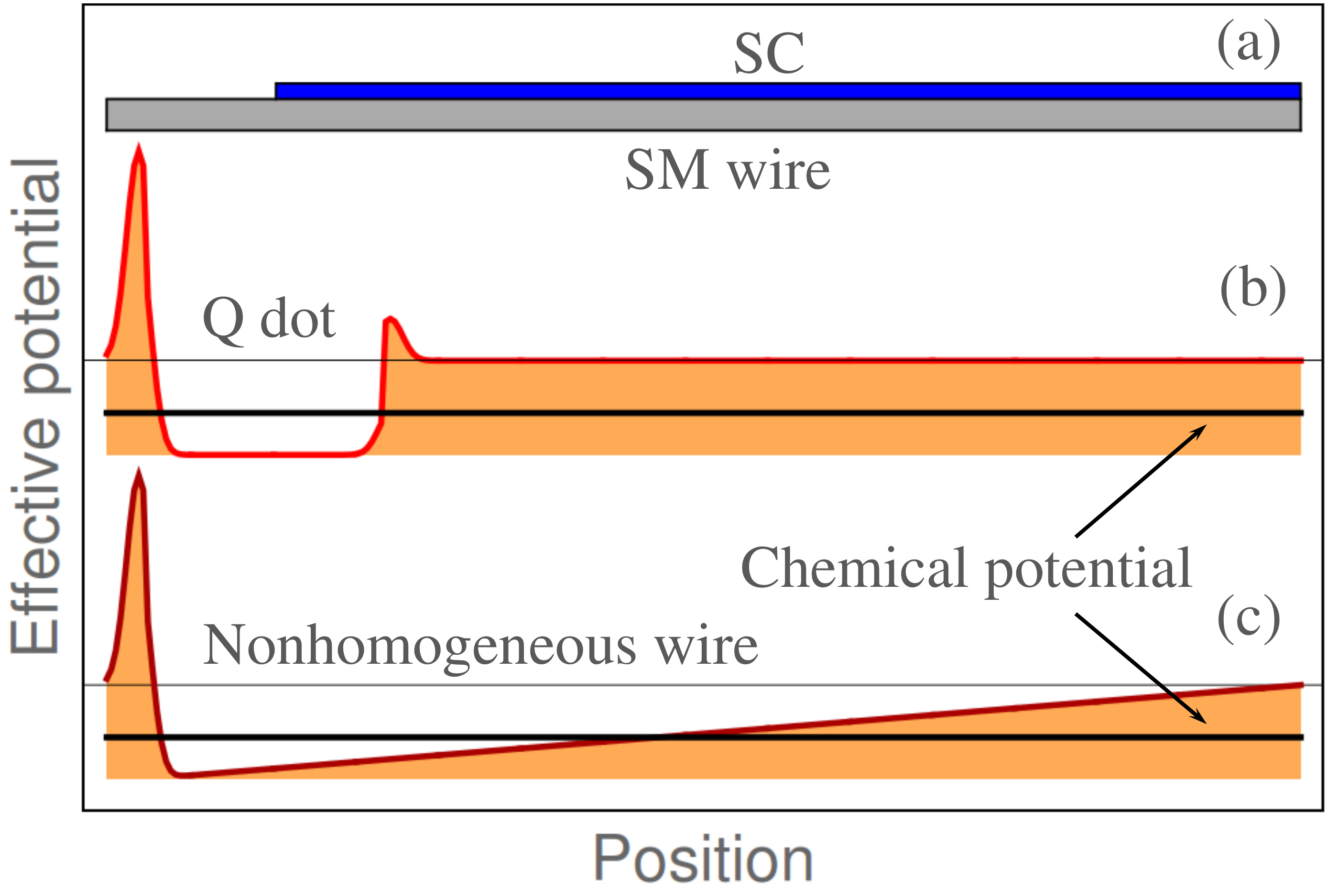}
\vspace{-3mm}
\end{center}
\caption{(color online). (a) Schematic representation of hybrid structure.  (b) Effective potential as a function of position for a wire with a quantum dot near its left end.  In the calculations the length of the quantum dot region is $250~$nm, while the rest of the wire is $1~\mu$m long. Note that the length parent superconductor (SC) can be varied, so that the quantum dot region can be uncovered, partially covered, or completely covered by the SC. (c) Smooth non-homogeneous effective potential. The peak at the left end of the wire represents the tunnel barrier.}
\vspace{-2mm}
\label{Fig_V1}
\end{figure}
%%%%%%%%%%%%%%%%%%%%%%

In Fig. \ref{Fig_V1}, we represent schematically the hybrid structure [panel (a)] and the effective potential [panel (b)] corresponding to three different situations that we consider explicitly in this section using exactly the same model parameters: dot entirely outside the proximitized segment of the nanowire, dot completely inside the nanowire (i.e., the whole dot is superconducting), and dot partially covered by the parent superconductor. The depth of the potential well in the quantum dot region is about $1~$meV and its length is $250~$nm.  The coupling between the quantum dot and the rest of the wire is controlled by the height of the corresponding potential barrier [see panel (b) in Fig.  \ref{Fig_V1}]. In addition, the coupling depends on how much of the dot is covered by the superconductor. The parameters used in our calculations correspond to intermediate and strong coupling regimes. We note that replacing the potential well from Fig. \ref{Fig_V1} (b) with potential barrier of a height several times larger than the induced gap $\Delta_{ind}$  leads to low-energy features similar to those described below for the potential well. Finally, for comparison we also consider a nanowire with a smoothly varying non-homogeneous potential [panel (c) in Fig.  \ref{Fig_V1}]. 

%%%%%%%%%%%%%%%%%%%%%%%%%%%%
\begin{figure}[t]
\begin{center}
\includegraphics[width=0.48\textwidth]{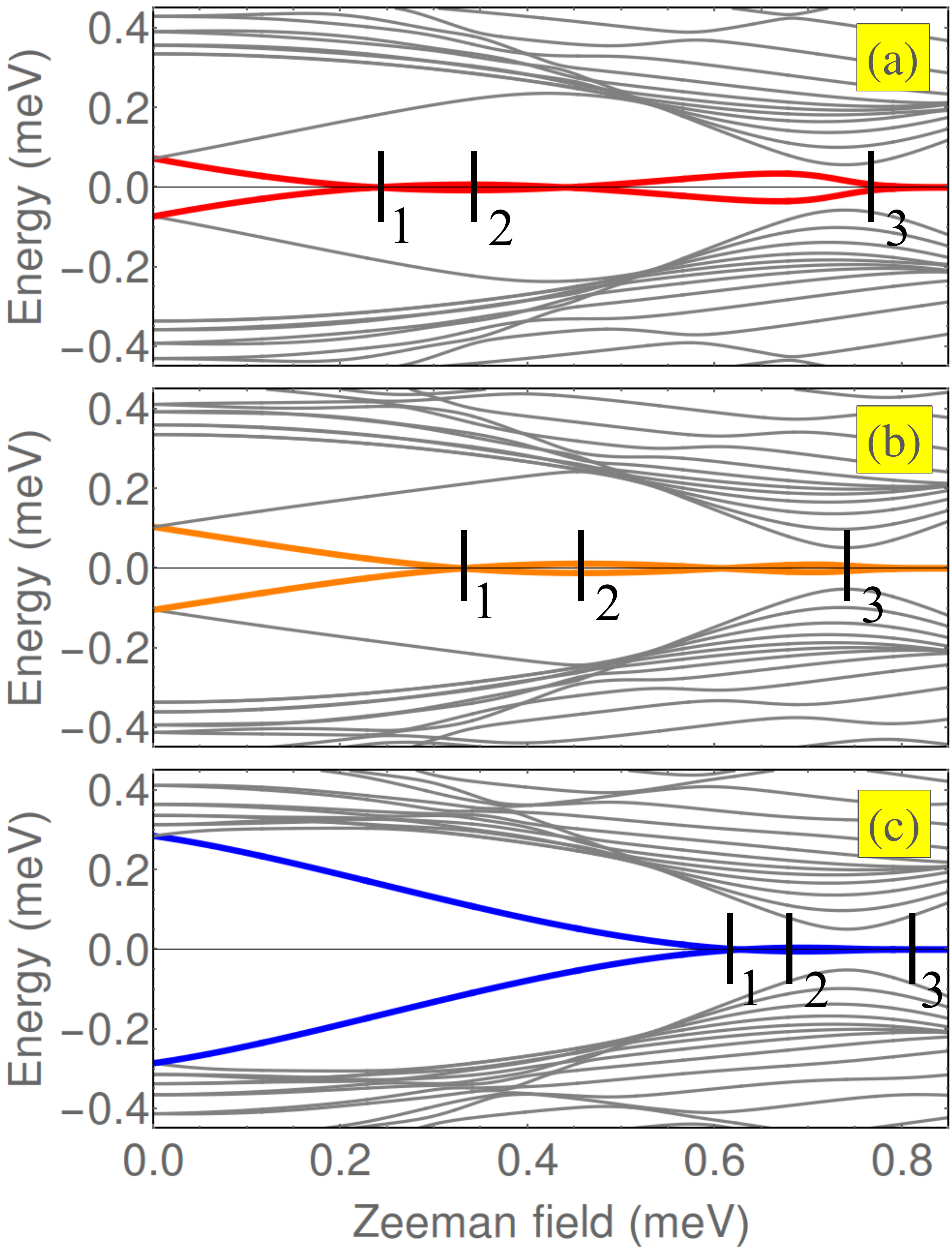}
\vspace{-3mm}
\end{center}
\caption{(color online). Dependence of the low-energy spectrum on the applied Zeeman field for a nanowire with a quantum dot near the left end (see Fig. \ref{Fig_V1}). (a) Quantum dot outside the superconducting region. (b) Quantum dot half-covered by the parent superconductor. (c) Completely covered quantum dot.   The induced gap is $\Delta_{ind}=0.25~$meV and the chemical potential $\mu=-2.83\Delta_{ind}$, which corresponds to a critical Zeeman field of about $0.75~$meV. The zero-temperature conductance along various constant field cuts marked ``1'', ``2'', and ``3'' are shown in Fig. \ref{Fig_V3}.}
\vspace{-2mm}
\label{Fig_V2}
\end{figure}
%%%%%%%%%%%%%%%%%%%%%%	

In Fig. \ref{Fig_V2} we show the calculated low lying energy spectra for three cases: (a) normal dot (i.e. uncovered by the SC), (b) half-covered dot, and (c) fully-covered dot. The system is characterized by an induced gap $\Delta_{ind}=0.25~$meV and a chemical potential $\mu=-2.83\Delta_{ind}$. The corresponding critical field associated with the topological quantum phase transition, $V_{Zc}\approx 3\Delta_{ind}=0.75~$ meV, is signaled by a minimum of the quasiparticle gap, as expected in a finite length system.    
First, we note that  all three situations illustrated in Fig. \ref{Fig_V2} clearly show trivial almost-zero-energy ABSs in a certain range of Zeeman field (lower than the critical field). However, the Zeeman field $V_Z^*$ associated with the first zero-energy crossing is significantly lower in the case of an uncovered dot [panel (a)] as compared to the partially-covered dot [panel (b)] and especially the fully covered dot [panel (c)]. Consequently, the range of Zeeman field corresponding to almost-zero-energy ABSs gets reduced with increasing the coverage of the quantum dot by the SC. Another key feature is the dependence of the energy of the ABS at $V_Z=0$ on the dot coverage. For the fully covered dot [panel (c)], this energy is practically $\Delta_{ind}$. In fact, by proximity effect, all the states that ``reside'' entirely under the parent superconductor  have energies (at $V_Z=0$) equal or larger than the induced gap for the corresponding band. By contrast, the zero-field energy of the ABSs in the half-covered  [panel (b)]
 and uncovered [panel (a)] dots is significantly lower that induced gap. To obtain such a state it is required that a significant fraction of the corresponding wave function be localized outside the proximitized segment of the wire. We find that, quite generically, strongly coupled dots that are uncovered or partially covered (when the uncovered fraction is significant) can support ABSs that i) have energies at $V_Z=0$ much smaller than the induced gap and ii) are characterized by ``merging fields'' $V_Z^*$ significantly lower than the critical value $V_{Zc}$. Consequently, in  hybrid systems having strongly coupled dots at the end it is rather straightforward to obtain low-energy Andreev bound states that merge toward zero and generate MZM-like zero-bias conductance peaks in the {\em topologically-trivial} regime, way before the topological quantum phase transition. In a real system it is possible that superconductivity be suppressed by the magnetic field before reaching the critical value $V_{Zc}$. In such a scenario, a robust ZBCP that sticks to zero energy over a significant field  range is entirely caused by (topologically trivial) merging ABSs, rather than (non-Abelian) MZMs. We speculate that the the rigid zero-energy state shown in Fig. S6 of Ref.~\cite{Deng2016Majorana} is an example of such a trivial (nearly) zero-energy state.  

%%%%%%%%%%%%%%%%%%%%%%%%%%%%
\begin{figure}[t]
\begin{center}
\includegraphics[width=0.48\textwidth]{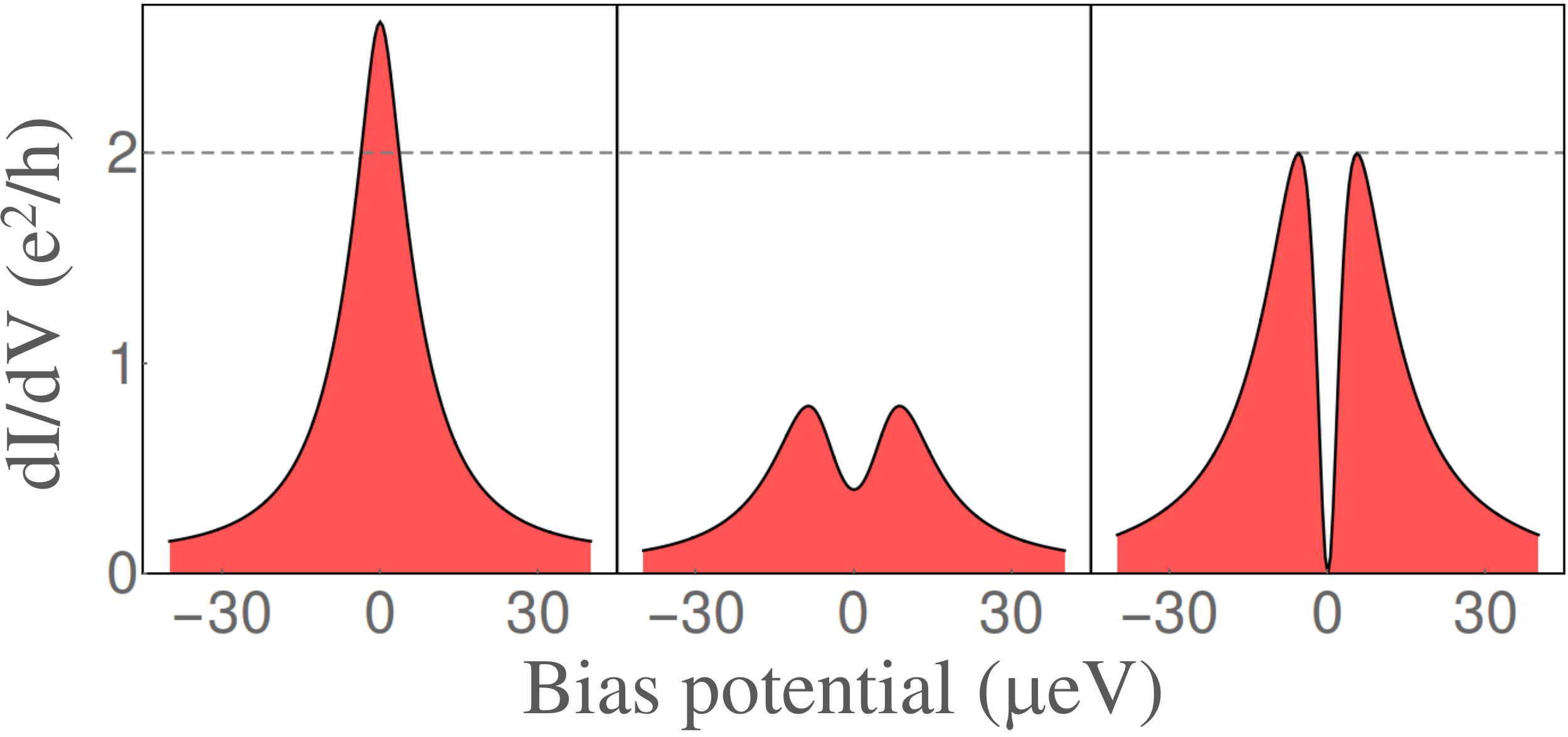}
\includegraphics[width=0.48\textwidth]{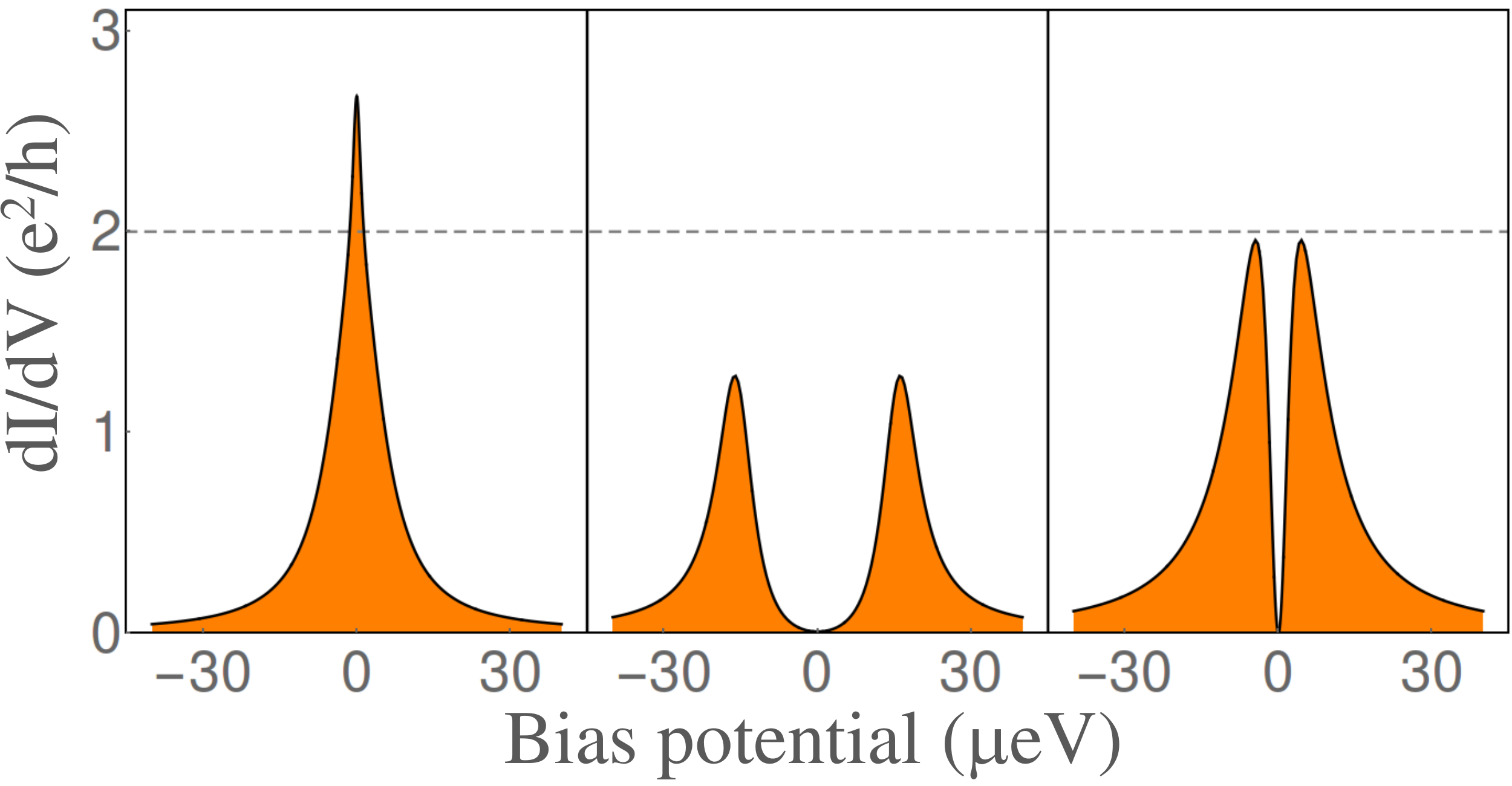}
\includegraphics[width=0.48\textwidth]{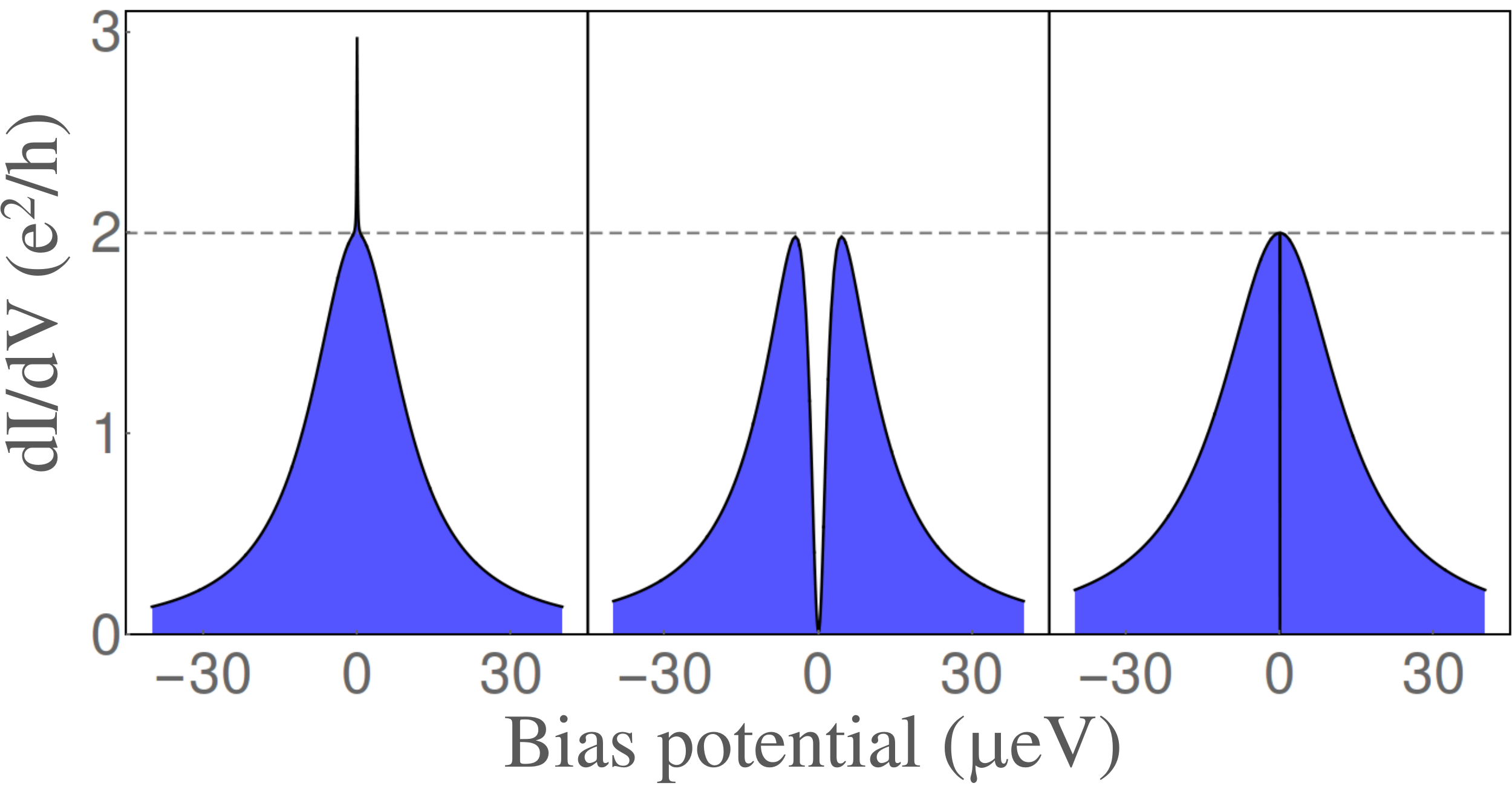}
\vspace{-3mm}
\end{center}
\caption{(color online). Differential conductance as function of the bias voltage for a quantum dot not covered by the superconductor (top panel), a half-covered dot (middle panel), and a fully-covered quantum dot (bottom panel). Each panel shows low-energy conductance peaks for three different values of the Zeeman field marked ``1'', ``2'', and ``3'' in the corresponding panel of Fig. \ref{Fig_V2}.}
\vspace{-2mm}
\label{Fig_V3}
\end{figure}
%%%%%%%%%%%%%%%%%%%%%%		
   
Next, we address the following question: can one discriminate between a MZM-induced zero-bias conductance peak and a trivial, ABS-induced ZBCP based on the height of the peak at zero temperature? More specifically, does the observation of a quantized peak guarantee its MZM nature? In short, the answer is {\em no}. However, observing a quantized ZBCP that is robust against small variations of parameters such as the Zeeman field, the chemical potential, and external gate potentials provides  strong indication  that the peak is probably not generated by merging ABSs partially localized outside the proximitized segment of the wire, i.e. scenarios (a) and (b) in Fig. \ref{Fig_V2}. The results that support this conclusion are shown in Fig. \ref{Fig_V3}. 
Each panel in Fig.  \ref{Fig_V3} shows the (low-energy) differential conductance at $T=0$ for  three different values of the Zeeman field marked ``1'', ``2'', and ``3'' in the corresponding panel of Fig. \ref{Fig_V2}. Generally, the largest value of the ZBCP obtains for Zeeman fields corresponding to the first zero-energy crossing, $V_Z^*$, marked ``1'' in Fig. \ref{Fig_V2}. In this case, the maximum height exceeds $2e^2/h$. However, for the fully covered dot (bottom panel) the excess conductance consists of a very narrow secondary peak that would be practically unobservable at finite temperature. In fact, we find that in the case of a fully covered dot, at  low-temperature, the conductance peak height is practically quantized in both the trivial regime (field cuts ``1'' and ``2'') and the topological regime   (field cut ``3''), regardless of whether the ZBCP is split or not. By contrast, for the uncovered and the half-covered dots (top and middle panels, respectively) the peak height can have any value between $0$ and $4e^2/h$ in the trivial regime and becomes quantized in topological regime. Of course, a quantized ZBCP can be obtained even in the trivial regime at certain specific values of the Zeeman field, but its quantization is not robust against small variations of the control parameters (e.g., Zeeman splitting, chemical potential, SC gap).
		
%%%%%%%%%%%%%%%%%%%%%%%%%%%%
\begin{figure}[t]
\begin{center}
\includegraphics[width=0.48\textwidth]{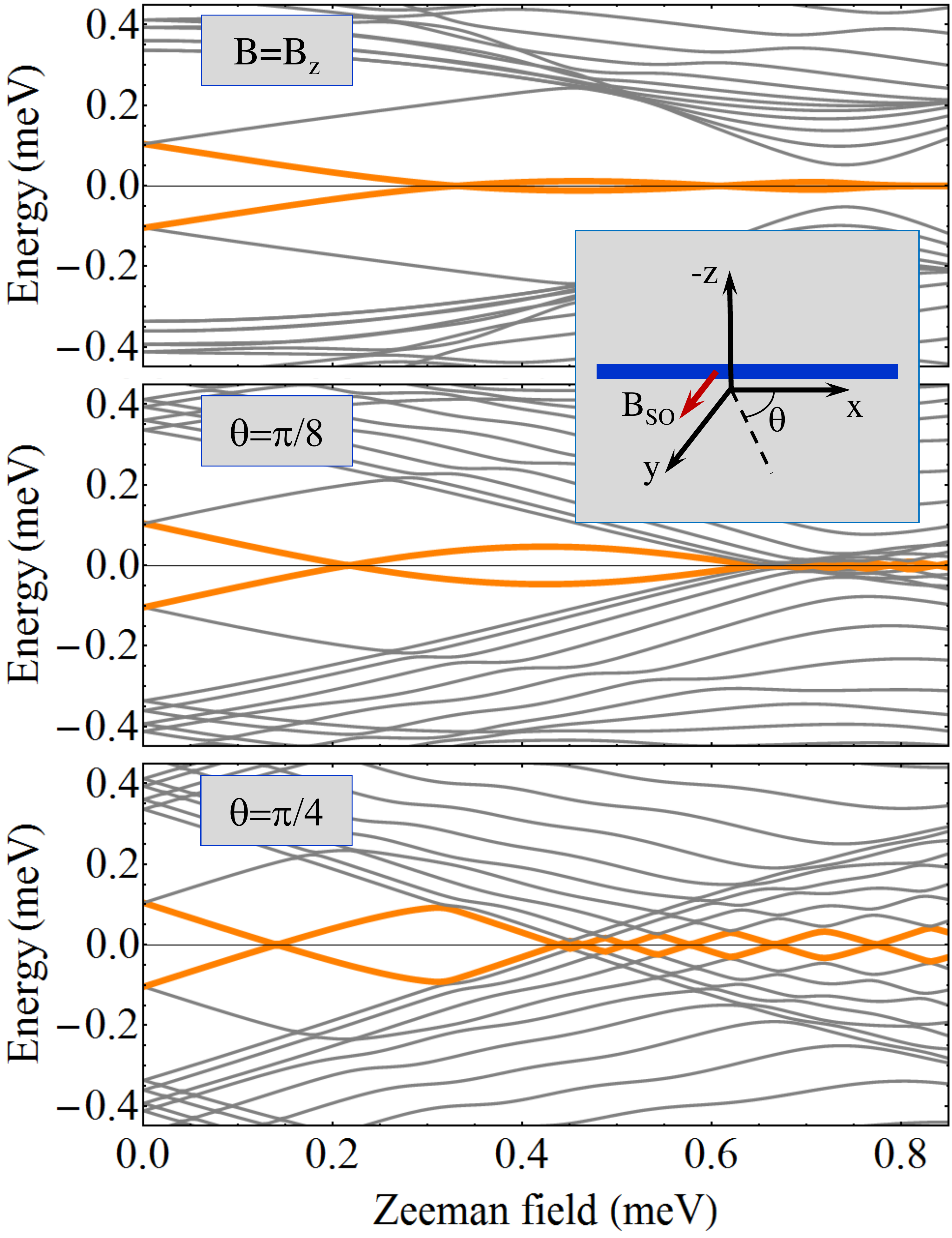}
\vspace{-3mm}
\end{center}
\caption{(color online).  Dependence of the low-energy spectrum from Fig. \ref{Fig_V2} (b) on the orientation of the applied magnetic field. {\em Top}: Magnetic field oriented along the $z$ axis (i.e. perpendicular to the wire and the effective SO field, see inset). The spectrum is identical to panel (b) from Fig.  \ref{Fig_V2}. {\em Middle and bottom}:  Rotating the field in the x-y plane destroys the property of the ABSs to coalesce into stable nearly zero energy modes. In addition, the spectrum becomes gapless above a certain (angle-dependent) value of the Zeeman splitting. }
\vspace{-2mm}
\label{Fig_V4}
\end{figure}
%%%%%%%%%%%%%%%%%%%%%%	
		
\begin{figure}[t]
\begin{center}
\includegraphics[width=0.48\textwidth]{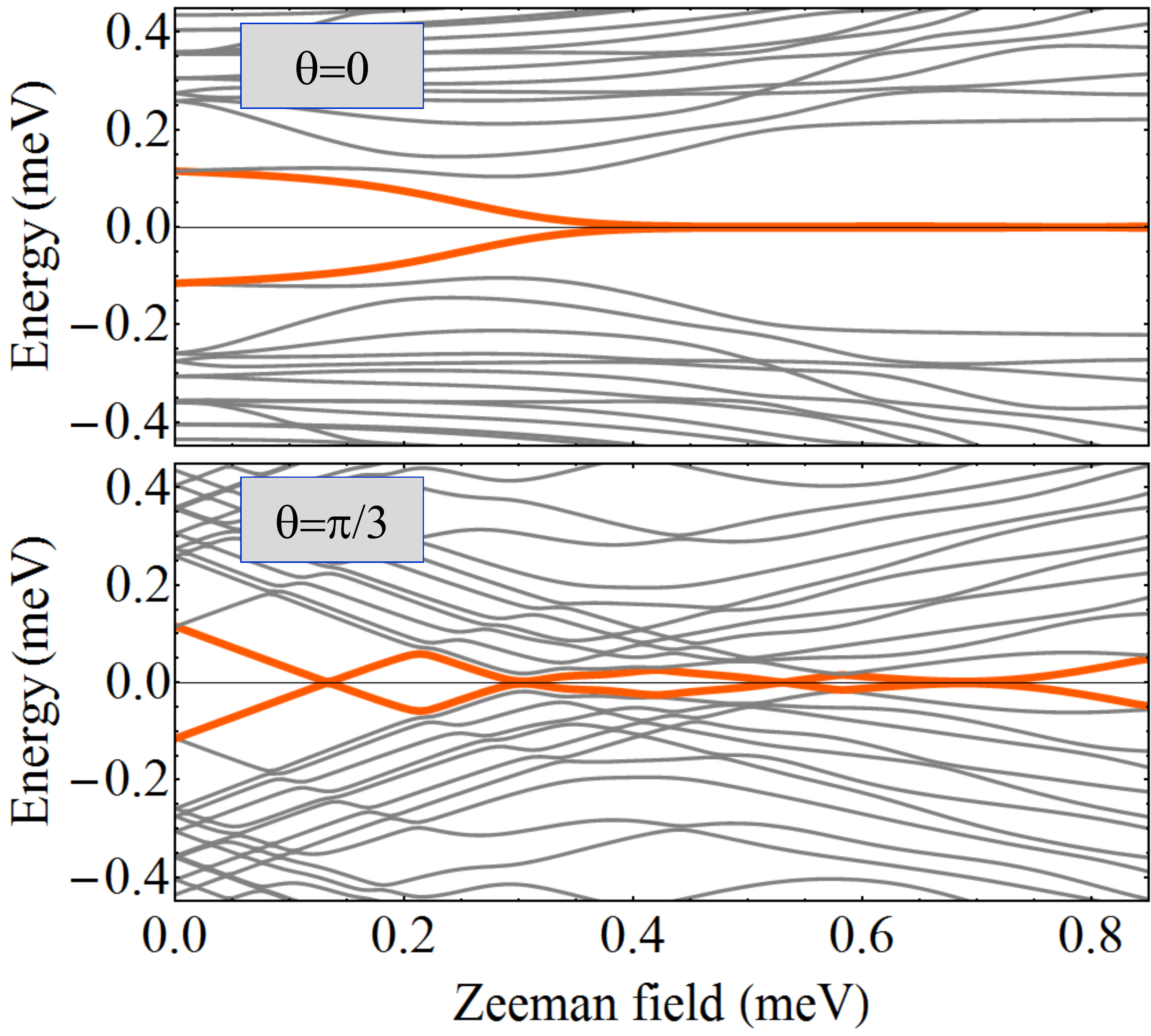}
\vspace{-3mm}
\end{center}
\caption{(color online). Dependence of the low-energy spectrum on the field orientation for a wire-dot system in the Majorana regime.  The model parameters are the same as in Fig.  \ref{Fig_V2} (b), except the chemical potential, which is set to  $\mu=-0.25\Delta_{ind}$. The top panel corresponds to a field oriented along the wire (or any other direction in the x-z plane), while the bottom panel corresponds to an angle $\theta=\pi/3$ in the x-y plane (see inset of Fig. \ref{Fig_V4}). Note the similarity with the bottom panel from Fig. \ref{Fig_V4}.}
\vspace{-2mm}
\label{Fig_V5}
\end{figure}

A key requirement for the  realization of  topological superconductivity and Majorana zero modes in semiconductor-superconductor hybrid structures is that the applied magnetic field be perpendicular to the effective Rashba spin-orbit (SO) field. More specifically, the MZMs  are robust against rotations of the applied field  in the plane perpendicular to the SO field, but become unstable as the angle between the applied and the SO fields (which corresponds to $\pi/2-\theta$ in the inset of Fig. \ref{Fig_V4}) is reduced. The natural question is whether the nearly-zero ABS modes induced by a quantum dot (or other type of inhomogeneity) show a similar behavior. We find that the coalescing ABSs (and, more generally, the low-energy spectrum) are insensitive to rotations of the applied field in the plane perpendicular to the effective SO field (i.e. the  x-z plane in Fig.  \ref{Fig_V4}). This property is illustrated by the spectrum shown in the top panel of Fig. \ref{Fig_V4} corresponding to a field oriented along the z-axis. Note that this spectrum is identical to  Fig. \ref{Fig_V2} (b), which corresponds to a field oriented along the x-axis. By contrast, when the field is rotated in the x-y plane, the nearly-zero ABS mode becomes unstable (see the middle and bottom panels in \ref{Fig_V4}). In addition, the spectrum becomes gapless above a certain (angle-dependent)   value of the Zeeman splitting. We conclude that the coalescing ABSs behave qualitatively similar to the MZMs with respect to rotations of the field orientation.  To further support this conclusion, we calculate the low-energy spectra of the wire-dot system in the Majorana regime for two different orientations of the applied magnetic field. The results are shown in Fig. \ref{Fig_V5}. We note that rotating the field in the x-z plane (i.e. the plane perpendicular to the SO field) does not affect the spectrum. By contrast, rotating the field in the x-y plane changes the low-energy features in a manner similar to that discussed in the context of coalescing ABSs.

%%%%%%%%%%%%%%%%%%%%%%%%%%%%
%%%%%%%%%%%%%%%%%%%%%%			

\begin{figure}[t]
\begin{center}
\includegraphics[width=0.48\textwidth]{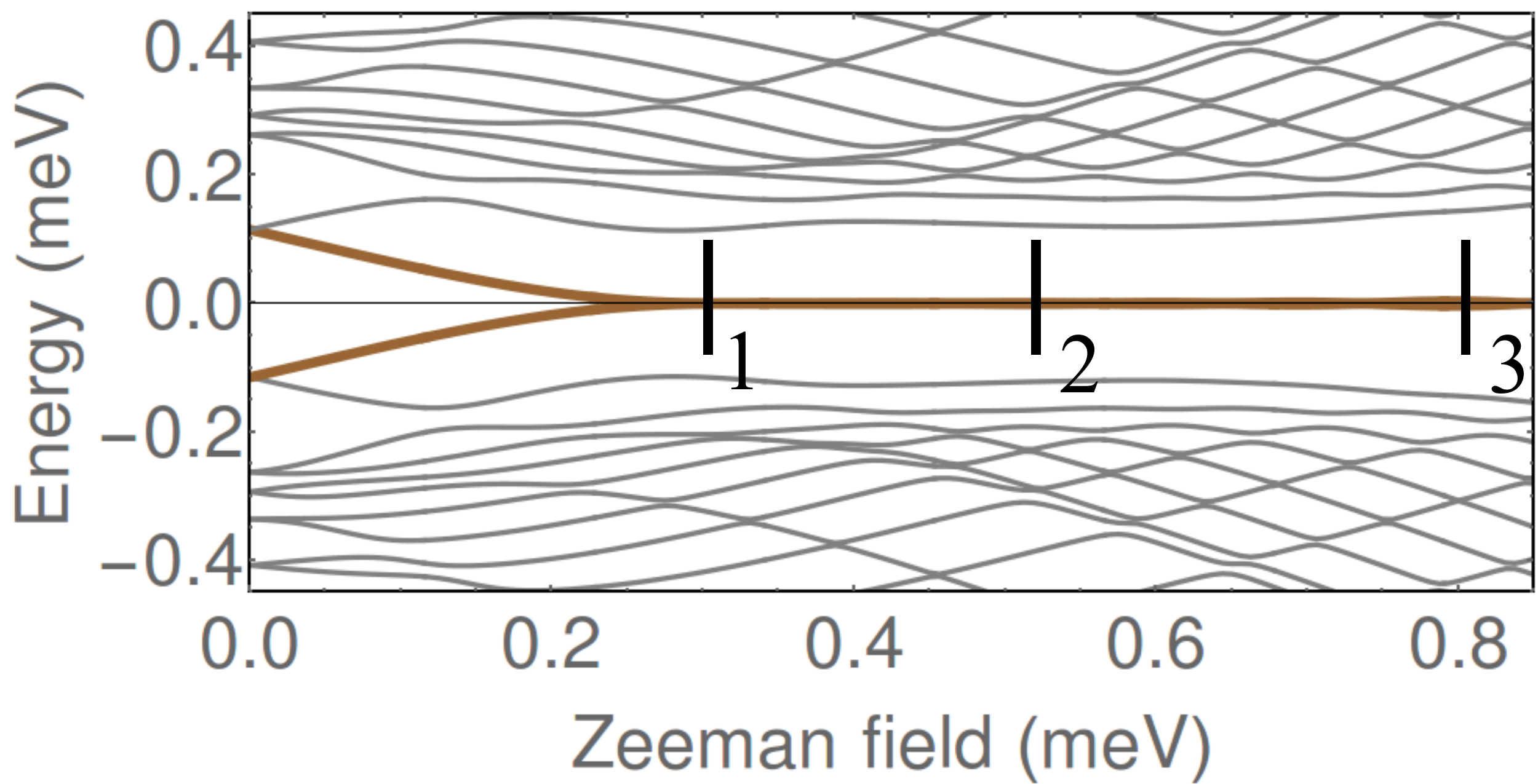}
\vspace{-3mm}
\end{center}
\caption{(color online). Low-energy spectrum as function of the applied Zeeman field for a system with smooth non-homogeneous effective potential [see Fig. \ref{Fig_V1}, panel (c)]. The length of the parent SC is the same as in the case of half-covered quantum dot (i.e. a segment of the wire of about 125 nm is not covered). Note the robust (nearly) zero-mode and the absence of a well defined minimum of the quasiparticle gap corresponding to the crossover between the trivial and the ``topological'' regimes.}
\vspace{-2mm}
\label{Fig_V6}
\end{figure}

\begin{figure}[t]
\begin{center}
\includegraphics[width=0.48\textwidth]{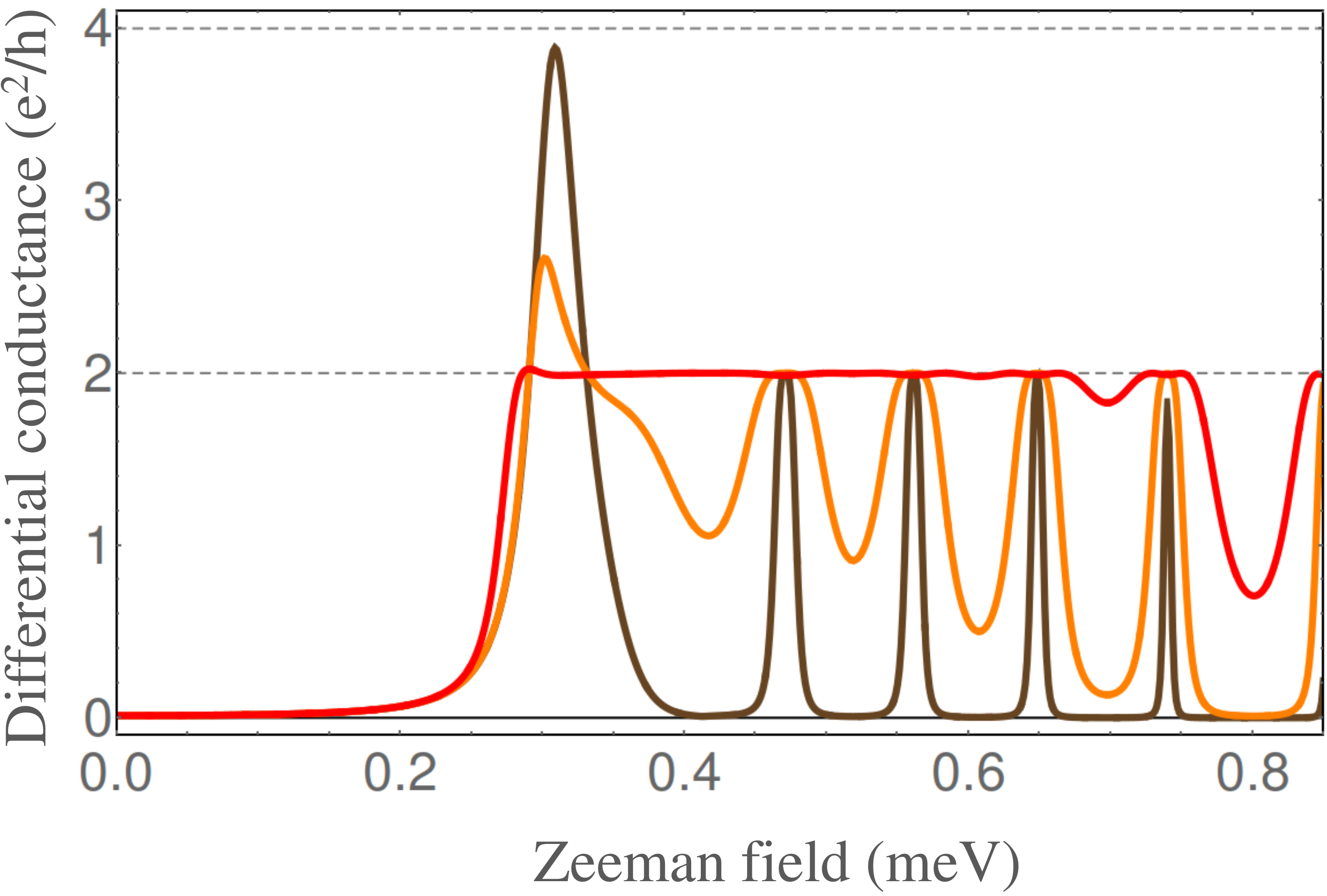}
\vspace{-3mm}
\end{center}
\caption{(color online). Dependence of the low-energy differential conductance on the Zeeman splitting for the non-homogeneous wire with the spectrum shown in Fig. \ref{Fig_V6}.   The black, orange (light gray), and red (gray) lines correspond to a bias voltage $V_{bias} =0.05, 0.15$, and $0.75~\mu$V, respectively.}
\vspace{-2mm}
\label{Fig_V7}
\end{figure}

\begin{figure}[t]
\begin{center}
\includegraphics[width=0.48\textwidth]{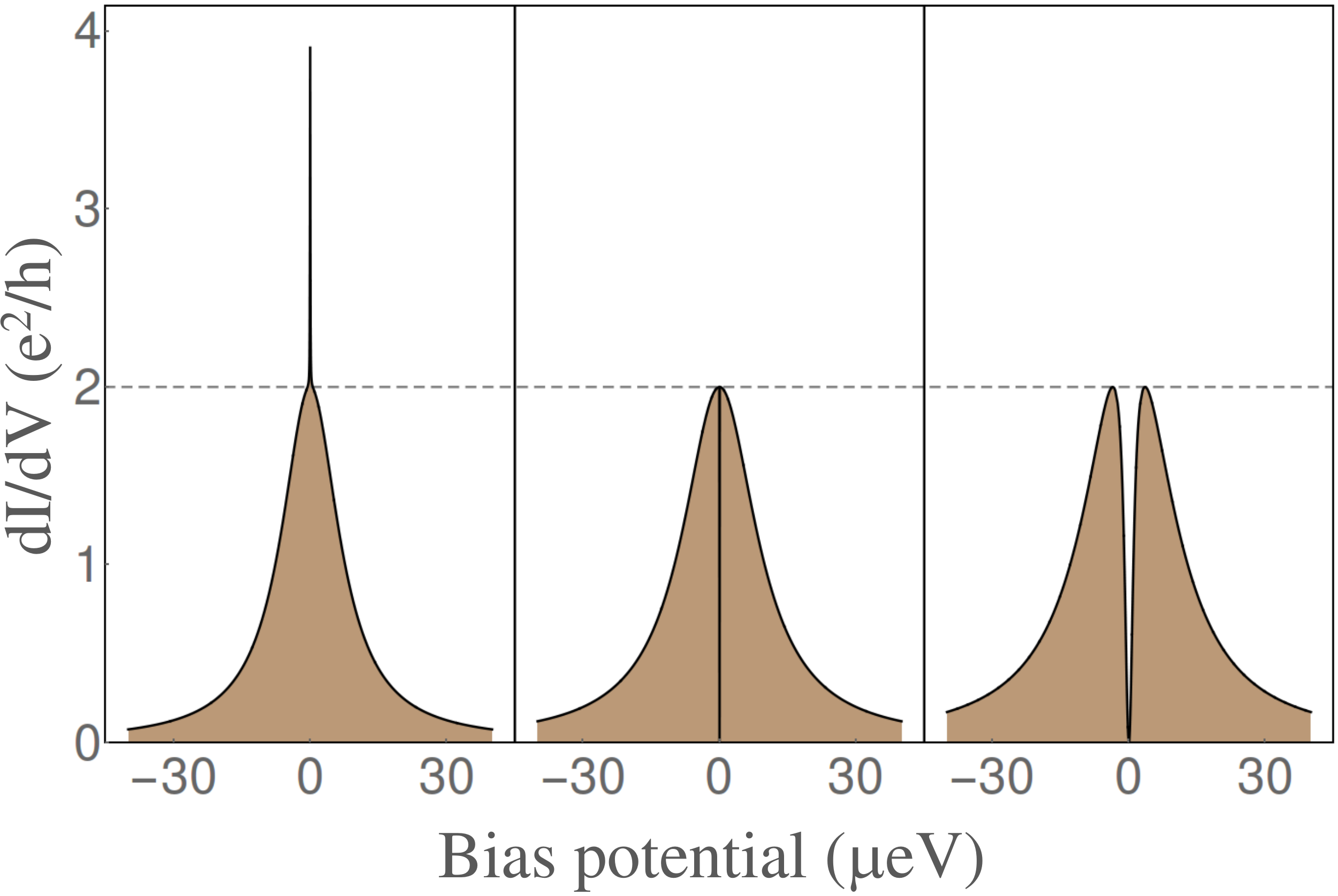}
\vspace{-3mm}
\end{center}
\caption{(color online). Zero temperature differential conductance as function of the bias voltage for three different values of the Zeeman field marked  ``1'', ``2'', and ``3'' in Fig. \ref{Fig_V6}.}
\vspace{-2mm}
\label{Fig_V8}
\end{figure}

Before concluding this section, we compare a hybrid system having a (strongly coupled) quantum dot near one end with an inhomogeneous system with a smooth effective potential as shown in Fig. \ref{Fig_V1} (c). In the language of Ref.~\cite{Moore2016Majorana}, this would correspond to a long-range inhomogeneity, in contrast to the quantum dots which can be viewed as short-range inhomogeneities. The low-energy spectrum of the non-homogeneous system is shown in Fig. \ref{Fig_V6}. At zero field, the energy of the ABS is lower than the induced gap as a result of the nanowire being only partially covered (about $90\%$) by the parent superconductor, as discussed above. Note the striking  absence of a minimum of the quasiparticle gap, which would signal the topological quantum phase transition in a homogeneous system. The merging ABSs form a very robust nearly-zero mode, which, according the analysis  in Ref.~\cite{Moore2016Majorana}, consists of partially overlapping Majorana bound states. 
%%%%%%%%%%%%%%%%%%%%%%%%%%%%
%%%%%%%%%%%%%%%%%%%%%%	
The low-energy differential conductance corresponding to the nearly-zero mode in Fig. \ref{Fig_V6} is shown in Fig. \ref{Fig_V7} (as function of the Zeeman field for three different values of the bias voltage) and  Fig. \ref{Fig_V8} (as function of the bias voltage for three different Zeeman fields marked ``1'', ``2'', and ``3'' in Fig. \ref{Fig_V6}).  The low-bias differential conductance traces  shown in Fig. \ref{Fig_V7} have values between $0$ and (almost) $4e^2/h$. In particular, the differential conductance exceeds $2e^2/h$ in the vicinity of the first zero-energy crossing, $V_Z \approx 0.3$ meV (see Fig. \ref{Fig_V6}). However, in practice it would be extremely difficult to observe a ZBCP larger than $2e^2/h$ at finite temperature. This is due to the fact that the contribution exceeding the quantized value forms a very narrow secondary peak (see Fig.~\ref{Fig_V8}, left panel), similar to the completely covered dot shown in Fig.~\ref{Fig_V3}. We interpret the double-peak structure of the ZBCP as resulting from the partially-overlapping Majorana bound state (MBS)  that form the ABS. The broad peak is generated by the MBS localized closer to the wire end (which is strongly coupled to the metallic lead), while the narrow additional peak is due to the MBS localized further away from the end (which is weakly coupled to the lead). 
Finally, we note that the low conductance values in Fig.~\ref{Fig_V7} are due to the splitting of the ZBCP. However, the maximum value of the ZBCP is practically quantized at very low (but finite) temperature, as evident from the results shown in Fig.~\ref{Fig_V8}. 

 %%%%%%%%%%%%%%%%%%%%%%%%%%%%
%%%%%%%%%%%%%%%%%%%%%%			

In summary, the results presented in this section lead us to the following conclusions. First, semiconductor-superconductor hybrid systems having strongly-coupled quantum dots at the end of the wire, which can be viewed as systems with short-range potential inhomogeneities,  generate ABSs that, quite generically, tend to merge at zero energy with increasing Zeeman field, but still within the topologically-trivial regime. Second, ABSs with energies at $V_Z=0$ significantly lower than the induced gap and low values of the merging field $V_Z^*$  are likely to generate extremely robust topologically-trivial ZBCPs. Third, measuring a quantized ( to $2e^2/h$) ZBCP does not provide definitive evidence for Majorana zero modes (although finding ZBCP quantization which is robust over variations in many parameters, e.g., magnetic field, chemical potential, tunnel barrier, carrier density, would be very strong evidence for the existence of MZMsas emphasized already in this paper). However, trivial conductance peaks generated by merging ABSs having wave functions partially localized outside the superconducting region are generally expected to produce ZBCPs with heights between  $0$ and $4e^2/h$. In this regime, an accidental quantized peak will not be robust against small variations of the control parameters. By contrast, if the wave function is entirely inside the proximitized region, the ZBCP is (practically) quantized and cannot be distinguished from a MZM-induced conduction peak by a local tunneling measurement. In this case, a minimal requirement for the Majorana scenario is to be able to reproduce the (robust) ZBCP by  performing a tunneling measurement at the opposite end of the wire, in the spirit of Ref.~\cite{DasSarma2012Splitting}. Finally, our fourth conclusion is that very similar phenomenologies can be generated using rather different effective potentials(i.e., the effective 'quantum dot' leading to the ABS could arise from many different physical origins and could lie inside or outside the nanowire).  A better understanding of the profile of the effective potential along the wire (which can be obtained, for example, by performing detailed Poisson-Schrodinger calculations) represents a critical task in this field. 

%%%%%%%%%%%%%%%%%%%%%%%%%%%%
%%%%%%%%%%%%%%%%%%%%%%	

\section{Understanding near-zero-energy Andreev bound states from reflection matrix theory}\label{sec:understanding}
The absence of level repulsion in symmetry class D enhances the likelihood of a pair of levels sticking together at zero energy as some parameter such as the Zeeman splitting or the chemical potential is varied as discussed throughout this manuscript. Despite this generic fact associated with symmetry class D that describes systems containing Zeeman splitting, spin-orbit coupling and superconductivity, the range of Zeeman splitting over which the spectrum sticks is not guaranteed to be large. In fact, the range of Zeeman field is typically not large for most disordered Hamiltonian~\cite{Liu2012Zero}. In the experiment~\cite{Deng2016Majorana} and in our simulations (with quantum dots, but no disorder), however, the zero-sticking propensity of trivial ABSs extends over a large range of Zeeman splitting ($V_Z$).

A more specific mechanism that provides a relatively robust (compared to the usual disordered class D) near-zero-energy states within symmetry class D involves the so-called smooth confinement~\cite{Kells2012Near, Moore2016Majorana}. The essential idea is that large Zeeman splitting ($V_Z$) compared to SC pairing ($\Delta$) suppresses conventional $s$-wave pairing compared to $p$-wave pairing leading to a tendency for the formation of Majorana states at the end of the system for each spin-polarized channel in the nanowire. However, the end potential typically scatters between the different channels and gaps the Majorana fermions out, i.e., an MZM splitting develops. If the inter-channel scattering between different channels is weak then this Majorana splitting is small and there is a near-zero-energy state in such a potential. This near-zero-energy mode is, however, nontopological as it is arising from split Majorana modes at the wire end. Thus, the ABS producing the ZBCP is a composite of two MZMs, only one of which contributes to tunneling, leading to a robust almost-zero mode in the trivial regime.

In subsection~\ref{ssec:energy}, we will first show the energy spectra for the quantum dot-proximitized nanowire hybrid structure using various parameters (e.g., chemical potential $\mu$, nanowire length $L$, dot length $l$, etc.) in order to show the trend of zero-energy sticking in the parameter regime. Second in subsection~\ref{ssec:understanding}, we use reflection matrix theory to explain why such zero-sticking bound states exist in the relevant parameter regime.

\subsection{Energy spectra for hybrid structures with various parameters}\label{ssec:energy}

\begin{figure}[!htb]
\includegraphics[width=0.49\textwidth]{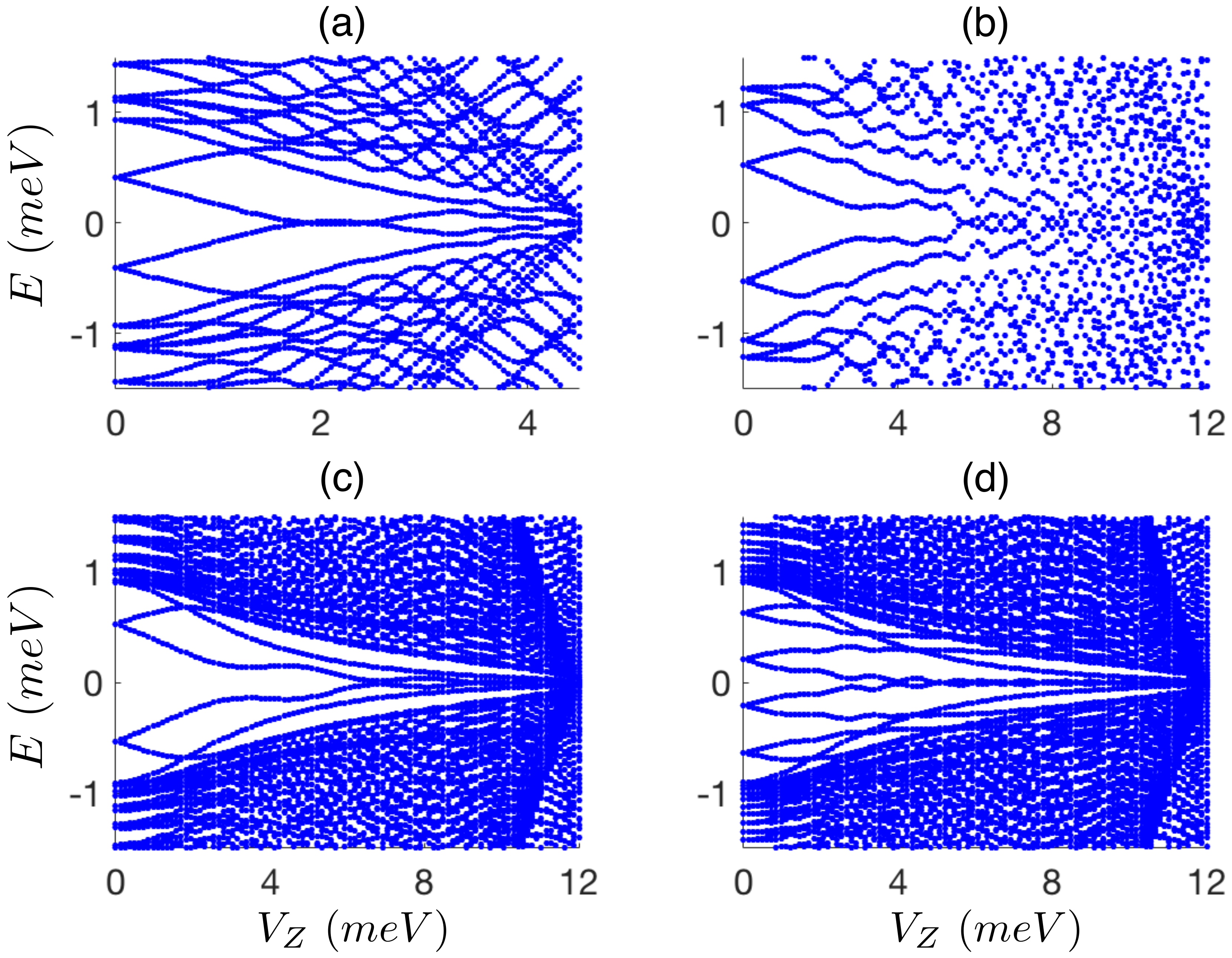}
\caption{(color online). Energy spectra for hybrid structures with various parameters. (a): $\mu=4.5~$meV, nanowire length $L=1.0~\mu$m, dot length $l=0.3~\mu$m. (b): $\mu=12.0~$meV, $L=1.0~\mu$m, $l=0.3~\mu$m. (c): $\mu=12.0~$meV, $L=4.0~\mu$m, $l=0.3~\mu$m. (d): $\mu=12.0~$meV, $L=4.0~\mu$m, $l=1.0~\mu$m. }
\label{fig:spectraSmooth}
\end{figure}

We show the energy spectra for various hybrid structures in Fig.~\ref{fig:spectraSmooth}. The few relevant parameters we focus on and thus vary between panels are chemical potential $\mu$, length of the nanowire $L$, length of the quantum dot $l$, while all other parameters, e.g. pairing potential $\Delta_0=0.9~$meV and etc., are kept the same as the default values introduced in the previous sections. Fig.~\ref{fig:spectraSmooth}(a) shows the energy spectrum of a typical hybrid structure discussed in the previous sections, with the parameters conforming to the known values in the realistic experimental setup. There is a finite range of Zeeman splitting over which the energy of the topologically trivial ABSs stick around zero. Through Fig.~\ref{fig:spectraSmooth}(b) to (d), we step by step increase the chemical potential $\mu$, the length of the semiconductor-superconductor nanowire $L$, and the length of the quantum dot $l$. Finally with all the three parameters $\mu, L, l$ large in Fig.~\ref{fig:spectraSmooth}(d), the energy of the trivial ABS is even closer to zero energy, and even more strikingly, the range of Zeeman splitting for such near-zero-energy ABSs becomes extremely large, starting from a few times the pairing potential up to the chemical potential. The trend of decreasing ABS energy and increasing range of zero-energy sticking shown by Fig.~\ref{fig:spectraSmooth}(a) to (d) indicates that Fig.~\ref{fig:spectraSmooth}(a) and Fig.~\ref{fig:spectraSmooth}(d) are essentially adiabatically connected. In the following subsection, we will discuss why there exist such near-zero-energy ABSs over such a large range of Zeeman field in large $\mu, L, l$ limit using reflection matrix theory. Since realistic situation is adiabatically connected to this large $\mu, L, l$ limit, our understanding will also apply to most of the hybrid structures discussed in previous sections. Note that this discussion also explains why the zero-sticking of ABSs mostly arises in the large chemical potential regime.

\subsection{Understanding zero-energy sticking from reflection matrix theory}\label{ssec:understanding}

\begin{figure}[t]
\includegraphics[width=0.49\textwidth]{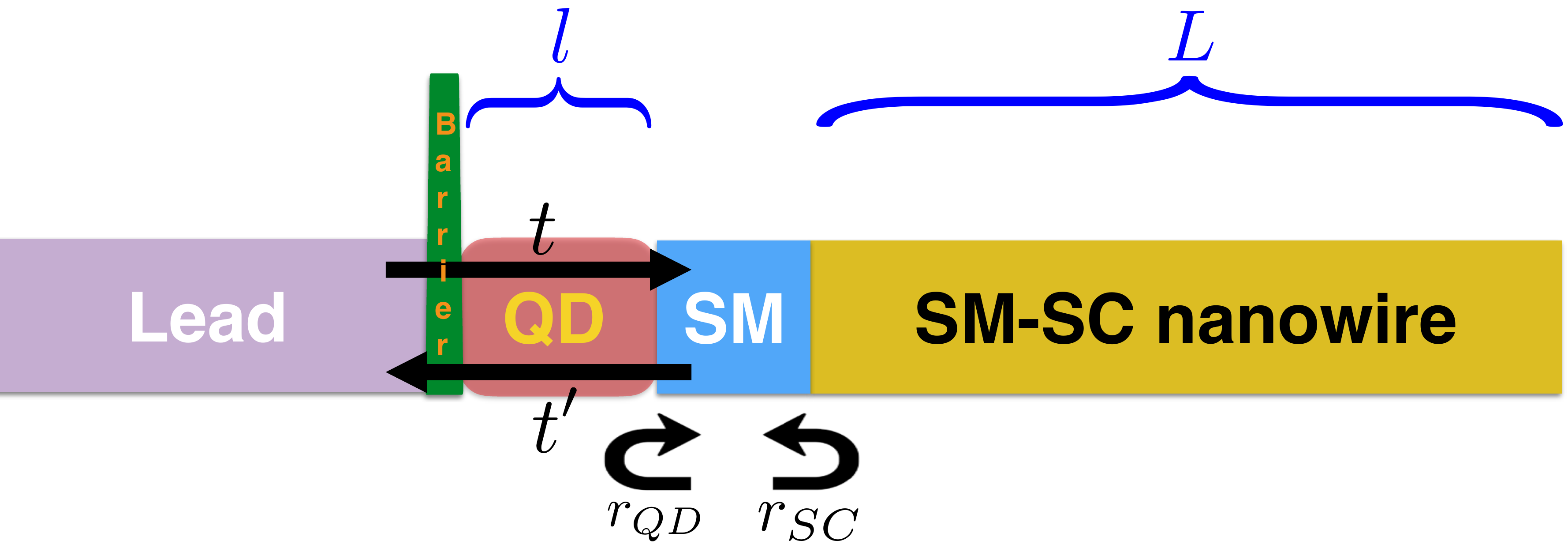}
\caption{(color online). A schematic for the NS junction setup. Although the setup is exactly the same as that shown in Fig.~\ref{fig:schematic}, an imaginary piece of semiconductor is added between the quantum dot and the semiconductor-superconductor nanowire for the discussion of reflection matrix theory. }
\label{fig:schematicReflection}
\end{figure}

In the previous subsection, numerical simulations show strong evidence that the energy of the ABSs approaches zero energy and the range of such near-zero-energy sticking increases with increasing chemical potential, increasing nanowire length, and increasing quantum dot length. Thus, here we try to understand this phenomenon using reflection matrix theory. The setup is shown in Fig.~\ref{fig:schematicReflection}. Although the NS junction setup is exactly the same as that shown in Fig.~\ref{fig:schematic}, an imaginary piece of semiconductor is added between the quantum dot and the semiconductor-superconductor nanowire for the discussion of the reflection matrix theory. This imaginary semiconductor can also be regarded as a part of the quantum dot but with nearly homogeneous potential. The total reflection matrix from the hybrid structure is
\begin{align}
r &= r_b + t' \left( r_{SC} + r_{SC}r_{QD}r_{SC} + ... \right) t \nn
&= r_b + t' \left(1 - r_{SC}r_{QD} \right)^{-1} r_{SC} t,
\label{eq:r}
\end{align} 
where $r_b$ is the reflection matrix for the incoming modes in the lead reflected by the barrier, $t$ is the transmission matrix for the lead modes transmitting to the semiconductor, $r_{SC}$ is the reflection matrix for the semiconductor modes reflected by the proximitized nanowire, $r_{QD}$ is the reflection matrix for the semiconductor modes reflected by the quantum dot, and $t'$ is the is the transmission matrix for the semiconductor modes transmitted to the lead. The near-zero-energy differential conductance is
\begin{align}
G = \frac{e^2}{h} Tr \left( \hat{1} - r^{\dagger}_{ee} r_{ee} + r^{\dagger}_{he} r_{he} \right) = \frac{2e^2}{h} Tr \left( r^{\dagger}_{he} r_{he} \right),
\end{align}
where $r_{he}$ is the Andreev reflection matrix from the hybrid structure. The last step holds due to the unitarity of the total reflection matrix when bias voltage is below the superconducting gap. The Andreev reflection is contained in the second term of Eq.~\eqref{eq:r}, and the pole of $\left(1 - r_{SC}r_{QD} \right)^{-1}$ corresponds to the peak of the differential conductance. On the other hand, the pole of the reflection matrix is also the condition for the formation of a bound state, i.e., a bound state forms when
\begin{align}
Det \left( 1 - r_{SC}r_{QD} \right)=0
\label{eq:bound}
\end{align}
is satisfied.

\begin{figure}[t]
\includegraphics[width=0.49\textwidth]{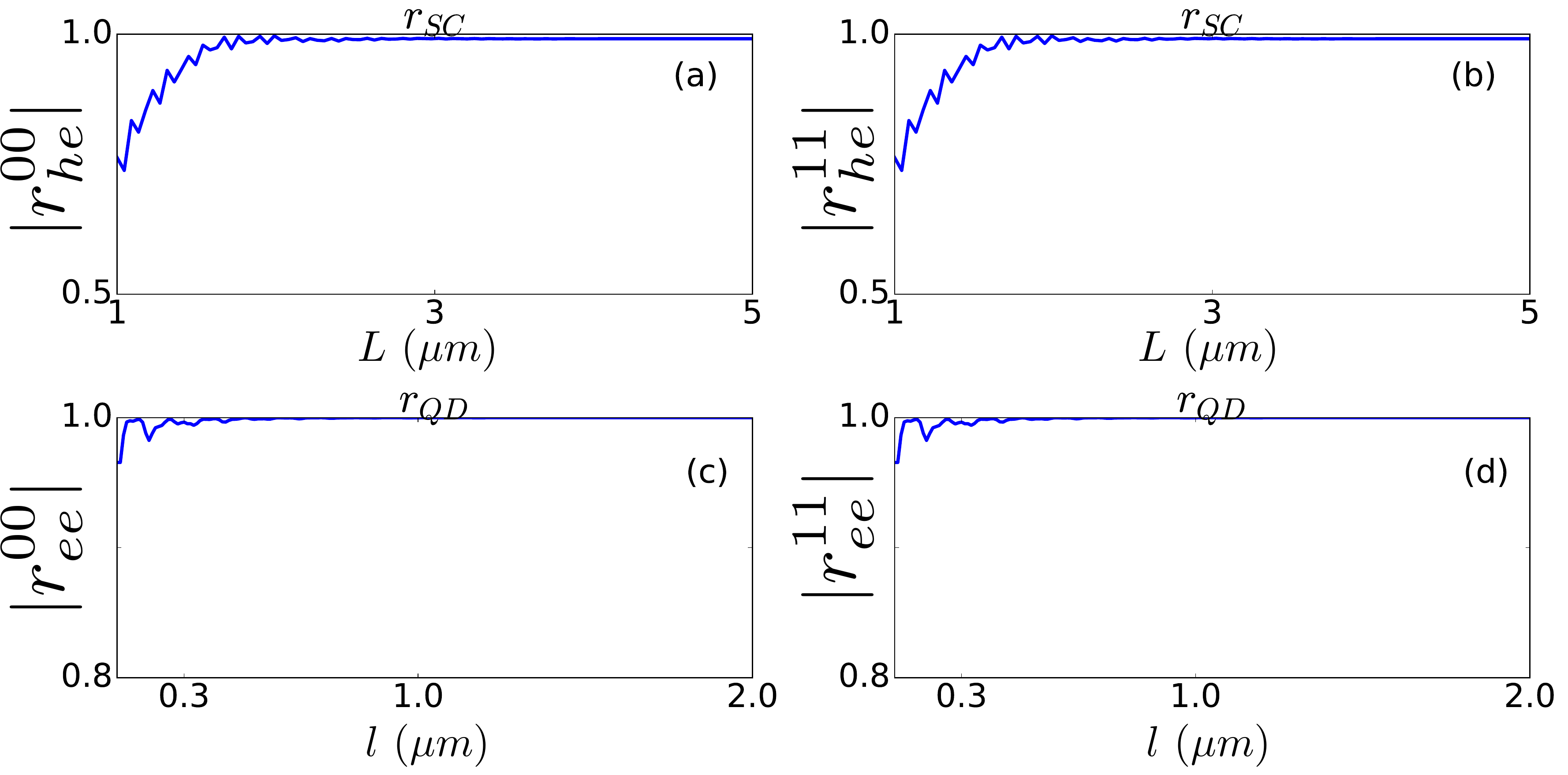}
\caption{(color online). Matrix elements for the reflection matrices from the semiconductor-superconductor nanowire and the quantum dot, with chemical potential $\mu=12~$meV, $V_Z=8~$meV. The upper panels are the Andreev reflection between each spinful channel with in index 0 and 1 (i.e., the $|e^{i\alpha}|$ in Eq.~\eqref{eq:rSC}) as a function of nanowire length. In the long nanowire limit, the Andreev reflection becomes perfect. The lower panels are the normal reflection between each spinful channel (i.e., the $|e^{i\beta}|$ in Eq.~\eqref{eq:rQD}) as a function of dot length.  }
\label{fig:matrixElements}
\end{figure}

In the large Zeeman field limit, i.e., $V_Z \gg \Delta, \alpha_R$, the spin-orbit-coupled nanowire can be thought of as two spin-polarized bands with a large difference in chemical potential and Fermi momenta. When considering the scattering process between the effectively spin-polarized semiconductor and the semi-infinite superconductor, the momentum must be conserved in the limit of Andreev approximation $\Delta \ll \mu$. The constraint of momentum conservation prohibits the normal reflection between either the same or the other spinful channel due to the large difference in Fermi momenta between two channels. Thus the scattering process between semiconductor and the superconductor can be thought of as effectively two independent perfect Andreev reflection processes among each spin-polarized channel. So the reflection matrix for each channel can be written as
\begin{align}
r_{SC} = 
\begin{pmatrix}
0 & e^{i\alpha} \\
e^{-i\alpha} & 0
\end{pmatrix}.
\label{eq:rSC}
\end{align}
For the scattering process between the semiconductor and the quantum dot, when the dot potential is smooth, the normal reflection only connects the Fermi level within the same spinful channel, and thus again the two spin-polarized bands of the semiconductor can be thought of as independent of each other. So the reflection matrix for each band can be written as
\begin{align}
r_{QD} = 
\begin{pmatrix}
e^{i\beta} & 0 \\
0 & e^{-i\beta} 
\end{pmatrix}.
\label{eq:rQD}
\end{align}
The numerical evidence for the form of $r_{SC}$ and $r_{QD}$ are shown in Fig.~\ref{fig:matrixElements}, which is consistent with our argument in the large Zeeman field and Andreev approximation limit. It is easy to see that such zero-bias reflection matrices satisfy the condition for the formation of a bound state, i.e., Eq.~\eqref{eq:bound}. It indicates that in the large Zeeman field and Andreev approximation limit, the semiconductor-superconductor nanowire can be seen as consisting of two nearly spin-polarized $p$-wave superconductors, and each of them holds a MZM at the wire end. Since the interchannel coupling between the two $p$-wave superconductors is weak in the presence of a smooth dot potential at the wire end, the two MZMs from two channels do not gap out each other, they form a near-zero-energy ABS.

Although the above discussion assumes large chemical potential, long semiconductor-superconductor nanowire, and long quantum dot, the conclusion well applies to the realistic situation with intermediate value of chemical potential, finite length of the nanowire and quantum dot, since these two situations are adiabatically connected with each other. This conclusion is explicitly verified by the extensive numerical results presented in this work.

%%%%%%%%%%%%%%%%%%%%%%%%%%%%

\section{conclusion}\label{sec:conclusion}

We have developed a non-interacting theory for the low-lying energy spectra and the associated tunneling transport properties of quantum dot-nanowire-superconductor hybrid structures focusing on quantum dots strongly coupled to the proximitized wire. The theory is motivated by a striking recent experiment~\cite{Deng2016Majorana} reporting intriguing coalescence of Andreev bound states into zero-energy states characterized by zero-bias conductance peaks that mimic the predicted Majorana zero mode behavior. 
The specific question we address in our work is whether the midgap coalescence of Andreev bound states and their sticking together propensity at zero energy necessarily imply a metamorphosis of Andreev states into topological Majorana modes in the presence of spin-orbit coupling and Zeeman splitting. The topological Majorana bound states are operationally defined as the pairs of well-separated Majorana zero modes localized at the opposite ends of the wire, while the Andreev bound states, which can be viewed as pairs of overlapping (or partially overlapping) Majorana zero modes, are localized near one end of the hybrid system. Our numerical simulations produce essentially exact qualitative agreement with the data of Ref.~\cite{Deng2016Majorana}, reproducing the observed features of the Andreev states as functions of Zeeman splitting and chemical potential, although a quantitative comparison (and hence, a definitive conclusion) is impossible because the experimental parameters to be used in the theory are mostly unknown.

Our conclusion is that in strongly-coupled dot-nanowire hybrid structures (and in the presence of superconductivity, Zeeman splitting, and spin-orbit coupling) Andreev states generically coalesce around zero energy producing zero-bias tunneling conductance values that mimic Majorana properties, although the physics is non-topological. In fact, the transport properties of such ``accidental'' almost zero-energy trivial Andreev states in class D systems are (locally) difficult to distinguish from the conductance behavior of topological Majorana zero modes. We show that this zero-energy-sticking behavior of trivial Andreev bound states (superficially mimicking topological Majorana behavior) persists all the way from an isolated (i.e. non-superconducting) quantum dot at the end of the nanowire to a quantum dot completely immersed inside the nanowire (i.e. superconducting) as long as finite Zeeman splitting and spin-orbit coupling are present. Our theory thus connects the recent observations of Deng \textit{et al}.~\cite{Deng2016Majorana} to the earlier observations of Lee \textit{et al}.~\cite{Lee2014Spin}, who studied Andreev bound states in a superconducting dot (not attached to a long  nanowire), establishing that the physics in these two situations interpolates smoothly. In both theses cases zero-bias conductance peaks may arise from trivial Andreev bound states in the presence of superconductivity, spin-orbit coupling, and Zeeman splitting. Of course, in a small quantum dot, the concept of MZMs does not apply because of strong overlap between the two ends whereas in the Deng \textit{et al}. experiment (i.e. in a dot-nanowire hybrid system) the ZBCP may arise from either trivial ABS or topological MZM. We establish, however, that in both cases the ABS can be thought of as overlapping MZMs, and hence the generic zero-sticking property of the ABS arises from the combination of spin-orbit coupling, spin splitting, and superconductivity. An immediate (and distressing) conclusion of our work is that the observation of a zero-bias conductance peak (even if the conductance value is close to the expected $2e^2/h$ quantization) cannot by itself be construed as evidence supporting the existence of topological Majorana zero modes. In particular, both trivial Andreev bound states and topological Majorana bound states may give rise to zero-bias peaks, and there is no simple way of distinguishing them just by looking at the tunneling spectra. Since the possibility that a given experimental nanowire may contain inside it some kind of accidental quantum dot can never be ruled out, the tunneling conductance exhibiting zero-bias peaks in any nanowire may simply be the result of the existence of almost-zero-energy Andreev bound states in the system. Our work shows this generic trivial situation to be a compelling scenario, bringing into question whether any of the observed zero-bias conductance peaks in various experiments \textit{by themselves} can  be taken as strong evidence in favor of the existence of Majorana zero modes since the possibility that these ZBCPs arising from accidental trivial ABSs cannot a priori be ruled out. Consequently, a zero-bias conductance peak obtained by tunneling from one end of the wire cannot be accepted as a compelling topological Majorana signature (even when the height of the peak is quantized at $2e^2/h$), since a likely alternative scenario is that the zero-bias peak is, in fact, a signature of a trivial Andreev bound state associated with a strongly coupled quantum dot or other type of inhomogeneity (unintentionally) present in the system. One must carry out careful additional consistency checks on the observed ZBCPs in order to carefully distinguish between ABS and MZM.

Therefore, to be more decisive, transport experiments must demonstrate the robustness of the quantization to all possible variations in the barrier. One possibility is to study avoided crossings between levels in the quantum dot and a potential Majorana state~\cite{Clarke2017Experimentally, Prada2017Measuring} that essentially eliminate the quantum dot. This can be done for example by extending the normal region in the semiconductor wire in between the metallic and superconducting lead shown in Fig.~\ref{fig:schematic}. By such an extension, one can enhance gate control so as to be able to create a single-channel quantum point contact. The quantization of the conductance (at low enough temperature compared to the transmission of the point contact) is still a topological invariant~\cite{Wimmer2011Quantum}. In addition, one should always check (by using suitable externally controlled gate potentials) the stability of any observed ZBCP to variations in the tunnel barrier as well as the electrostatic environment near the wire ends (as in Sec.~\ref{sec:distinguish}). This test is absolutely essential in our opinion since the ABS-induced trivial ZBCP should manifest splitting as the dot potential is tuned strongly. Despite these checks, it is still likely that transport measurements will need to include additional consistency tests to confirm the nonlocal nature of the Majorana modes (e.g, observing the ZBCPs from both ends of the wire, measuring nonlocal correlations) and their robustness (e.g.,  robustness of the ZBCP quantization against variations of the  barrier height, Zeeman splitting, chemical potential, and other variables). Any type of hybrid structure that is not capable of passing these relatively straightforward tests of ZBCP robustness would not be suitable for more complex experiments involving interferometry, fusion, or braiding. In short, a ZBCP is only a necessary condition for an MZM, and could easily arise also for non-topological zero-energy ABSs in class D systems.

The obvious consistency test is of course the robustness of the ZBCP to variations in all controllable experimental parameters. The topological MZM-induced ZBCP should show stable robustness whereas the ABS-induced ZBCP will not. In particular, we discuss in Sec.~\ref{sec:distinguish} that varying the dot potential will lead to splitting or possibly even disappearance of the trivial ABS-induced ZBCP, but the MZM-induced ZBCP should be relatively stable. This, in principle, enables a unique distinction between the two cases, but in reality this may not be as simple. Since the ``quantum dot'' is often not obvious, it is not clear how to vary its potential. Perhaps the most obvious check is to use additional gates with varying gate voltage to ensure a complete stability of the observed ZBCP.  Another possible test is rotating the magnetic field, but here both trivial and topological MZMs go away as the field is rotated toward the spin-orbit direction in the wire (and is unaffected by any rotation in the plane perpendicular to the spin-orbit direction).  Although there are quantitative differences between the two cases, it may not be easy to be definitive. Seeing correlations in the ZBCP while tunneling from the two ends of the wire separately may be quite definitive since it is unlikely that the same ABS can be operational at both ends of the wire (as it requires identical quantum dot confinements at the two ends), but this kind of correlated tunneling measurements from both wire ends have not yet been successfully performed in the laboratory.

We find that generically the ABS-induced ZBCPs require high values of chemical potential, $\mu> \Delta$, and for $\mu \gg \Delta$, the trivial zero-sticking region could extend over a very large Zeeman field range from $V_Z=\Delta$ to $\mu$, with the eventual topological phase emerging at a still higher field $\sqrt{\Delta^2 + \mu^2}$. But, some non-universal beating or apparent oscillation of the ZBCP around zero energy is likely since the ABSs do not stick precisely to zero energy as there is no exponential protection here unlike the corresponding MZM case in the long-wire limit. On the other hand, the MZM-induced ZBCPs also manifest an apparent beating around zero energy due to MZM splitting oscillations arising from Majorana overlap invariably present in any finite wire. (We note that the exponentially small MZM splitting can only happen in very long wires since at high magnetic field the induced gap is small making the SC coherence length very large.)  The question, therefore, arises if the oscillatory behaviors of the two situations (the ABS beating around zero energy in the trivial phase because of the zero-sticking in D class SC versus the MZM oscillating around zero energy in the topological phase due to the Majorana overlap from the two ends) can somehow be used to distinguish trivial and topological zeros. This question was addressed in a related, but somewhat different, context by Chiu \textit{et al}.~\cite{Chiu2017Conductance} in trying to understand the experiment of Albrecht \textit{et al}.~\cite{Albrecht2016Exponential}. In fact, Chiu \textit{et al}. showed~\cite{Chiu2017Conductance} that the data of Albrecht \textit{et al}. claiming exponential Majorana protection~\cite{Albrecht2016Exponential} can be understood entirely by invoking ABS physics, consistent with our findings in the current work. We show in Appendix~\ref{app:oscillations} our calculated low-lying energy spectra for both trivial ABS and topological MZM approximate zero-modes in simple nanowire and hybrid (i.e. nanowire + dot) structures respectively, keeping all the other parameters very similar. It is clear that the oscillatory or beating structures in the two cases are superficially similar except that the ABS oscillations are non-universal whereas the MZM oscillations always manifest increasing amplitude with increasing $V_Z$ by virtue of the decreasing induced gap with increasing $V_Z$.

We mention that although we have used the terminology `class D' to describe the system and the physics studied in the current work, the standard terminology for class D systems~\cite{Liu2012Zero, Bagrets2012Class, Pikulin2012Zero, Altland1997Nonstandard, Sau2013Density} specifically invokes disorder and discusses random or chaotic systems whereas we are discussing clean systems with no disorder. We only mean the simultaneous presence of spin-orbit coupling, Zeeman splitting, and superconductivity when we mention `class D' , and as such our ABS-induced ZBCP is fundamentally distinct from those discussed in Refs.~\cite{Liu2012Zero, Bagrets2012Class, Pikulin2012Zero, Sau2013Density}.

Before concluding, we point out that, although Ref.~\cite{Deng2016Majorana} contains some of the most compelling experimental evidence for the existence of stable almost-zero-energy Andreev bound states in quantum dot-nanowire hybrid structures, there have been several earlier experiments hinting at the underlying Andreev physics discussed in our work. The foremost in this group is, of course, the experiment by Lee \textit{et al}.~\cite{Lee2014Spin} who studied zero-bias peaks induced by Andreev bound states in quantum dots in the presence of spin-orbit coupling, Zeeman splitting, and superconductivity.  But a re-evaluation of the experimental data in the InAs-Al system by Das \textit{et al}.~\cite{Das2012Zero}, where the nanowires were typically very short (i.e., almost dot-like), indicates that the zero-bias peak in this experiment is most likely a precursor of the Deng \textit{et al}. experiment with Andreev bound states coming together and coalescing around midgap with increasing Zeeman splitting. Of course, in a very short nanowire the midgap state is an operational  Andreev bound state {\em by construction}, since the condition of ``well separated'' Majorana bound states cannot be satisfied due to the short wire length. By contrast, in long wires with quantum dots (engineered or unintentional) and other types of inhomogeneities, the emergence  of topological Majorana modes is possible (and may very well have happened for some of the ZBCPs observed in Ref.~\cite{Deng2016Majorana}), but the observation of a robust zero-bias peak does not guarantee their presence (since trivial coalescing Andreev bound states are a likely alternative). Recent theoretical work by Chiu \textit{et al}.~\cite{Chiu2017Conductance} provides support to the idea that the experimental observation of Coulomb blockaded zero-bias peaks by Albrecht \textit{et al}.~\cite{Albrecht2016Exponential} in a quantum dot-nanowire hybrid structure most likely arises from the presence of Andreev bound states in the system(in combination with MZMs). Finally, very recent unpublished work from Delft and Copenhagen~\cite{Kouwenhoven2017PrivateCom, Marcus2017PrivateCom} hint at the possibility that zero-bias conductance peaks manifesting conductance values $2e^2/h$ may have now been observed in nanowire systems. The peaks could be Majorana-induced, but  (trivial) Andreev bound states generated  by unintentional quantum dots present in these structures represent a likely alternative scenario that a priori cannot be ruled out without a systematic study of the barrier dependence as discussed in the  last paragraph. To understand these brand new experiments in high-quality epitaxial semiconductor-superconductor hybrid structures, more work is necessary involving both experiment (i.e., performing the consistency tests) and theory (e.g., modeling the effective potential profiles). In particular, robustness of the ZBCP to variations in parameters (e.g., magnetic field, chemical potential, tunnel barrier, dot confinement) is essential before MZM claims can be taken seriously even when the ZBCP is quantized at $2e^2/h$.

The key message of our work is that Andreev bound states could coalesce in the trivial superconducting regime of nanowires producing surprisingly stable almost-zero-energy modes mimicking Majorana zero mode behavior even in completely clean disorder-free systems, thus making it difficult to differentiate between Andreev bound states and Majorana zero modes in some situations. Thus the existence of a zero-bias conductance peak is at best a necessary condition for the existence of Majorana zero modes.

%%%%%%%%%%%%%%%%%%%%%%%%%%%%

\begin{acknowledgements}
The authors thank Leo Kouwenhoven for a critical reading of the manuscript and for making several suggestions for improvement. This work is supported by Microsoft, JQI-NSF-PFC, and LPS-MPO-CMTC. TDS was supported in part by NSF DMR-1414683. 
\end{acknowledgements}

%%%%%%%%%%%%%%%%%%%%%%%%%%%%

\appendix

\section{Conductance of two-subband simple nanowire model}\label{app:2b}

\begin{figure}[!htb]
\includegraphics[width=0.49\textwidth]{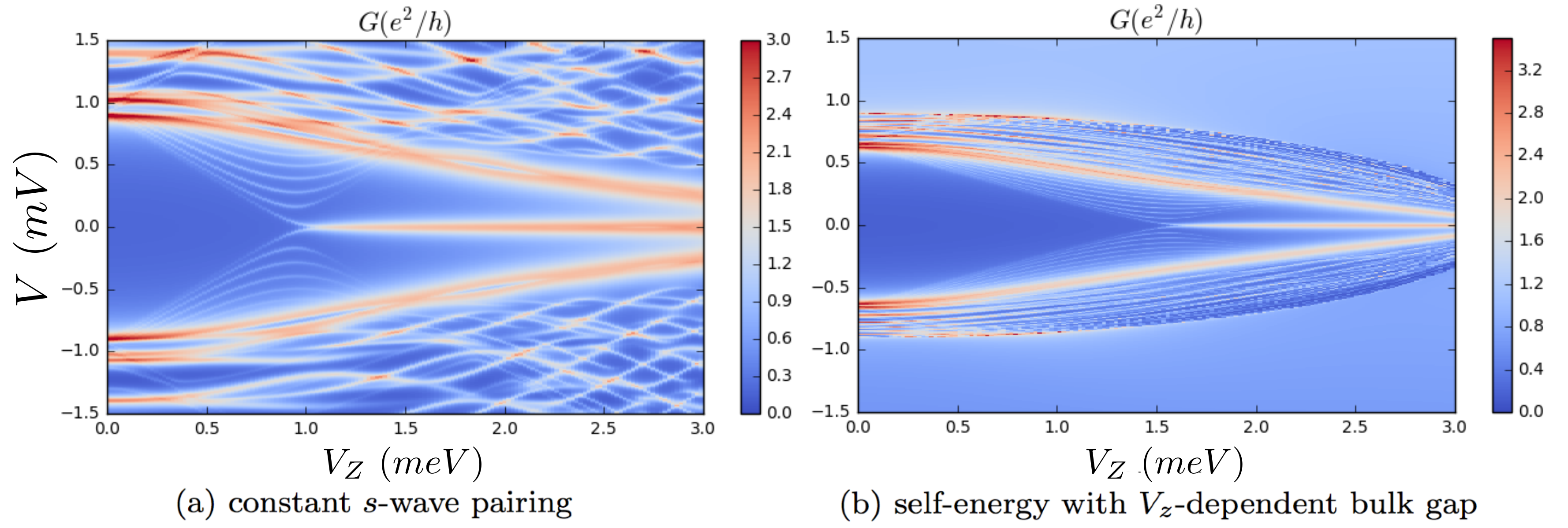}
\caption{(color online). Differential conductance through simple two-subband Majorana nanowires with chemical potential $ \mu=0~$meV, $\mu'=5.0~$meV, and length $L=1.3~\mu$m. (a): the proximity superconducting effect is introduced by a constant $s$-wave pairing $\Delta_0\tau_x$. (b): the proximity effect is introduced by a self-energy term in Eq.~\eqref{eq:SE} with a Zeeman-dependent bulk gap $\Delta(V_Z)$ as in Eq.~\eqref{eq:DeltaVZ}.}
\label{fig:G2b}
\end{figure}

For a quasi-one-dimensional nanowire with small cross section, a second subband may appear. We model the second band using the same Hamiltonian as Eq.~\eqref{eq:H_NW} but with a higher chemical potential $\mu'=5~$meV such that the second band is always non-topological within the range of Zeeman field of our interest. The total conductance through this two-subband model is approximately the sum of the individual conductance through each band. The differential conductance for two such nanowires are shown in Fig.~\ref{fig:G2b}. The most significant features are the $k_F$ peaks from the second band, in addition to all the other features already existing in the one-band model. In Fig.~\ref{fig:G2b}(a), the proximity superconducting effect is introduced by a constant $s$-wave pairing $\Delta_0\tau_x$. In Fig.~\ref{fig:G2b}(b), proximity effect is introduced by a self-energy term in Eq.~\eqref{eq:SE} with a Zeeman-dependent bulk gap $\Delta(V_Z)$ as in Eq.~\eqref{eq:DeltaVZ}. Thus the crucial difference between the two cases is that in Fig.~\ref{fig:G2b}(b), there is an edge of quasi-particle continuum above which conductance becomes smeared and featureless.

%%%%%%%%%%%%%%%%%%%%%%%%%%%%

\section{Conductance of hybrid structure with constant $s$-wave pairing}\label{app:hybrid_const}

\begin{figure}[!thb]
\includegraphics[width=0.49\textwidth]{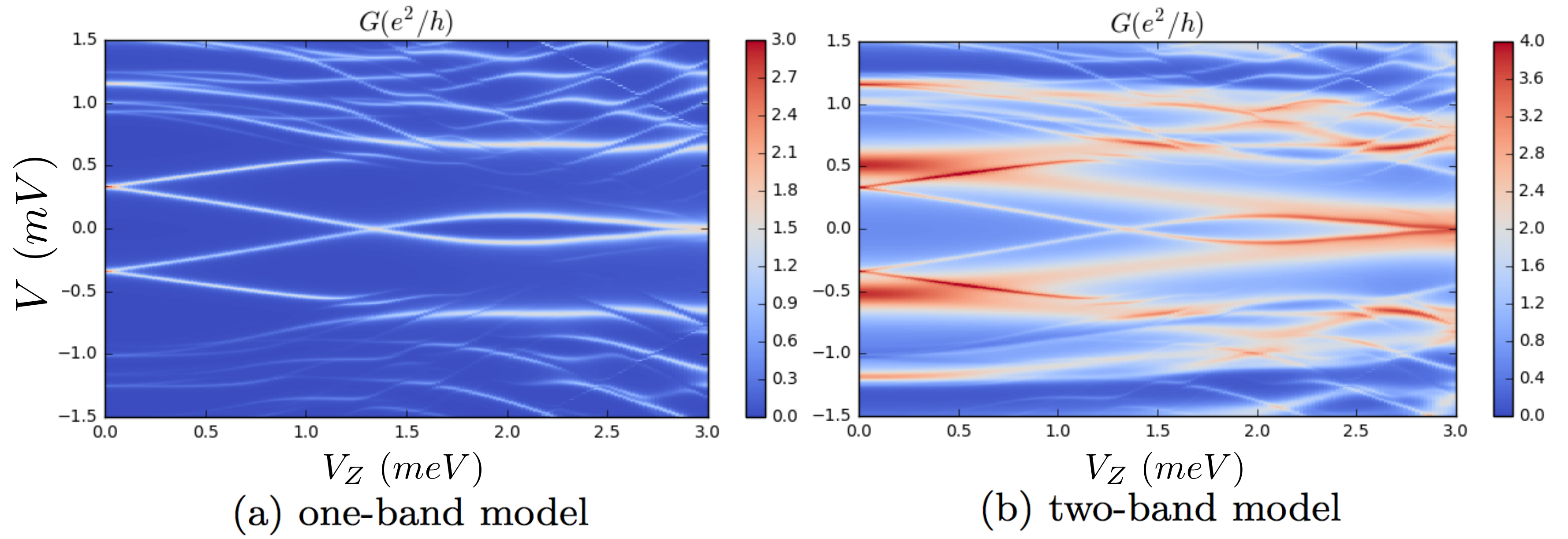}
\caption{(color online). Differential conductance for hybrid structures with constant $s$-wave pairing. (a): one-band model with $\mu=0~$meV. (b): Two-band model with a second band of larger chemical potential $\mu'=10~$meV.}
\label{fig:hybrid_const}
\end{figure}

The differential conductance for one-band and two-band hybrid structures with constant $s$-wave pairing are shown in Fig.~\ref{fig:hybrid_const}. In Fig.~\ref{fig:hybrid_const}(a), low-energy (small-bias) behavior of conductance is quite similar to the case with self-energy in the main text shown in Fig.~\ref{fig:scanVZ}(a), while high-energy (large-bias) behavior of conductance is quite different because there is no quasiparticle continuum in this case, leading to clear patterns in conductance. For two-band model with a second band with larger chemical potential $\mu'=10~$meV, the total conductance is approximated as the sum of the conductance of each band separately. The differential conductance is shown as Fig.~\ref{fig:hybrid_const}(b). In addition to almost the same behavior as the one-band model, a significant new feature is that the conductance from the lowest few eigenstates from the second band is much larger and broader than the first band. This is because a higher chemical potential is effectively lowering the tunneling barrier, thus enhancing conductance. 

%%%%%%%%%%%%%%%%%%%%%%%%%%%%

\section{Energy spectra with and without spin splitting and spin-orbit coupling}\label{app:with_without}

\begin{figure*}[!thb]
\includegraphics[width=0.99\textwidth]{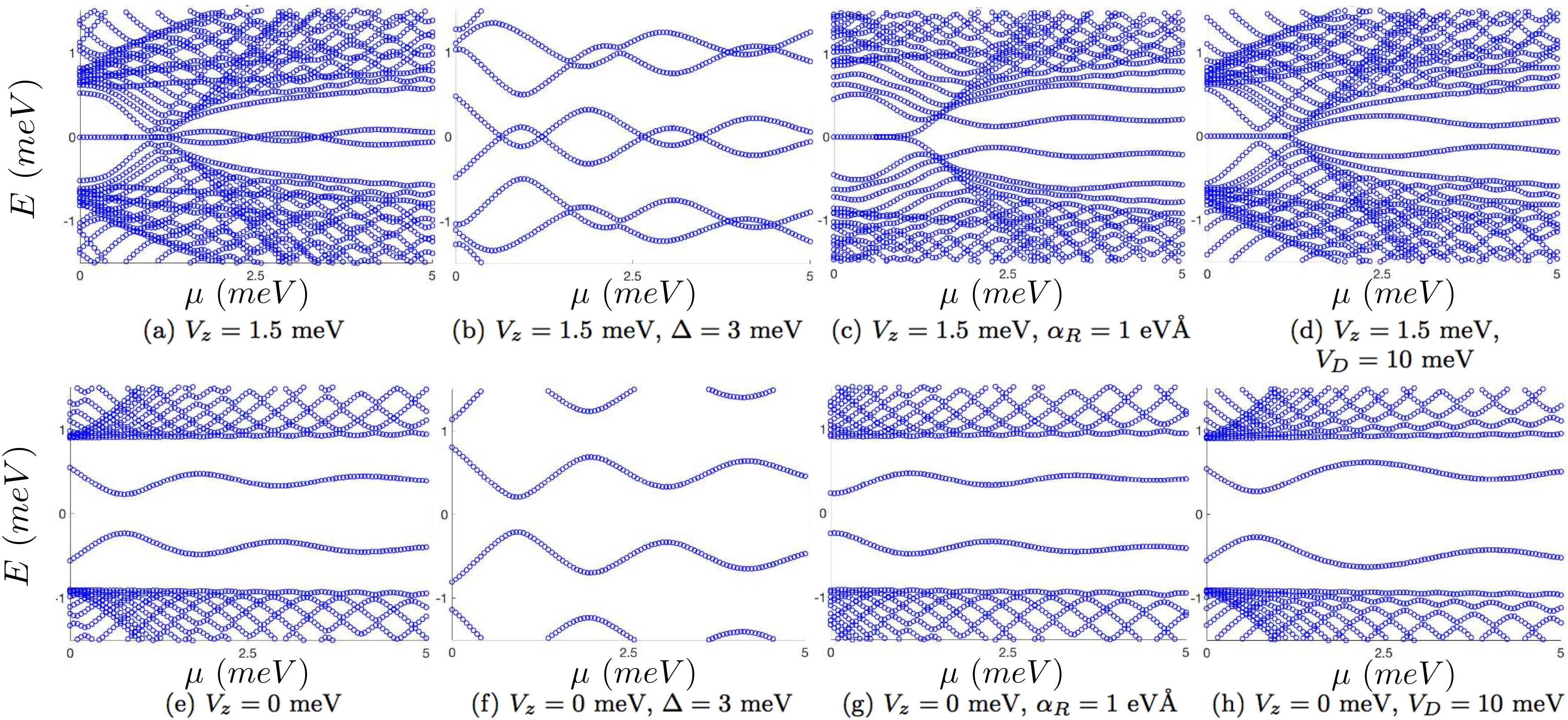}
\caption{(color online). Energy spectra of hybrid structures with and without Zeeman spin splitting. The four panels in the upper row (a)-(d) show energy spectra as function of chemical potential at finite Zeeman spin splitting $V_Z=1.5~$meV. The spectra generically cross zero energy. The four panels in the lower row (e)-(h) show energy spectra without Zeeman spin splitting, i.e., $V_Z=0$. The spectra have no zero energy states. }
\label{fig:with_withoutVZ}
\end{figure*}

\begin{figure*}[!htb]
\includegraphics[width=0.99\textwidth]{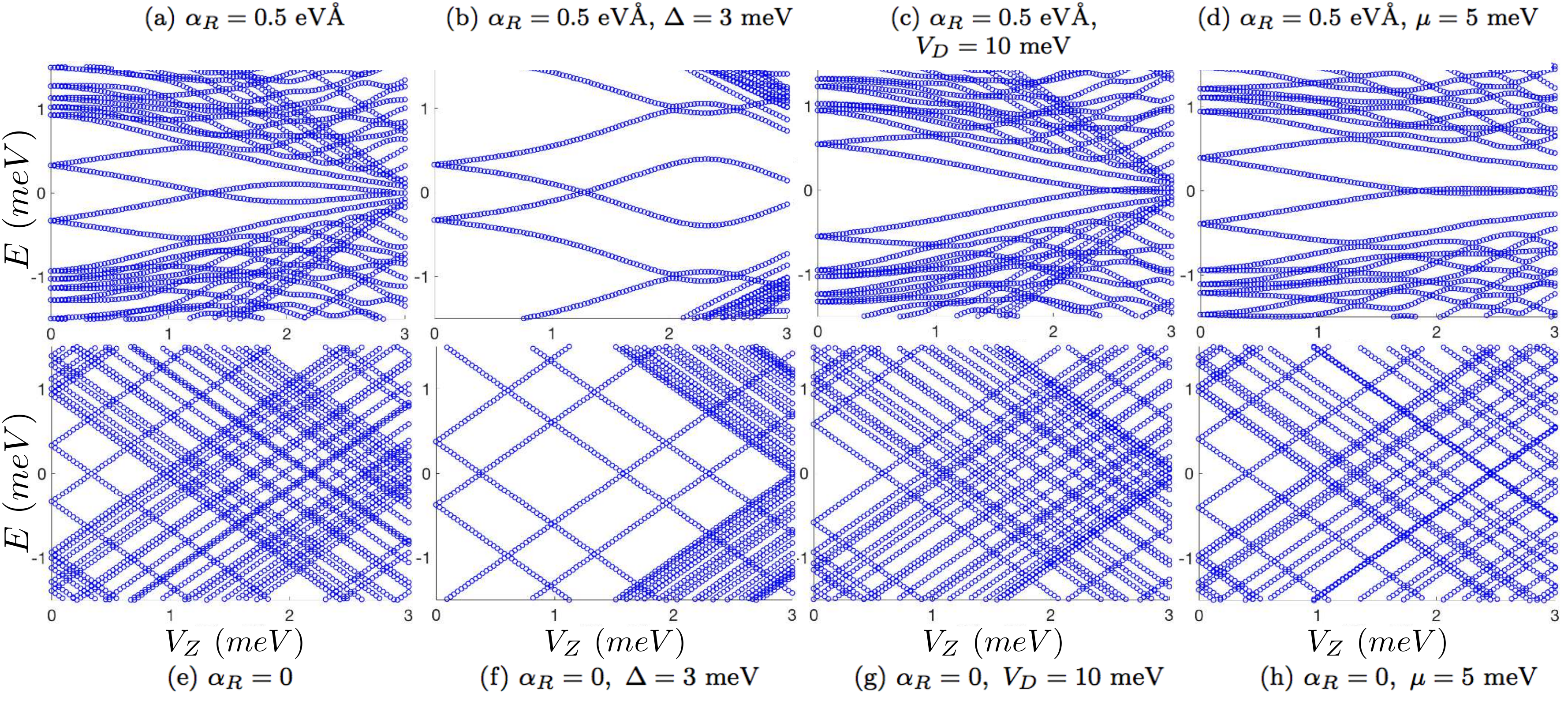}
\caption{(color online). Energy spectra of hybrid structures with and without spin-orbit coupling. The four panels in the upper row (a)-(d) show energy spectra as function of Zeeman field with SOC $\alpha_R=0.5~$eV$\angstrom$. Chemical potential is $\mu=3~$meV except that in (d) and (h) $\mu=5~$meV. The four panels in the lower row (e)-(h) show energy spectra without SOC, i.e., $\alpha_R=0$.}
\label{fig:with_withoutSOC}
\end{figure*}

Here we show the calculated energy spectra of hybrid structures with and without Zeeman spin splitting and spin-orbit coupling in Fig.~\ref{fig:with_withoutVZ} and~\ref{fig:with_withoutSOC}. As shown in the lower panels of Fig.~\ref{fig:with_withoutVZ}, spectra have no zero-energy states when the Zeeman splitting is turned off. On the other hand, as shown in the lower panels of Fig.~\ref{fig:with_withoutSOC}, the low-energy spectra without spin-orbit coupling are composed of straight lines. In these cases the energy spectra have a simple analytic form $E = V_Z \pm \sqrt{\epsilon^2 + \Delta^2}$, where $\epsilon$ is the eigen-energy of nanowire without Zeeman splitting and spin-orbit coupling, and the energy scales linearly with Zeeman field. It is clear that superconductivity along with both Zeeman splitting and spin-orbit coupling are necessary for obtaining low energy Andreev bound states sticking to the midgap.

%%%%%%%%%%%%%%%%%%%%%%%%%%%%

\section{Expansion of projected self-energy term in quantum dot subspace}\label{app:expansion}

We can constrain the form of the projected self-energy $F(\omega)$ making use of the particle-hole symmetry in the nanowire:
\begin{align}
P^{-1}H_{NW}P = - H_{NW},
\end{align}
where $P=\sigma_{y} \otimes \tau_y K $. Thus for any eigenstate $\ket{\psi_a}$ with eigenenergy $E$, there must be another state $\ket{\psi_{\bar{a}}}=P\ket{\psi_{a}}$ with eigenenergy $-E$. So applying particle-hole symmetry onto the projected self-energy $F(\omega)$ in the quantum dot subspace, we have
\begin{align}
F_{ab}(\omega) &= \braket{ \psi_a | u \frac{1}{\omega - H_{NW}} u^{\dagger} | \psi_b } \nn
&= \braket{ \psi_a | P^{-1}P u \frac{1}{\omega - H_{NW}} u^{\dagger} P^{-1}P | \psi_b } \nn
&= \braket{ \psi_{\bar{a}} | u \frac{1}{\omega + H_{NW}} u^{\dagger} | \psi_{\bar{b}} } \nn
&=-F_{\bar{a}\bar{b}}(-\omega).
\label{eq:symmetry_constraint}
\end{align}
If we expand the $2\times2$ matrix of $F(\omega)$ by Pauli matrices
\begin{align}
F(\omega) = f_0(\omega)\gamma_0 + f_x(\omega)\gamma_x + f_z(\omega)\gamma_z,
\end{align}
and it is easy to see that $f_{0,x}$ are odd functions of $\omega$, while $f_{z}$ is an even function of $\omega$ based on Eq.~\eqref{eq:symmetry_constraint}. The absence of $\gamma_y$ is due to the fact that the Hamiltonian $H_{NW}$ is accidentally real.

%%%%%%%%%%%%%%%%%%%%%%%%%%%%

\section{Spectra of ABSs and their wave-functions}\label{app:spectra_wf}

\begin{figure*}[!htb]
\includegraphics[width=0.99\textwidth]{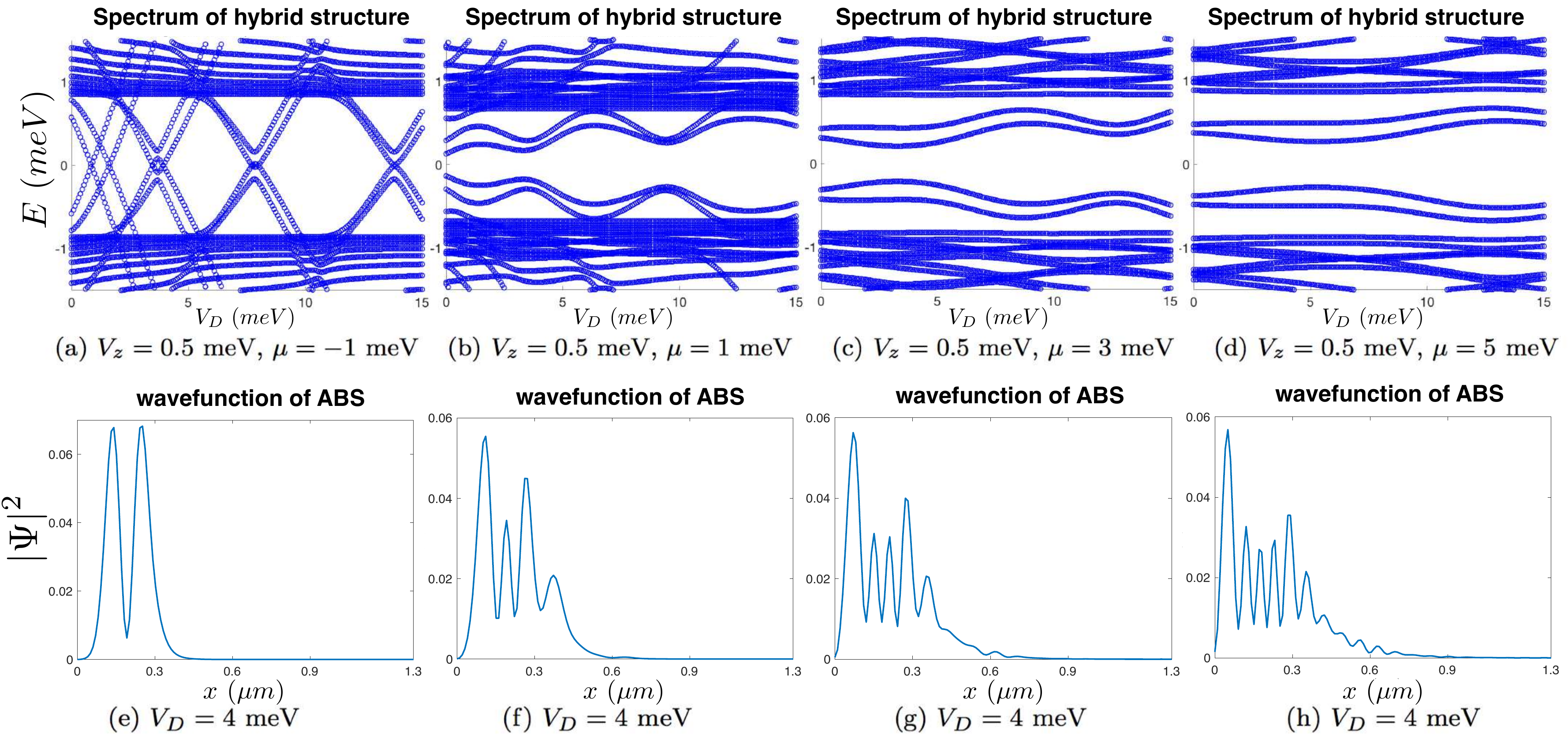}
\caption{(color online). Spectra of hybrid structures with unproximitized quantum dot (dot length $l \simeq 0.3~\mu$m ) as a function of the dot depth $V_D$ and the corresponding wave-functions of these ABSs. The upper panels are the spectra of increasing chemical potential (from left to right). The lower panels are the corresponding wave-functions.}
\label{fig:spectra_wf}
\end{figure*}

We show spectra of hybrid structures with unproximitized quantum dot (dot length $l \simeq 0.3~\mu$m ) as a function of the dot depth $V_D$ and the corresponding wave-functions of these ABSs in Fig.~\ref{fig:spectra_wf}. The upper panels are spectra, for which we focus on the spectra of ABSs within the induced SC gap. The trend is that at low chemical potential (upper left panels), the spectra of ABSs are quite sensitive to the depth of quantum dot, while at high chemical potential (upper right panels), the spectra of ABSs are insensitive to the depth of quantum dot. This can be understood by looking at the corresponding wave-functions, as shown in the lower panels. When the chemical potential is small, the wave-function (lower left panels) is quite localized inside the quantum dot, and therefore the ABS is easily affected by the dot depth. When the chemical potential is large, the wave-function becomes more extended, leaking well into the nanowire (lower right panels) due to a larger Fermi wave-vector. Thus a variation of dot depth affects only a fraction of the wave-function, leading to a minor change in the spectra.

%%%%%%%%%%%%%%%%%%%%%%%%%%%%

\section{Majorana oscillations and ABS oscillations}\label{app:oscillations}

\begin{figure*}[!htb]
\includegraphics[width=0.99\textwidth]{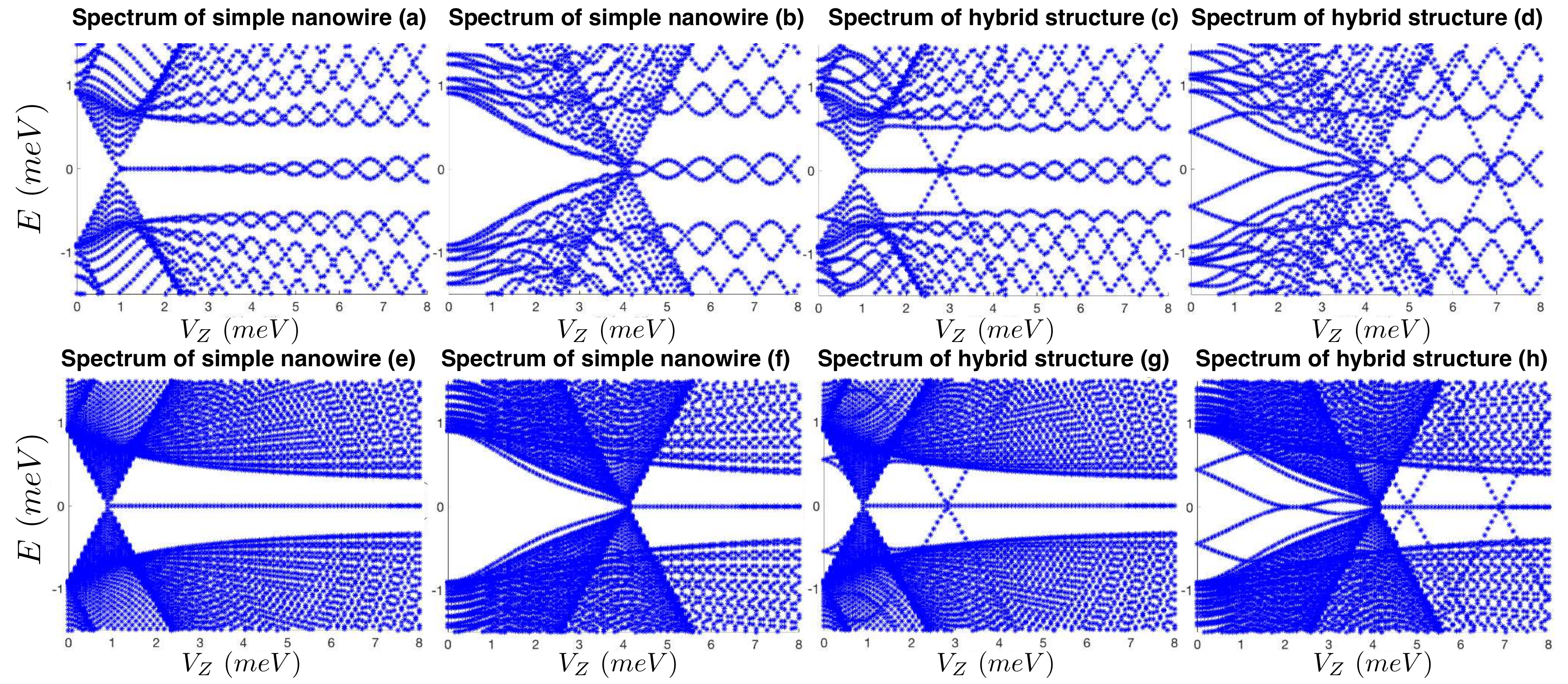}
\caption{(color online). Comparison of the energy spectra in simple nanowire and hybrid structure: the left four panels (a), (b), (e), (f) are energy spectra for simple nanowires and the right panels (c), (d), (g), (h) are energy spectra for hybrid structures with a quantum dot outside the SC nanowire. In all cases, topological MZM-induced ZBCPs form when $V_Z > \sqrt{\Delta^2+\mu^2}$, with $\mu=0$ for (a), (c), (e), (g), while $\mu=4~$meV for (b), (d), (f), (h). Upper panels (a)-(d) have shorter nanowire length $L=1~\mu$m, so we can see apparent Majorana oscillations due to Majorana overlap in contrast with lower panels (e)-(h) with longer wires $L=4~\mu$m and thus less Majorana overlap. For hybrid structures with large chemical potential (d), (f) we also see beating or oscillation patterns in the non-topological regime arising from Andreev bound states (e.g., $V_Z<\mu$). But these patterns are non-universal while the amplitude of Majorana oscillations has a universal trend of increasing with increasing $V_Z$(a)-(d). All parameters in the two systems (nanowire and hybrid) are the same except for the presence of a quantum dot outside the nanowire in the hybrid structure.}
\label{fig:oscillations}
\end{figure*}

We show here in Fig.~\ref{fig:oscillations} the calculated results for topological MZM and trivial ABS oscillations in simple nanowire and nanowire $+$ dot hybrid systems for two different wire lengths. In both cases, TQPT point is at $V_{Zc}=0.9~$meV for $\mu=0$ and at $V_{Zc} \simeq 4.1~$meV for $\mu=4.0~$meV. For hybrid structures with large chemical potential (Figs.~\ref{fig:oscillations}(d) and (f)) we see beating or oscillation patterns in the non-topological regime arising from Andreev bound states (e.g., $V_Z<\mu$). But these patterns are non-universal while the amplitude of Majorana oscillations (Figs.~\ref{fig:oscillations}(a)-(d)) has a universal trend of increasing with increasing $V_Z$. All parameters in the two systems (nanowire and hybrid) are the same except for the presence of a quantum dot outside the nanowire in the hybrid structure.

%%%%%%%%%%%%%%%%%%%%%%%%%%%%

\bibliography{BibMajorana}

\begin{thebibliography}{57}%
\makeatletter
\providecommand \@ifxundefined [1]{%
 \@ifx{#1\undefined}
}%
\providecommand \@ifnum [1]{%
 \ifnum #1\expandafter \@firstoftwo
 \else \expandafter \@secondoftwo
 \fi
}%
\providecommand \@ifx [1]{%
 \ifx #1\expandafter \@firstoftwo
 \else \expandafter \@secondoftwo
 \fi
}%
\providecommand \natexlab [1]{#1}%
\providecommand \enquote  [1]{``#1''}%
\providecommand \bibnamefont  [1]{#1}%
\providecommand \bibfnamefont [1]{#1}%
\providecommand \citenamefont [1]{#1}%
\providecommand \href@noop [0]{\@secondoftwo}%
\providecommand \href [0]{\begingroup \@sanitize@url \@href}%
\providecommand \@href[1]{\@@startlink{#1}\@@href}%
\providecommand \@@href[1]{\endgroup#1\@@endlink}%
\providecommand \@sanitize@url [0]{\catcode `\\12\catcode `\$12\catcode
  `\&12\catcode `\#12\catcode `\^12\catcode `\_12\catcode `\%12\relax}%
\providecommand \@@startlink[1]{}%
\providecommand \@@endlink[0]{}%
\providecommand \url  [0]{\begingroup\@sanitize@url \@url }%
\providecommand \@url [1]{\endgroup\@href {#1}{\urlprefix }}%
\providecommand \urlprefix  [0]{URL }%
\providecommand \Eprint [0]{\href }%
\providecommand \doibase [0]{http://dx.doi.org/}%
\providecommand \selectlanguage [0]{\@gobble}%
\providecommand \bibinfo  [0]{\@secondoftwo}%
\providecommand \bibfield  [0]{\@secondoftwo}%
\providecommand \translation [1]{[#1]}%
\providecommand \BibitemOpen [0]{}%
\providecommand \bibitemStop [0]{}%
\providecommand \bibitemNoStop [0]{.\EOS\space}%
\providecommand \EOS [0]{\spacefactor3000\relax}%
\providecommand \BibitemShut  [1]{\csname bibitem#1\endcsname}%
\let\auto@bib@innerbib\@empty
%</preamble>
\bibitem [{\citenamefont {Nayak}\ \emph {et~al.}(2008)\citenamefont {Nayak},
  \citenamefont {Simon}, \citenamefont {Stern}, \citenamefont {Freedman},\ and\
  \citenamefont {Das~Sarma}}]{Nayak2008Non-Abelian}%
  \BibitemOpen
  \bibfield  {author} {\bibinfo {author} {\bibfnamefont {C.}~\bibnamefont
  {Nayak}}, \bibinfo {author} {\bibfnamefont {S.~H.}\ \bibnamefont {Simon}},
  \bibinfo {author} {\bibfnamefont {A.}~\bibnamefont {Stern}}, \bibinfo
  {author} {\bibfnamefont {M.}~\bibnamefont {Freedman}}, \ and\ \bibinfo
  {author} {\bibfnamefont {S.}~\bibnamefont {Das~Sarma}},\ }\href {\doibase
  10.1103/RevModPhys.80.1083} {\bibfield  {journal} {\bibinfo  {journal} {Rev.
  Mod. Phys.}\ }\textbf {\bibinfo {volume} {80}},\ \bibinfo {pages} {1083}
  (\bibinfo {year} {2008})}\BibitemShut {NoStop}%
\bibitem [{\citenamefont {Sarma}\ \emph {et~al.}(2015)\citenamefont {Sarma},
  \citenamefont {Freedman},\ and\ \citenamefont
  {Nayak}}]{DasSarma2015Majorana}%
  \BibitemOpen
  \bibfield  {author} {\bibinfo {author} {\bibfnamefont {S.~D.}\ \bibnamefont
  {Sarma}}, \bibinfo {author} {\bibfnamefont {M.}~\bibnamefont {Freedman}}, \
  and\ \bibinfo {author} {\bibfnamefont {C.}~\bibnamefont {Nayak}},\ }\href
  {http://dx.doi.org/10.1038/npjqi.2015.1} {\bibfield  {journal} {\bibinfo
  {journal} {Npj Quantum Information}\ }\textbf {\bibinfo {volume} {1}},\
  \bibinfo {pages} {15001 EP } (\bibinfo {year} {2015})}\BibitemShut {NoStop}%
\bibitem [{\citenamefont {Alicea}(2012)}]{Alicea2012New}%
  \BibitemOpen
  \bibfield  {author} {\bibinfo {author} {\bibfnamefont {J.}~\bibnamefont
  {Alicea}},\ }\href {http://stacks.iop.org/0034-4885/75/i=7/a=076501}
  {\bibfield  {journal} {\bibinfo  {journal} {Rep. Prog. Phys.}\ }\textbf
  {\bibinfo {volume} {75}},\ \bibinfo {pages} {076501} (\bibinfo {year}
  {2012})}\BibitemShut {NoStop}%
\bibitem [{\citenamefont {Elliott}\ and\ \citenamefont
  {Franz}(2015)}]{Elliott2015Colloquium}%
  \BibitemOpen
  \bibfield  {author} {\bibinfo {author} {\bibfnamefont {S.~R.}\ \bibnamefont
  {Elliott}}\ and\ \bibinfo {author} {\bibfnamefont {M.}~\bibnamefont
  {Franz}},\ }\href {\doibase 10.1103/RevModPhys.87.137} {\bibfield  {journal}
  {\bibinfo  {journal} {Rev. Mod. Phys.}\ }\textbf {\bibinfo {volume} {87}},\
  \bibinfo {pages} {137} (\bibinfo {year} {2015})}\BibitemShut {NoStop}%
\bibitem [{\citenamefont {Stanescu}\ and\ \citenamefont
  {Tewari}(2013)}]{Stanescu2013Majorana}%
  \BibitemOpen
  \bibfield  {author} {\bibinfo {author} {\bibfnamefont {T.~D.}\ \bibnamefont
  {Stanescu}}\ and\ \bibinfo {author} {\bibfnamefont {S.}~\bibnamefont
  {Tewari}},\ }\href
  {http://iopscience.iop.org/article/10.1088/0953-8984/25/23/233201/meta}
  {\bibfield  {journal} {\bibinfo  {journal} {J. Phys.: Condens. Matter}\
  }\textbf {\bibinfo {volume} {25}},\ \bibinfo {pages} {233201} (\bibinfo
  {year} {2013})}\BibitemShut {NoStop}%
\bibitem [{\citenamefont {Leijnse}\ and\ \citenamefont
  {Flensberg}(2012)}]{Leijnse2012Introduction}%
  \BibitemOpen
  \bibfield  {author} {\bibinfo {author} {\bibfnamefont {M.}~\bibnamefont
  {Leijnse}}\ and\ \bibinfo {author} {\bibfnamefont {K.}~\bibnamefont
  {Flensberg}},\ }\href {http://stacks.iop.org/0268-1242/27/i=12/a=124003}
  {\bibfield  {journal} {\bibinfo  {journal} {Semicond. Sci. Technol.}\
  }\textbf {\bibinfo {volume} {27}},\ \bibinfo {pages} {124003} (\bibinfo
  {year} {2012})}\BibitemShut {NoStop}%
\bibitem [{\citenamefont {Beenakker}(2013)}]{Beenakker2013Search}%
  \BibitemOpen
  \bibfield  {author} {\bibinfo {author} {\bibfnamefont {C.}~\bibnamefont
  {Beenakker}},\ }\href {\doibase 10.1146/annurev-conmatphys-030212-184337}
  {\bibfield  {journal} {\bibinfo  {journal} {Annu. Rev. Condens. Matter
  Phys.}\ }\textbf {\bibinfo {volume} {4}},\ \bibinfo {pages} {113} (\bibinfo
  {year} {2013})}\BibitemShut {NoStop}%
\bibitem [{\citenamefont {Sau}\ \emph {et~al.}(2010{\natexlab{a}})\citenamefont
  {Sau}, \citenamefont {Lutchyn}, \citenamefont {Tewari},\ and\ \citenamefont
  {Das~Sarma}}]{Sau2010Generic}%
  \BibitemOpen
  \bibfield  {author} {\bibinfo {author} {\bibfnamefont {J.~D.}\ \bibnamefont
  {Sau}}, \bibinfo {author} {\bibfnamefont {R.~M.}\ \bibnamefont {Lutchyn}},
  \bibinfo {author} {\bibfnamefont {S.}~\bibnamefont {Tewari}}, \ and\ \bibinfo
  {author} {\bibfnamefont {S.}~\bibnamefont {Das~Sarma}},\ }\href {\doibase
  10.1103/PhysRevLett.104.040502} {\bibfield  {journal} {\bibinfo  {journal}
  {Phys. Rev. Lett.}\ }\textbf {\bibinfo {volume} {104}},\ \bibinfo {pages}
  {040502} (\bibinfo {year} {2010}{\natexlab{a}})}\BibitemShut {NoStop}%
\bibitem [{\citenamefont {Lutchyn}\ \emph {et~al.}(2010)\citenamefont
  {Lutchyn}, \citenamefont {Sau},\ and\ \citenamefont
  {Das~Sarma}}]{Lutchyn2010Majorana}%
  \BibitemOpen
  \bibfield  {author} {\bibinfo {author} {\bibfnamefont {R.~M.}\ \bibnamefont
  {Lutchyn}}, \bibinfo {author} {\bibfnamefont {J.~D.}\ \bibnamefont {Sau}}, \
  and\ \bibinfo {author} {\bibfnamefont {S.}~\bibnamefont {Das~Sarma}},\ }\href
  {\doibase 10.1103/PhysRevLett.105.077001} {\bibfield  {journal} {\bibinfo
  {journal} {Phys. Rev. Lett.}\ }\textbf {\bibinfo {volume} {105}},\ \bibinfo
  {pages} {077001} (\bibinfo {year} {2010})}\BibitemShut {NoStop}%
\bibitem [{\citenamefont {Oreg}\ \emph {et~al.}(2010)\citenamefont {Oreg},
  \citenamefont {Refael},\ and\ \citenamefont {von Oppen}}]{Oreg2010Helical}%
  \BibitemOpen
  \bibfield  {author} {\bibinfo {author} {\bibfnamefont {Y.}~\bibnamefont
  {Oreg}}, \bibinfo {author} {\bibfnamefont {G.}~\bibnamefont {Refael}}, \ and\
  \bibinfo {author} {\bibfnamefont {F.}~\bibnamefont {von Oppen}},\ }\href
  {\doibase 10.1103/PhysRevLett.105.177002} {\bibfield  {journal} {\bibinfo
  {journal} {Phys. Rev. Lett.}\ }\textbf {\bibinfo {volume} {105}},\ \bibinfo
  {pages} {177002} (\bibinfo {year} {2010})}\BibitemShut {NoStop}%
\bibitem [{\citenamefont {Sau}\ \emph {et~al.}(2010{\natexlab{b}})\citenamefont
  {Sau}, \citenamefont {Tewari}, \citenamefont {Lutchyn}, \citenamefont
  {Stanescu},\ and\ \citenamefont {Das~Sarma}}]{Sau2010Non}%
  \BibitemOpen
  \bibfield  {author} {\bibinfo {author} {\bibfnamefont {J.~D.}\ \bibnamefont
  {Sau}}, \bibinfo {author} {\bibfnamefont {S.}~\bibnamefont {Tewari}},
  \bibinfo {author} {\bibfnamefont {R.~M.}\ \bibnamefont {Lutchyn}}, \bibinfo
  {author} {\bibfnamefont {T.~D.}\ \bibnamefont {Stanescu}}, \ and\ \bibinfo
  {author} {\bibfnamefont {S.}~\bibnamefont {Das~Sarma}},\ }\href {\doibase
  10.1103/PhysRevB.82.214509} {\bibfield  {journal} {\bibinfo  {journal} {Phys.
  Rev. B}\ }\textbf {\bibinfo {volume} {82}},\ \bibinfo {pages} {214509}
  (\bibinfo {year} {2010}{\natexlab{b}})}\BibitemShut {NoStop}%
\bibitem [{\citenamefont {Mourik}\ \emph {et~al.}(2012)\citenamefont {Mourik},
  \citenamefont {Zuo}, \citenamefont {Frolov}, \citenamefont {Plissard},
  \citenamefont {Bakkers},\ and\ \citenamefont
  {Kouwenhoven}}]{Mourik2012Signatures}%
  \BibitemOpen
  \bibfield  {author} {\bibinfo {author} {\bibfnamefont {V.}~\bibnamefont
  {Mourik}}, \bibinfo {author} {\bibfnamefont {K.}~\bibnamefont {Zuo}},
  \bibinfo {author} {\bibfnamefont {S.~M.}\ \bibnamefont {Frolov}}, \bibinfo
  {author} {\bibfnamefont {S.}~\bibnamefont {Plissard}}, \bibinfo {author}
  {\bibfnamefont {E.~P. A.~M.}\ \bibnamefont {Bakkers}}, \ and\ \bibinfo
  {author} {\bibfnamefont {L.~P.}\ \bibnamefont {Kouwenhoven}},\ }\href
  {\doibase 10.1126/science.1222360} {\bibfield  {journal} {\bibinfo  {journal}
  {Science}\ }\textbf {\bibinfo {volume} {336}},\ \bibinfo {pages} {1003}
  (\bibinfo {year} {2012})}\BibitemShut {NoStop}%
\bibitem [{\citenamefont {Das}\ \emph {et~al.}(2012)\citenamefont {Das},
  \citenamefont {Ronen}, \citenamefont {Most}, \citenamefont {Oreg},
  \citenamefont {Heiblum},\ and\ \citenamefont {Shtrikman}}]{Das2012Zero}%
  \BibitemOpen
  \bibfield  {author} {\bibinfo {author} {\bibfnamefont {A.}~\bibnamefont
  {Das}}, \bibinfo {author} {\bibfnamefont {Y.}~\bibnamefont {Ronen}}, \bibinfo
  {author} {\bibfnamefont {Y.}~\bibnamefont {Most}}, \bibinfo {author}
  {\bibfnamefont {Y.}~\bibnamefont {Oreg}}, \bibinfo {author} {\bibfnamefont
  {M.}~\bibnamefont {Heiblum}}, \ and\ \bibinfo {author} {\bibfnamefont
  {H.}~\bibnamefont {Shtrikman}},\ }\href {http://dx.doi.org/10.1038/nphys2479}
  {\bibfield  {journal} {\bibinfo  {journal} {Nat. Phys.}\ }\textbf {\bibinfo
  {volume} {8}},\ \bibinfo {pages} {887} (\bibinfo {year} {2012})}\BibitemShut
  {NoStop}%
\bibitem [{\citenamefont {Deng}\ \emph {et~al.}(2012)\citenamefont {Deng},
  \citenamefont {Yu}, \citenamefont {Huang}, \citenamefont {Larsson},
  \citenamefont {Caroff},\ and\ \citenamefont {Xu}}]{Deng2012Anomalous}%
  \BibitemOpen
  \bibfield  {author} {\bibinfo {author} {\bibfnamefont {M.~T.}\ \bibnamefont
  {Deng}}, \bibinfo {author} {\bibfnamefont {C.~L.}\ \bibnamefont {Yu}},
  \bibinfo {author} {\bibfnamefont {G.~Y.}\ \bibnamefont {Huang}}, \bibinfo
  {author} {\bibfnamefont {M.}~\bibnamefont {Larsson}}, \bibinfo {author}
  {\bibfnamefont {P.}~\bibnamefont {Caroff}}, \ and\ \bibinfo {author}
  {\bibfnamefont {H.~Q.}\ \bibnamefont {Xu}},\ }\href {\doibase
  10.1021/nl303758w} {\bibfield  {journal} {\bibinfo  {journal} {Nano Lett.}\
  }\textbf {\bibinfo {volume} {12}},\ \bibinfo {pages} {6414} (\bibinfo {year}
  {2012})}\BibitemShut {NoStop}%
\bibitem [{\citenamefont {Churchill}\ \emph {et~al.}(2013)\citenamefont
  {Churchill}, \citenamefont {Fatemi}, \citenamefont {Grove-Rasmussen},
  \citenamefont {Deng}, \citenamefont {Caroff}, \citenamefont {Xu},\ and\
  \citenamefont {Marcus}}]{Churchill2013Superconductor}%
  \BibitemOpen
  \bibfield  {author} {\bibinfo {author} {\bibfnamefont {H.~O.~H.}\
  \bibnamefont {Churchill}}, \bibinfo {author} {\bibfnamefont {V.}~\bibnamefont
  {Fatemi}}, \bibinfo {author} {\bibfnamefont {K.}~\bibnamefont
  {Grove-Rasmussen}}, \bibinfo {author} {\bibfnamefont {M.~T.}\ \bibnamefont
  {Deng}}, \bibinfo {author} {\bibfnamefont {P.}~\bibnamefont {Caroff}},
  \bibinfo {author} {\bibfnamefont {H.~Q.}\ \bibnamefont {Xu}}, \ and\ \bibinfo
  {author} {\bibfnamefont {C.~M.}\ \bibnamefont {Marcus}},\ }\href {\doibase
  10.1103/PhysRevB.87.241401} {\bibfield  {journal} {\bibinfo  {journal} {Phys.
  Rev. B}\ }\textbf {\bibinfo {volume} {87}},\ \bibinfo {pages} {241401}
  (\bibinfo {year} {2013})}\BibitemShut {NoStop}%
\bibitem [{\citenamefont {Finck}\ \emph {et~al.}(2013)\citenamefont {Finck},
  \citenamefont {Van~Harlingen}, \citenamefont {Mohseni}, \citenamefont
  {Jung},\ and\ \citenamefont {Li}}]{Finck2013Anomalous}%
  \BibitemOpen
  \bibfield  {author} {\bibinfo {author} {\bibfnamefont {A.~D.~K.}\
  \bibnamefont {Finck}}, \bibinfo {author} {\bibfnamefont {D.~J.}\ \bibnamefont
  {Van~Harlingen}}, \bibinfo {author} {\bibfnamefont {P.~K.}\ \bibnamefont
  {Mohseni}}, \bibinfo {author} {\bibfnamefont {K.}~\bibnamefont {Jung}}, \
  and\ \bibinfo {author} {\bibfnamefont {X.}~\bibnamefont {Li}},\ }\href
  {\doibase 10.1103/PhysRevLett.110.126406} {\bibfield  {journal} {\bibinfo
  {journal} {Phys. Rev. Lett.}\ }\textbf {\bibinfo {volume} {110}},\ \bibinfo
  {pages} {126406} (\bibinfo {year} {2013})}\BibitemShut {NoStop}%
\bibitem [{\citenamefont {Albrecht}\ \emph {et~al.}(2016)\citenamefont
  {Albrecht}, \citenamefont {Higginbotham}, \citenamefont {Madsen},
  \citenamefont {Kuemmeth}, \citenamefont {Jespersen}, \citenamefont
  {Nyg{\aa}rd}, \citenamefont {Krogstrup},\ and\ \citenamefont
  {Marcus}}]{Albrecht2016Exponential}%
  \BibitemOpen
  \bibfield  {author} {\bibinfo {author} {\bibfnamefont {S.}~\bibnamefont
  {Albrecht}}, \bibinfo {author} {\bibfnamefont {A.}~\bibnamefont
  {Higginbotham}}, \bibinfo {author} {\bibfnamefont {M.}~\bibnamefont
  {Madsen}}, \bibinfo {author} {\bibfnamefont {F.}~\bibnamefont {Kuemmeth}},
  \bibinfo {author} {\bibfnamefont {T.}~\bibnamefont {Jespersen}}, \bibinfo
  {author} {\bibfnamefont {J.}~\bibnamefont {Nyg{\aa}rd}}, \bibinfo {author}
  {\bibfnamefont {P.}~\bibnamefont {Krogstrup}}, \ and\ \bibinfo {author}
  {\bibfnamefont {C.}~\bibnamefont {Marcus}},\ }\href
  {http://dx.doi.org/10.1038/nature17162} {\bibfield  {journal} {\bibinfo
  {journal} {Nature}\ }\textbf {\bibinfo {volume} {531}},\ \bibinfo {pages}
  {206} (\bibinfo {year} {2016})}\BibitemShut {NoStop}%
\bibitem [{\citenamefont {Chen}\ \emph {et~al.}(2016)\citenamefont {Chen},
  \citenamefont {Yu}, \citenamefont {Stenger}, \citenamefont {Hocevar},
  \citenamefont {Car}, \citenamefont {Plissard}, \citenamefont {Bakkers},
  \citenamefont {Stanescu},\ and\ \citenamefont
  {Frolov}}]{Chen2016Experimental}%
  \BibitemOpen
  \bibfield  {author} {\bibinfo {author} {\bibfnamefont {J.}~\bibnamefont
  {Chen}}, \bibinfo {author} {\bibfnamefont {P.}~\bibnamefont {Yu}}, \bibinfo
  {author} {\bibfnamefont {J.}~\bibnamefont {Stenger}}, \bibinfo {author}
  {\bibfnamefont {M.}~\bibnamefont {Hocevar}}, \bibinfo {author} {\bibfnamefont
  {D.}~\bibnamefont {Car}}, \bibinfo {author} {\bibfnamefont {S.~R.}\
  \bibnamefont {Plissard}}, \bibinfo {author} {\bibfnamefont {E.~P.}\
  \bibnamefont {Bakkers}}, \bibinfo {author} {\bibfnamefont {T.~D.}\
  \bibnamefont {Stanescu}}, \ and\ \bibinfo {author} {\bibfnamefont {S.~M.}\
  \bibnamefont {Frolov}},\ }\href {https://arxiv.org/abs/1610.04555} {\bibfield
   {journal} {\bibinfo  {journal} {arXiv:1610.04555}\ } (\bibinfo {year}
  {2016})}\BibitemShut {NoStop}%
\bibitem [{\citenamefont {Zhang}\ \emph {et~al.}(2016)\citenamefont {Zhang},
  \citenamefont {G{\"u}l} \emph {et~al.}}]{Zhang2016Ballistic}%
  \BibitemOpen
  \bibfield  {author} {\bibinfo {author} {\bibfnamefont {H.}~\bibnamefont
  {Zhang}}, \bibinfo {author} {\bibfnamefont {{\"O}.}~\bibnamefont {G{\"u}l}},
  \emph {et~al.},\ }\href {https://arxiv.org/abs/1603.04069} {\bibfield
  {journal} {\bibinfo  {journal} {arXiv:1603.04069}\ } (\bibinfo {year}
  {2016})}\BibitemShut {NoStop}%
\bibitem [{\citenamefont {Deng}\ \emph {et~al.}(2016)\citenamefont {Deng},
  \citenamefont {Vaitiekenas}, \citenamefont {Hansen}, \citenamefont {Danon},
  \citenamefont {Leijnse}, \citenamefont {Flensberg}, \citenamefont
  {Nyg{\aa}rd}, \citenamefont {Krogstrup},\ and\ \citenamefont
  {Marcus}}]{Deng2016Majorana}%
  \BibitemOpen
  \bibfield  {author} {\bibinfo {author} {\bibfnamefont {M.~T.}\ \bibnamefont
  {Deng}}, \bibinfo {author} {\bibfnamefont {S.}~\bibnamefont {Vaitiekenas}},
  \bibinfo {author} {\bibfnamefont {E.~B.}\ \bibnamefont {Hansen}}, \bibinfo
  {author} {\bibfnamefont {J.}~\bibnamefont {Danon}}, \bibinfo {author}
  {\bibfnamefont {M.}~\bibnamefont {Leijnse}}, \bibinfo {author} {\bibfnamefont
  {K.}~\bibnamefont {Flensberg}}, \bibinfo {author} {\bibfnamefont
  {J.}~\bibnamefont {Nyg{\aa}rd}}, \bibinfo {author} {\bibfnamefont
  {P.}~\bibnamefont {Krogstrup}}, \ and\ \bibinfo {author} {\bibfnamefont
  {C.~M.}\ \bibnamefont {Marcus}},\ }\href {\doibase 10.1126/science.aaf3961}
  {\bibfield  {journal} {\bibinfo  {journal} {Science}\ }\textbf {\bibinfo
  {volume} {354}},\ \bibinfo {pages} {1557} (\bibinfo {year}
  {2016})}\BibitemShut {NoStop}%
\bibitem [{\citenamefont {Sengupta}\ \emph {et~al.}(2001)\citenamefont
  {Sengupta}, \citenamefont {\ifmmode \check{Z}\else
  \v{Z}\fi{}uti\ifmmode~\acute{c}\else \'{c}\fi{}}, \citenamefont {Kwon},
  \citenamefont {Yakovenko},\ and\ \citenamefont
  {Das~Sarma}}]{Sengupta2001Midgap}%
  \BibitemOpen
  \bibfield  {author} {\bibinfo {author} {\bibfnamefont {K.}~\bibnamefont
  {Sengupta}}, \bibinfo {author} {\bibfnamefont {I.}~\bibnamefont {\ifmmode
  \check{Z}\else \v{Z}\fi{}uti\ifmmode~\acute{c}\else \'{c}\fi{}}}, \bibinfo
  {author} {\bibfnamefont {H.-J.}\ \bibnamefont {Kwon}}, \bibinfo {author}
  {\bibfnamefont {V.~M.}\ \bibnamefont {Yakovenko}}, \ and\ \bibinfo {author}
  {\bibfnamefont {S.}~\bibnamefont {Das~Sarma}},\ }\href {\doibase
  10.1103/PhysRevB.63.144531} {\bibfield  {journal} {\bibinfo  {journal} {Phys.
  Rev. B}\ }\textbf {\bibinfo {volume} {63}},\ \bibinfo {pages} {144531}
  (\bibinfo {year} {2001})}\BibitemShut {NoStop}%
\bibitem [{\citenamefont {Akhmerov}\ \emph {et~al.}(2009)\citenamefont
  {Akhmerov}, \citenamefont {Nilsson},\ and\ \citenamefont
  {Beenakker}}]{Akhmerov2009Electrically}%
  \BibitemOpen
  \bibfield  {author} {\bibinfo {author} {\bibfnamefont {A.~R.}\ \bibnamefont
  {Akhmerov}}, \bibinfo {author} {\bibfnamefont {J.}~\bibnamefont {Nilsson}}, \
  and\ \bibinfo {author} {\bibfnamefont {C.~W.~J.}\ \bibnamefont {Beenakker}},\
  }\href {\doibase 10.1103/PhysRevLett.102.216404} {\bibfield  {journal}
  {\bibinfo  {journal} {Phys. Rev. Lett.}\ }\textbf {\bibinfo {volume} {102}},\
  \bibinfo {pages} {216404} (\bibinfo {year} {2009})}\BibitemShut {NoStop}%
\bibitem [{\citenamefont {Law}\ \emph {et~al.}(2009)\citenamefont {Law},
  \citenamefont {Lee},\ and\ \citenamefont {Ng}}]{Law2009Majorana}%
  \BibitemOpen
  \bibfield  {author} {\bibinfo {author} {\bibfnamefont {K.~T.}\ \bibnamefont
  {Law}}, \bibinfo {author} {\bibfnamefont {P.~A.}\ \bibnamefont {Lee}}, \ and\
  \bibinfo {author} {\bibfnamefont {T.~K.}\ \bibnamefont {Ng}},\ }\href
  {\doibase 10.1103/PhysRevLett.103.237001} {\bibfield  {journal} {\bibinfo
  {journal} {Phys. Rev. Lett.}\ }\textbf {\bibinfo {volume} {103}},\ \bibinfo
  {pages} {237001} (\bibinfo {year} {2009})}\BibitemShut {NoStop}%
\bibitem [{\citenamefont {Flensberg}(2010)}]{Flensberg2010Tunneling}%
  \BibitemOpen
  \bibfield  {author} {\bibinfo {author} {\bibfnamefont {K.}~\bibnamefont
  {Flensberg}},\ }\href {\doibase 10.1103/PhysRevB.82.180516} {\bibfield
  {journal} {\bibinfo  {journal} {Phys. Rev. B}\ }\textbf {\bibinfo {volume}
  {82}},\ \bibinfo {pages} {180516} (\bibinfo {year} {2010})}\BibitemShut
  {NoStop}%
\bibitem [{\citenamefont {Lin}\ \emph {et~al.}(2012)\citenamefont {Lin},
  \citenamefont {Sau},\ and\ \citenamefont {Das~Sarma}}]{Lin2012Zero}%
  \BibitemOpen
  \bibfield  {author} {\bibinfo {author} {\bibfnamefont {C.-H.}\ \bibnamefont
  {Lin}}, \bibinfo {author} {\bibfnamefont {J.~D.}\ \bibnamefont {Sau}}, \ and\
  \bibinfo {author} {\bibfnamefont {S.}~\bibnamefont {Das~Sarma}},\ }\href
  {\doibase 10.1103/PhysRevB.86.224511} {\bibfield  {journal} {\bibinfo
  {journal} {Phys. Rev. B}\ }\textbf {\bibinfo {volume} {86}},\ \bibinfo
  {pages} {224511} (\bibinfo {year} {2012})}\BibitemShut {NoStop}%
\bibitem [{\citenamefont {Das~Sarma}\ \emph {et~al.}(2016)\citenamefont
  {Das~Sarma}, \citenamefont {Nag},\ and\ \citenamefont
  {Sau}}]{DasSarma2016How}%
  \BibitemOpen
  \bibfield  {author} {\bibinfo {author} {\bibfnamefont {S.}~\bibnamefont
  {Das~Sarma}}, \bibinfo {author} {\bibfnamefont {A.}~\bibnamefont {Nag}}, \
  and\ \bibinfo {author} {\bibfnamefont {J.~D.}\ \bibnamefont {Sau}},\ }\href
  {\doibase 10.1103/PhysRevB.94.035143} {\bibfield  {journal} {\bibinfo
  {journal} {Phys. Rev. B}\ }\textbf {\bibinfo {volume} {94}},\ \bibinfo
  {pages} {035143} (\bibinfo {year} {2016})}\BibitemShut {NoStop}%
\bibitem [{\citenamefont {Liu}\ \emph {et~al.}(2017)\citenamefont {Liu},
  \citenamefont {Sau},\ and\ \citenamefont {Das~Sarma}}]{Liu2017Role}%
  \BibitemOpen
  \bibfield  {author} {\bibinfo {author} {\bibfnamefont {C.-X.}\ \bibnamefont
  {Liu}}, \bibinfo {author} {\bibfnamefont {J.~D.}\ \bibnamefont {Sau}}, \ and\
  \bibinfo {author} {\bibfnamefont {S.}~\bibnamefont {Das~Sarma}},\ }\href
  {\doibase 10.1103/PhysRevB.95.054502} {\bibfield  {journal} {\bibinfo
  {journal} {Phys. Rev. B}\ }\textbf {\bibinfo {volume} {95}},\ \bibinfo
  {pages} {054502} (\bibinfo {year} {2017})}\BibitemShut {NoStop}%
\bibitem [{\citenamefont {Lee}\ \emph {et~al.}(2012)\citenamefont {Lee},
  \citenamefont {Jiang}, \citenamefont {Aguado}, \citenamefont {Katsaros},
  \citenamefont {Lieber},\ and\ \citenamefont {De~Franceschi}}]{Lee2012Zero}%
  \BibitemOpen
  \bibfield  {author} {\bibinfo {author} {\bibfnamefont {E.~J.~H.}\
  \bibnamefont {Lee}}, \bibinfo {author} {\bibfnamefont {X.}~\bibnamefont
  {Jiang}}, \bibinfo {author} {\bibfnamefont {R.}~\bibnamefont {Aguado}},
  \bibinfo {author} {\bibfnamefont {G.}~\bibnamefont {Katsaros}}, \bibinfo
  {author} {\bibfnamefont {C.~M.}\ \bibnamefont {Lieber}}, \ and\ \bibinfo
  {author} {\bibfnamefont {S.}~\bibnamefont {De~Franceschi}},\ }\href {\doibase
  10.1103/PhysRevLett.109.186802} {\bibfield  {journal} {\bibinfo  {journal}
  {Phys. Rev. Lett.}\ }\textbf {\bibinfo {volume} {109}},\ \bibinfo {pages}
  {186802} (\bibinfo {year} {2012})}\BibitemShut {NoStop}%
\bibitem [{\citenamefont {Liu}\ \emph {et~al.}(2012)\citenamefont {Liu},
  \citenamefont {Potter}, \citenamefont {Law},\ and\ \citenamefont
  {Lee}}]{Liu2012Zero}%
  \BibitemOpen
  \bibfield  {author} {\bibinfo {author} {\bibfnamefont {J.}~\bibnamefont
  {Liu}}, \bibinfo {author} {\bibfnamefont {A.~C.}\ \bibnamefont {Potter}},
  \bibinfo {author} {\bibfnamefont {K.~T.}\ \bibnamefont {Law}}, \ and\
  \bibinfo {author} {\bibfnamefont {P.~A.}\ \bibnamefont {Lee}},\ }\href
  {\doibase 10.1103/PhysRevLett.109.267002} {\bibfield  {journal} {\bibinfo
  {journal} {Phys. Rev. Lett.}\ }\textbf {\bibinfo {volume} {109}},\ \bibinfo
  {pages} {267002} (\bibinfo {year} {2012})}\BibitemShut {NoStop}%
\bibitem [{\citenamefont {Bagrets}\ and\ \citenamefont
  {Altland}(2012)}]{Bagrets2012Class}%
  \BibitemOpen
  \bibfield  {author} {\bibinfo {author} {\bibfnamefont {D.}~\bibnamefont
  {Bagrets}}\ and\ \bibinfo {author} {\bibfnamefont {A.}~\bibnamefont
  {Altland}},\ }\href {\doibase 10.1103/PhysRevLett.109.227005} {\bibfield
  {journal} {\bibinfo  {journal} {Phys. Rev. Lett.}\ }\textbf {\bibinfo
  {volume} {109}},\ \bibinfo {pages} {227005} (\bibinfo {year}
  {2012})}\BibitemShut {NoStop}%
\bibitem [{\citenamefont {Pikulin}\ \emph {et~al.}(2012)\citenamefont
  {Pikulin}, \citenamefont {Dahlhaus}, \citenamefont {Wimmer}, \citenamefont
  {Schomerus},\ and\ \citenamefont {Beenakker}}]{Pikulin2012Zero}%
  \BibitemOpen
  \bibfield  {author} {\bibinfo {author} {\bibfnamefont {D.}~\bibnamefont
  {Pikulin}}, \bibinfo {author} {\bibfnamefont {J.}~\bibnamefont {Dahlhaus}},
  \bibinfo {author} {\bibfnamefont {M.}~\bibnamefont {Wimmer}}, \bibinfo
  {author} {\bibfnamefont {H.}~\bibnamefont {Schomerus}}, \ and\ \bibinfo
  {author} {\bibfnamefont {C.}~\bibnamefont {Beenakker}},\ }\href
  {http://iopscience.iop.org/article/10.1088/1367-2630/14/12/125011/meta}
  {\bibfield  {journal} {\bibinfo  {journal} {New J. Phys.}\ }\textbf {\bibinfo
  {volume} {14}},\ \bibinfo {pages} {125011} (\bibinfo {year}
  {2012})}\BibitemShut {NoStop}%
\bibitem [{\citenamefont {Lee}\ \emph {et~al.}(2014)\citenamefont {Lee},
  \citenamefont {Jiang}, \citenamefont {Houzet}, \citenamefont {Aguado},
  \citenamefont {Lieber},\ and\ \citenamefont {De~Franceschi}}]{Lee2014Spin}%
  \BibitemOpen
  \bibfield  {author} {\bibinfo {author} {\bibfnamefont {E.~J.~H.}\
  \bibnamefont {Lee}}, \bibinfo {author} {\bibfnamefont {X.}~\bibnamefont
  {Jiang}}, \bibinfo {author} {\bibfnamefont {M.}~\bibnamefont {Houzet}},
  \bibinfo {author} {\bibfnamefont {R.}~\bibnamefont {Aguado}}, \bibinfo
  {author} {\bibfnamefont {C.~M.}\ \bibnamefont {Lieber}}, \ and\ \bibinfo
  {author} {\bibfnamefont {S.}~\bibnamefont {De~Franceschi}},\ }\href
  {http://dx.doi.org/10.1038/nnano.2013.267} {\bibfield  {journal} {\bibinfo
  {journal} {Nat. Nanotechnol.}\ }\textbf {\bibinfo {volume} {9}},\ \bibinfo
  {pages} {79} (\bibinfo {year} {2014})}\BibitemShut {NoStop}%
\bibitem [{\citenamefont {Mi}\ \emph {et~al.}(2014)\citenamefont {Mi},
  \citenamefont {Pikulin}, \citenamefont {Marciani},\ and\ \citenamefont
  {Beenakker}}]{Mi2014X}%
  \BibitemOpen
  \bibfield  {author} {\bibinfo {author} {\bibfnamefont {S.}~\bibnamefont
  {Mi}}, \bibinfo {author} {\bibfnamefont {D.}~\bibnamefont {Pikulin}},
  \bibinfo {author} {\bibfnamefont {M.}~\bibnamefont {Marciani}}, \ and\
  \bibinfo {author} {\bibfnamefont {C.}~\bibnamefont {Beenakker}},\ }\href
  {https://inis.iaea.org/search/search.aspx?orig_q=RN:47042298} {\bibfield
  {journal} {\bibinfo  {journal} {JETP}\ }\textbf {\bibinfo {volume} {119}},\
  \bibinfo {pages} {1018} (\bibinfo {year} {2014})}\BibitemShut {NoStop}%
\bibitem [{\citenamefont {Kells}\ \emph {et~al.}(2012)\citenamefont {Kells},
  \citenamefont {Meidan},\ and\ \citenamefont {Brouwer}}]{Kells2012Near}%
  \BibitemOpen
  \bibfield  {author} {\bibinfo {author} {\bibfnamefont {G.}~\bibnamefont
  {Kells}}, \bibinfo {author} {\bibfnamefont {D.}~\bibnamefont {Meidan}}, \
  and\ \bibinfo {author} {\bibfnamefont {P.~W.}\ \bibnamefont {Brouwer}},\
  }\href {\doibase 10.1103/PhysRevB.86.100503} {\bibfield  {journal} {\bibinfo
  {journal} {Phys. Rev. B}\ }\textbf {\bibinfo {volume} {86}},\ \bibinfo
  {pages} {100503} (\bibinfo {year} {2012})}\BibitemShut {NoStop}%
\bibitem [{\citenamefont {Stanescu}\ and\ \citenamefont
  {Tewari}(2014)}]{Stanescu2014Nonlocality}%
  \BibitemOpen
  \bibfield  {author} {\bibinfo {author} {\bibfnamefont {T.~D.}\ \bibnamefont
  {Stanescu}}\ and\ \bibinfo {author} {\bibfnamefont {S.}~\bibnamefont
  {Tewari}},\ }\href {\doibase 10.1103/PhysRevB.89.220507} {\bibfield
  {journal} {\bibinfo  {journal} {Phys. Rev. B}\ }\textbf {\bibinfo {volume}
  {89}},\ \bibinfo {pages} {220507} (\bibinfo {year} {2014})}\BibitemShut
  {NoStop}%
\bibitem [{\citenamefont {Moore}\ \emph {et~al.}(2016)\citenamefont {Moore},
  \citenamefont {Stanescu},\ and\ \citenamefont {Tewari}}]{Moore2016Majorana}%
  \BibitemOpen
  \bibfield  {author} {\bibinfo {author} {\bibfnamefont {C.}~\bibnamefont
  {Moore}}, \bibinfo {author} {\bibfnamefont {T.~D.}\ \bibnamefont {Stanescu}},
  \ and\ \bibinfo {author} {\bibfnamefont {S.}~\bibnamefont {Tewari}},\ }\href
  {https://arxiv.org/abs/1611.07058} {\bibfield  {journal} {\bibinfo  {journal}
  {arXiv:1611.07058}\ } (\bibinfo {year} {2016})}\BibitemShut {NoStop}%
\bibitem [{\citenamefont {Chiu}\ \emph {et~al.}(2017)\citenamefont {Chiu},
  \citenamefont {Sau},\ and\ \citenamefont {Sarma}}]{Chiu2017Conductance}%
  \BibitemOpen
  \bibfield  {author} {\bibinfo {author} {\bibfnamefont {C.-K.}\ \bibnamefont
  {Chiu}}, \bibinfo {author} {\bibfnamefont {J.~D.}\ \bibnamefont {Sau}}, \
  and\ \bibinfo {author} {\bibfnamefont {S.~D.}\ \bibnamefont {Sarma}},\ }\href
  {https://arxiv.org/abs/1702.04357} {\bibfield  {journal} {\bibinfo  {journal}
  {arXiv:1702.04357}\ } (\bibinfo {year} {2017})}\BibitemShut {NoStop}%
\bibitem [{Fle()}]{Flensberg2017Note}%
  \BibitemOpen
  \href@noop {} {}\bibinfo {note} {See the supplementary material in Ref.
  20}\BibitemShut {NoStop}%
\bibitem [{\citenamefont {Schuray}\ \emph {et~al.}(2017)\citenamefont
  {Schuray}, \citenamefont {Weithofer},\ and\ \citenamefont
  {Recher}}]{Schuray2017Fano}%
  \BibitemOpen
  \bibfield  {author} {\bibinfo {author} {\bibfnamefont {A.}~\bibnamefont
  {Schuray}}, \bibinfo {author} {\bibfnamefont {L.}~\bibnamefont {Weithofer}},
  \ and\ \bibinfo {author} {\bibfnamefont {P.}~\bibnamefont {Recher}},\ }\href
  {https://arxiv.org/abs/1702.03909} {\bibfield  {journal} {\bibinfo  {journal}
  {arXiv:1702.03909}\ } (\bibinfo {year} {2017})}\BibitemShut {NoStop}%
\bibitem [{\citenamefont {Clarke}(2017)}]{Clarke2017Experimentally}%
  \BibitemOpen
  \bibfield  {author} {\bibinfo {author} {\bibfnamefont {D.~J.}\ \bibnamefont
  {Clarke}},\ }\href {https://arxiv.org/abs/1702.01740} {\bibfield  {journal}
  {\bibinfo  {journal} {arXiv:1702.01740}\ } (\bibinfo {year}
  {2017})}\BibitemShut {NoStop}%
\bibitem [{\citenamefont {Prada}\ \emph {et~al.}(2017)\citenamefont {Prada},
  \citenamefont {Aguado},\ and\ \citenamefont {San-Jose}}]{Prada2017Measuring}%
  \BibitemOpen
  \bibfield  {author} {\bibinfo {author} {\bibfnamefont {E.}~\bibnamefont
  {Prada}}, \bibinfo {author} {\bibfnamefont {R.}~\bibnamefont {Aguado}}, \
  and\ \bibinfo {author} {\bibfnamefont {P.}~\bibnamefont {San-Jose}},\ }\href
  {https://arxiv.org/abs/1702.02525} {\bibfield  {journal} {\bibinfo  {journal}
  {arXiv:1702.02525}\ } (\bibinfo {year} {2017})}\BibitemShut {NoStop}%
\bibitem [{\citenamefont {Stanescu}\ \emph {et~al.}(2010)\citenamefont
  {Stanescu}, \citenamefont {Sau}, \citenamefont {Lutchyn},\ and\ \citenamefont
  {Das~Sarma}}]{Stanescu2010Proximity}%
  \BibitemOpen
  \bibfield  {author} {\bibinfo {author} {\bibfnamefont {T.~D.}\ \bibnamefont
  {Stanescu}}, \bibinfo {author} {\bibfnamefont {J.~D.}\ \bibnamefont {Sau}},
  \bibinfo {author} {\bibfnamefont {R.~M.}\ \bibnamefont {Lutchyn}}, \ and\
  \bibinfo {author} {\bibfnamefont {S.}~\bibnamefont {Das~Sarma}},\ }\href
  {\doibase 10.1103/PhysRevB.81.241310} {\bibfield  {journal} {\bibinfo
  {journal} {Phys. Rev. B}\ }\textbf {\bibinfo {volume} {81}},\ \bibinfo
  {pages} {241310} (\bibinfo {year} {2010})}\BibitemShut {NoStop}%
\bibitem [{\citenamefont {Groth}\ \emph {et~al.}(2014)\citenamefont {Groth},
  \citenamefont {Wimmer}, \citenamefont {Akhmerov},\ and\ \citenamefont
  {Waintal}}]{Kwant}%
  \BibitemOpen
  \bibfield  {author} {\bibinfo {author} {\bibfnamefont {C.~W.}\ \bibnamefont
  {Groth}}, \bibinfo {author} {\bibfnamefont {M.}~\bibnamefont {Wimmer}},
  \bibinfo {author} {\bibfnamefont {A.~R.}\ \bibnamefont {Akhmerov}}, \ and\
  \bibinfo {author} {\bibfnamefont {X.}~\bibnamefont {Waintal}},\ }\href
  {http://stacks.iop.org/1367-2630/16/i=6/a=063065} {\bibfield  {journal}
  {\bibinfo  {journal} {New J. Phys.}\ }\textbf {\bibinfo {volume} {16}},\
  \bibinfo {pages} {063065} (\bibinfo {year} {2014})}\BibitemShut {NoStop}%
\bibitem [{\citenamefont {Blonder}\ \emph {et~al.}(1982)\citenamefont
  {Blonder}, \citenamefont {Tinkham},\ and\ \citenamefont
  {Klapwijk}}]{Blonder1982Transition}%
  \BibitemOpen
  \bibfield  {author} {\bibinfo {author} {\bibfnamefont {G.~E.}\ \bibnamefont
  {Blonder}}, \bibinfo {author} {\bibfnamefont {M.}~\bibnamefont {Tinkham}}, \
  and\ \bibinfo {author} {\bibfnamefont {T.~M.}\ \bibnamefont {Klapwijk}},\
  }\href {\doibase 10.1103/PhysRevB.25.4515} {\bibfield  {journal} {\bibinfo
  {journal} {Phys. Rev. B}\ }\textbf {\bibinfo {volume} {25}},\ \bibinfo
  {pages} {4515} (\bibinfo {year} {1982})}\BibitemShut {NoStop}%
\bibitem [{\citenamefont {Setiawan}\ \emph {et~al.}(2015)\citenamefont
  {Setiawan}, \citenamefont {Brydon}, \citenamefont {Sau},\ and\ \citenamefont
  {Das~Sarma}}]{Setiawan2015Conductance}%
  \BibitemOpen
  \bibfield  {author} {\bibinfo {author} {\bibfnamefont {F.}~\bibnamefont
  {Setiawan}}, \bibinfo {author} {\bibfnamefont {P.~M.~R.}\ \bibnamefont
  {Brydon}}, \bibinfo {author} {\bibfnamefont {J.~D.}\ \bibnamefont {Sau}}, \
  and\ \bibinfo {author} {\bibfnamefont {S.}~\bibnamefont {Das~Sarma}},\ }\href
  {\doibase 10.1103/PhysRevB.91.214513} {\bibfield  {journal} {\bibinfo
  {journal} {Phys. Rev. B}\ }\textbf {\bibinfo {volume} {91}},\ \bibinfo
  {pages} {214513} (\bibinfo {year} {2015})}\BibitemShut {NoStop}%
\bibitem [{\citenamefont {Sau}\ \emph {et~al.}(2010{\natexlab{c}})\citenamefont
  {Sau}, \citenamefont {Lutchyn}, \citenamefont {Tewari},\ and\ \citenamefont
  {Das~Sarma}}]{Sau2010Robustness}%
  \BibitemOpen
  \bibfield  {author} {\bibinfo {author} {\bibfnamefont {J.~D.}\ \bibnamefont
  {Sau}}, \bibinfo {author} {\bibfnamefont {R.~M.}\ \bibnamefont {Lutchyn}},
  \bibinfo {author} {\bibfnamefont {S.}~\bibnamefont {Tewari}}, \ and\ \bibinfo
  {author} {\bibfnamefont {S.}~\bibnamefont {Das~Sarma}},\ }\href {\doibase
  10.1103/PhysRevB.82.094522} {\bibfield  {journal} {\bibinfo  {journal} {Phys.
  Rev. B}\ }\textbf {\bibinfo {volume} {82}},\ \bibinfo {pages} {094522}
  (\bibinfo {year} {2010}{\natexlab{c}})}\BibitemShut {NoStop}%
\bibitem [{\citenamefont {Reeg}\ and\ \citenamefont
  {Maslov}(2017)}]{Reeg2017Transport}%
  \BibitemOpen
  \bibfield  {author} {\bibinfo {author} {\bibfnamefont {C.}~\bibnamefont
  {Reeg}}\ and\ \bibinfo {author} {\bibfnamefont {D.~L.}\ \bibnamefont
  {Maslov}},\ }\href {\doibase 10.1103/PhysRevB.95.205439} {\bibfield
  {journal} {\bibinfo  {journal} {Phys. Rev. B}\ }\textbf {\bibinfo {volume}
  {95}},\ \bibinfo {pages} {205439} (\bibinfo {year} {2017})}\BibitemShut
  {NoStop}%
\bibitem [{\citenamefont {Stanescu}\ and\ \citenamefont
  {Das~Sarma}(2017)}]{Stanescu2017Proximity}%
  \BibitemOpen
  \bibfield  {author} {\bibinfo {author} {\bibfnamefont {T.~D.}\ \bibnamefont
  {Stanescu}}\ and\ \bibinfo {author} {\bibfnamefont {S.}~\bibnamefont
  {Das~Sarma}},\ }\href {\doibase 10.1103/PhysRevB.96.014510} {\bibfield
  {journal} {\bibinfo  {journal} {Phys. Rev. B}\ }\textbf {\bibinfo {volume}
  {96}},\ \bibinfo {pages} {014510} (\bibinfo {year} {2017})}\BibitemShut
  {NoStop}%
\bibitem [{\citenamefont {Cheng}\ \emph {et~al.}(2009)\citenamefont {Cheng},
  \citenamefont {Lutchyn}, \citenamefont {Galitski},\ and\ \citenamefont
  {Das~Sarma}}]{Cheng2009Splitting}%
  \BibitemOpen
  \bibfield  {author} {\bibinfo {author} {\bibfnamefont {M.}~\bibnamefont
  {Cheng}}, \bibinfo {author} {\bibfnamefont {R.~M.}\ \bibnamefont {Lutchyn}},
  \bibinfo {author} {\bibfnamefont {V.}~\bibnamefont {Galitski}}, \ and\
  \bibinfo {author} {\bibfnamefont {S.}~\bibnamefont {Das~Sarma}},\ }\href
  {\doibase 10.1103/PhysRevLett.103.107001} {\bibfield  {journal} {\bibinfo
  {journal} {Phys. Rev. Lett.}\ }\textbf {\bibinfo {volume} {103}},\ \bibinfo
  {pages} {107001} (\bibinfo {year} {2009})}\BibitemShut {NoStop}%
\bibitem [{\citenamefont {Das~Sarma}\ \emph {et~al.}(2012)\citenamefont
  {Das~Sarma}, \citenamefont {Sau},\ and\ \citenamefont
  {Stanescu}}]{DasSarma2012Splitting}%
  \BibitemOpen
  \bibfield  {author} {\bibinfo {author} {\bibfnamefont {S.}~\bibnamefont
  {Das~Sarma}}, \bibinfo {author} {\bibfnamefont {J.~D.}\ \bibnamefont {Sau}},
  \ and\ \bibinfo {author} {\bibfnamefont {T.~D.}\ \bibnamefont {Stanescu}},\
  }\href {\doibase 10.1103/PhysRevB.86.220506} {\bibfield  {journal} {\bibinfo
  {journal} {Phys. Rev. B}\ }\textbf {\bibinfo {volume} {86}},\ \bibinfo
  {pages} {220506} (\bibinfo {year} {2012})}\BibitemShut {NoStop}%
\bibitem [{\citenamefont {Akhmerov}\ \emph {et~al.}(2011)\citenamefont
  {Akhmerov}, \citenamefont {Dahlhaus}, \citenamefont {Hassler}, \citenamefont
  {Wimmer},\ and\ \citenamefont {Beenakker}}]{Akhmerov2011Quantized}%
  \BibitemOpen
  \bibfield  {author} {\bibinfo {author} {\bibfnamefont {A.~R.}\ \bibnamefont
  {Akhmerov}}, \bibinfo {author} {\bibfnamefont {J.~P.}\ \bibnamefont
  {Dahlhaus}}, \bibinfo {author} {\bibfnamefont {F.}~\bibnamefont {Hassler}},
  \bibinfo {author} {\bibfnamefont {M.}~\bibnamefont {Wimmer}}, \ and\ \bibinfo
  {author} {\bibfnamefont {C.~W.~J.}\ \bibnamefont {Beenakker}},\ }\href
  {\doibase 10.1103/PhysRevLett.106.057001} {\bibfield  {journal} {\bibinfo
  {journal} {Phys. Rev. Lett.}\ }\textbf {\bibinfo {volume} {106}},\ \bibinfo
  {pages} {057001} (\bibinfo {year} {2011})}\BibitemShut {NoStop}%
\bibitem [{\citenamefont {Altland}\ and\ \citenamefont
  {Zirnbauer}(1997)}]{Altland1997Nonstandard}%
  \BibitemOpen
  \bibfield  {author} {\bibinfo {author} {\bibfnamefont {A.}~\bibnamefont
  {Altland}}\ and\ \bibinfo {author} {\bibfnamefont {M.~R.}\ \bibnamefont
  {Zirnbauer}},\ }\href {\doibase 10.1103/PhysRevB.55.1142} {\bibfield
  {journal} {\bibinfo  {journal} {Phys. Rev. B}\ }\textbf {\bibinfo {volume}
  {55}},\ \bibinfo {pages} {1142} (\bibinfo {year} {1997})}\BibitemShut
  {NoStop}%
\bibitem [{\citenamefont {Sau}\ and\ \citenamefont
  {Das~Sarma}(2013)}]{Sau2013Density}%
  \BibitemOpen
  \bibfield  {author} {\bibinfo {author} {\bibfnamefont {J.~D.}\ \bibnamefont
  {Sau}}\ and\ \bibinfo {author} {\bibfnamefont {S.}~\bibnamefont
  {Das~Sarma}},\ }\href {\doibase 10.1103/PhysRevB.88.064506} {\bibfield
  {journal} {\bibinfo  {journal} {Phys. Rev. B}\ }\textbf {\bibinfo {volume}
  {88}},\ \bibinfo {pages} {064506} (\bibinfo {year} {2013})}\BibitemShut
  {NoStop}%
\bibitem [{\citenamefont {Beenakker}(1997)}]{Beenakker1997Random}%
  \BibitemOpen
  \bibfield  {author} {\bibinfo {author} {\bibfnamefont {C.~W.~J.}\
  \bibnamefont {Beenakker}},\ }\href {\doibase 10.1103/RevModPhys.69.731}
  {\bibfield  {journal} {\bibinfo  {journal} {Rev. Mod. Phys.}\ }\textbf
  {\bibinfo {volume} {69}},\ \bibinfo {pages} {731} (\bibinfo {year}
  {1997})}\BibitemShut {NoStop}%
\bibitem [{\citenamefont {Wimmer}\ \emph {et~al.}(2011)\citenamefont {Wimmer},
  \citenamefont {Akhmerov}, \citenamefont {Dahlhaus},\ and\ \citenamefont
  {Beenakker}}]{Wimmer2011Quantum}%
  \BibitemOpen
  \bibfield  {author} {\bibinfo {author} {\bibfnamefont {M.}~\bibnamefont
  {Wimmer}}, \bibinfo {author} {\bibfnamefont {A.}~\bibnamefont {Akhmerov}},
  \bibinfo {author} {\bibfnamefont {J.}~\bibnamefont {Dahlhaus}}, \ and\
  \bibinfo {author} {\bibfnamefont {C.}~\bibnamefont {Beenakker}},\ }\href@noop
  {} {\bibfield  {journal} {\bibinfo  {journal} {New Journal of Physics}\
  }\textbf {\bibinfo {volume} {13}},\ \bibinfo {pages} {053016} (\bibinfo
  {year} {2011})}\BibitemShut {NoStop}%
\bibitem [{Kou()}]{Kouwenhoven2017PrivateCom}%
  \BibitemOpen
  \href@noop {} {}\bibinfo {note} {L. Kouwenhoven, APS March Meeting talk at
  New Orleans, March 2017}\BibitemShut {NoStop}%
\bibitem [{Mar()}]{Marcus2017PrivateCom}%
  \BibitemOpen
  \href@noop {} {}\bibinfo {note} {C. Marcus, APS March Meeting talk at New
  Orleans, March 2017}\BibitemShut {NoStop}%
\end{thebibliography}%

%%%%%%%%%%%%%%%%%%%%%%%%%%%%

\end{document}